\documentclass[letterpaper,11pt]{article}
\usepackage{jheppub} 
\usepackage{lineno}
\usepackage[utf8]{inputenc}

\pdfoutput=1 
\usepackage{jheppub}
\usepackage{graphicx,verbatim}
\usepackage{blindtext}
\usepackage[T1]{fontenc}
\usepackage{epsf}
\usepackage{mathtools}
\usepackage{lmodern}
\usepackage[mathscr]{euscript}
\usepackage{ytableau}
\usepackage{youngtab}

\usepackage{amsmath}
\usepackage{amsfonts}
\usepackage{amssymb}
\usepackage{mathrsfs}
\usepackage{slashed}
\usepackage{xcolor}
\usepackage{tikz, float}
\usetikzlibrary{patterns,shapes.misc}

\def\kq{\mathfrak{q}}

\def\fR{\mathfrak{R}}

\def\rx{\mathrm{x}}

\def\ri{\mathrm{i}}
\def\rj{\mathrm{j}}

\def\bz{\mathbf{z}}

\def\bx{\mathbf{x}}
\def\ba{\mathbf{a}}

\def\bT{\mathbf{T}}

\def\bN{\mathbf{N}}
\def\bK{\mathbf{K}}
\def\bS{\mathbf{S}}
\def\bX{\mathbf{X}}
\def\bC{\mathbf{C}}

\def\bL{\mathbf{L}}

\def\bV{\mathbf{V}}

\def\BC{\mathbb{C}}
\def\BR{\mathbb{R}}
\def\BE{\mathbb{E}}
\def\BZ{\mathbb{Z}}

\def\CalN{\mathcal{N}}

\def\CalR{\mathcal{R}}
\def\CalZ{\mathcal{Z}}

\def\CalW{\mathcal{W}}

\def\CalO{\mathcal{O}}

\def\CalC{\mathcal{C}}

\def\Tr{{\rm Tr}}

\def\ve{{\varepsilon}}

\def\tx{\mathtt{x}}

\def\ti{\mathtt{i}}

\def\tp{\mathtt{p}}
\def\tq{\mathtt{q}}

\def\tn{\mathtt{n}}

\def\sT{\mathsf{T}}

\def\sk{\mathsf{k}}
\def\sw{\mathsf{w}}
\def\sb{\mathsf{b}}
\def\sv{\mathsf{v}}

\def\sU{\mathsf{U}}
\def\sH{\mathsf{H}}
\def\sV{\mathsf{V}}

\def\sG{\mathsf{G}}

\def\EQ{\EuScript{Q}}

\def\EY{\EuScript{Y}}
\def\EX{\EuScript{X}}

\def\EH{\EuScript{H}}

\def\ES{\EuScript{S}}

\def\ET{\EuScript{T}}

\def\ED{\EuScript{D}}
\def\EK{\EuScript{K}}
\def\EW{\EuScript{W}}

 \def\p{\partial}
 
 \def\a{\alpha}
 
 \def\g{\gamma}
 \def\d{\delta}

 \def\o{\omega }

  \def\sh{\text{sh}}

\def\beq{\begin{equation}}
\def\eeq{\end{equation}}

\newcommand{\NL}[1]{{\textcolor{blue}{(NL: #1)}}}

\title{New dimer integrable systems and defects in five dimensional gauge theory}
\author[{a}]{Norton Lee}
\affiliation[a]{Center for Geometry and Physics, Institute for Basic Science (IBS), \\ Pohang, 37673, Korea}
\emailAdd{norton.lee@ibs.re.kr}

\preprint{CGP-23022}

\vspace{2cm}
\abstract{
We study the relation between the quantum integrable systems derived from the dimer graphs and five dimensional $\CalN=1$ supersymmetric gauge theories on $S^1 \times \BR^4$. 
We construct integrable systems based on new dimer graphs obtained from modification of hexagon dimer diagram. 
We study the gauge theories in correspondence to the newly proposed integrable systems. By examining three types of defects -- a line defect, a canonical co-dimensional two defect and a monodromy defect -- in five-dimensional gauge theory with $\CalN=1$ supersymmetry and $\Omega_{\ve_1,\ve_2}$-background. We identify, in the $\ve_2 \to 0$ limit, the canonical co-dimensional two defect satisfying the Baxter T-Q equation of the generalized $A$-type dimer integrable system, and the monodromy defect as its common eigenstate of the commuting Hamiltonians, with the eigenvalues being the expectation value of the BPS Wilson loop in the anti-symmetric representation of the bulk gauge group. 
}

\begin{document}

\maketitle

\section{Introduction}

The intimite relation between the supersymmetric gauge theory and integrable system with finite number of degrees of freedom has been a fertile research ground over the past few decades. 
The physical quantities (coupling, masses) turn out to be parameters of these finite dimensional system. This connection is known as the \emph{Bethe/gauge correspondence}.
The most well-known in this story is that the Seiberg-Witten curve of $\CalN=2$ four dimensional gauge theory can be identified as the spectral curve of the integrable system. 



This remarkable correspondence was promoted to the relation between the quantum integrable systems and the $\Omega$-deformed four dimensional supersymmetric gauge theories with two parameters $\ve_1$ and $\ve_2$. The localization technique \cite{Nikita-Shatashvili, NRS2011} allows analytic computation of the partition function and certain BPS observables in a large class of $\CalN=2$ supersymmetric gauge theories in four dimension \cite{Nikita:I}. The $\ve_1,\ve_2 \to 0$ limit reveals the classical integrable system whose phase space is identified as the moduli space of some partial differential equation of gauge theoretic origin \cite{Nekrasov:2012xe}. In the Nekrasov-Shatashvili limit (NS-limit) $\ve_2 \to 0$ and $\ve_1=\hbar$, one expects to find the quantum version of the integrable system. The quantum Hamiltonians are identified with the twisted chiral ring of the two dimensional $\CalN=(2,2)$ gauge theories. 

This correspondence can be easily extended to supersymmetric gauge theory in 5 dimension with eight supercharges compactify on a circle. 
The Seiberg-Witten curve corresponds to relativistic integrable system where the compactified circle radius $R$ acting as inverse of speed of light \cite{Nekrasov:1996cz}. 
In the four-dimensional limit $R \to 0$
In particular, the Bethe/gauge correspondence is the tool to determine the Seiberg-Witten curve of the five dimensional gauge theories.

\subsection*{Dimer models and integrable systems} 
The integrable systems in correspondence to five dimensional $\CalN=1$ gauge theories are often relativistic due the the exponential kinetic and potential terms in the Hamiltonians from quantizing the mirror curve of the local Calabi-Yau geometry. 

In this paper we consider a class of cluster integrable systems proposed by Goncharov and Kenyon \cite{goncharov2011dimers}: integrable systems in correspondence with the dimer models on a torus. These integrable systems are the generalization of the affine $A$-type relativistic Toda lattice \cite{Eager:2011dp}. According to the correspondence, every dimer model defines an integrable system, whose conserving charges can be systematically calculated based on the perfect matching of the bipartite graph. 

The dual graph of the dimer model is a planar, periodic quiver. The quiver gauge theory arises from the worldvolume of stack of D3 probling a singular, toric Calabi-Yau (CY) 3-fold \cite{Benvenuti:2004wx}. There is a string dual in type IIB on $AdS_5 \times X_5$. $X_5$ is the Einstein base of a six dimensional cone, which is Calabi-Yau if $X_5$ is Sasaki-Einstein. The Sasaki-Einstein manifold $Y^{N,\sk}$ metric depends on two positive integer $(N,\sk)$, $\sk<N$. As such, the quiver gauge theory and the dimer model are also characterized by $Y^{N,\sk}$ system. 
For the $Y^{N,N}$ system it is well known Hexagon tiling \cite{Eager:2011dp}. 
The $Y^{N,\sk}$, $\sk<N$, systems are obtained by introducing impurity to the $Y^{N,N}$ quiver. The dimer graph then is obtained by taking the dual of the planar quiver \cite{Franco:2005rj,Benvenuti:2004dy}.

We find that by working with the dimer graph directly, new $Y^{N,k}[S]$ system associated to dimer graph different from the standard one from the planar quiver can be generated. These new dimer graphs are characterized by a finite set $S$ indicated by how the impurities are introduced to the $Y^{N,N}$ graph. 


\subsection*{Defects in 5d gauge theory}

In this work we focus on the 5d $\CalN=1$ super-Yang-Mills theory compactify on a circle with gauge group $SU(N)$ at Chern-Simon level $\sk$. 
We show matching between the gauge theory the Seiberg-Witten curve and the spectral curve of the $Y^{N,\sk}$ dimer system.
The quantization is achieved by subjecting the gauge theory to $\Omega_{\ve_1,\ve_2}$-background. 
We construct a half-BPS co-dimension two defect called $Q$-observable, which is defined by coupling three dimensional $\CalN=2$ gauged linear sigma model (supported on $S^1_R \times \BC_{\ve_1} \times \{0\} \subset S^1_R \times \BC_{\ve_1} \times \BC_{\ve_2}$) in the Coulomb phase.  
In the NS-limit with one of the $\Omega$-deformation is turned off $\ve_2 \to 0$, the $\CalN=2$ super-Poincar\'{e} symmetry is restored and the five dimensional $\CalN=1$ theory is now effectively described by the three dimensional $\CalN=2$ theory living on the co-dimensional two defect. Precisely speaking, the twisted superpotential of the two theories coincide when evaluated on their corresponding vacua \cite{Chen:2012we,HYC:2011,Dorey:2011pa}. 
We show that the vacuum expectation value of the $Q$-observable is the Baxter $Q$-operator of the $Y^{N,\sk}$ system. Its zeros satisfy the Bethe ansatz equation.

The second half-BPS defect we study is the Wilson line warpping the compactified $S^1_R$ direction. 
In the Bethe/gauge correspondence between the non-relativistic integrable system and the 4d $\CalN=2$ supersymmetric gauge theory, the quantum Hamiltonian of the integrable system is identified with the vacuum expectation value of the twisted chiral ring of the $\CalN=(2,2)$ theory, which describes the 4d theory in the NS-limit. 
In a relativistic case, the quantum Hamiltonians is identified as the expectation value of the Wilson loop of the 3d $\CalN=4$ Super Yang-Mills theory warping the compactified $S^1_R$ \cite{Kim:2016qqs,Tong:2014cha}. 

The generating function of the Wilson loops is a co-dimensional four line defect in the 5d gauge theory. It is the \emph{fundamental $qq$-character observable} first introduced in \cite{Nikita:I}. The main statement of \cite{Nikita:II} proves certain vanishing condition for the $qq$-character observables, known as the \emph{non-perturbative Dyson-Schwinger equations}.


The stationary states of the quantum integrable system, under Bethe/Gauge correspondence \cite{Nikita-Shatashvili,Nekrasov:2009ui}, are the vacua of the effective two-dimensional $\CalN=(2,2)$ gauge theory for non-relativistic integrable system, or three-dimensional $\CalN=2$ in the case of relativistic system. 
In order to obtain the wavefunction of the integrable system, we calculate the special local observable in the effecive three dimensional gauge theory -- a co-dimensional two defect in the five dimensional theory. 
This leads us to another half-BPS co-dimensional two defect: the regular surface defect \cite{nekrasov20042d}, also known as the Gukov-Witten monodromy-type defect \cite{Gukov:2008sn,Gukov:2006jk} defined by singular boundary condition along a surface, which can be modelled by the orbifold construction. 
The parameters of the co-dimensional two defect becomes the coordinates which the wavefunction depends on. 

We show that the vacuum expectation value of the regular monodromy defect in 5d $\CalN=1$ is the common eigenfunction of the commuting Hamiltonians of the $Y^{N,\sk}[S]$ dimer integrable system. 
This is done by introducing multiple defects to the gauge theory.
The regular defect is placed in parallel to the $Q$-observable defect, extending along $\BC_{\ve_1}$ at $\bz_2=0$. 
The correlators of these defects also obey the non-perturbative Dyson-Schwinger equation which comes from the compactness of the spike instanton \cite{Nikita:II,Nikita:III}, implying non-trivial equations on the gauge correlation functions \cite{Nikita:V,Jeong:2023qdr,jeong2021intersecting}. 
In the NS-limit, these Dyson-Schwinger equations either become Schr\"{o}dinger-type equations \cite{Chen:2019vvt, Chen:2020rxu} or can be used to construct the Lax matrices of the integrable system \cite{Lee:2020hfu,Jeong:2017pai,Jeong:2024hwf,Jeong:2024mxr},
leading to direct proof of the Bethe/guage correspondence and proving systematical methods for constructing the wave functions.




\subsection*{Outline}

This paper is organized as follows: 
In Section.~\ref{sec:RTL} we will give a quick review of the relativistic Toda lattice integrable system. 
In Section.~\ref{sec:Dimer} we review the construction of the dimer models, and derive the conserving Hamiltonians of the relativistic integrable system.  
In particular, we construct a family of new dimer models different from the standard ones \cite{goncharov2011dimers,Franco:2005rj}. 
In Section.~\ref{sec:gauge} we study the five dimensional $\CalN=1$ supersymmetric gauge theory and study the $qq$-character observable in the presence of regular surface defect. We prove, using the non-perturbative Dyson-Schwinger equations, that the defect partition function in the NS-limit is the eigenfunction of one of the conserving Hamiltonian of the integrable models constructed from the Dimer model.  
In Section.~\ref{sec:Baxter} we introduce both the orbifold defect and folded brane defect placed in parallel. 
We find the the correlators of the inserting surface defects obeying the non-perturbative Dyson-Schwinger equations. We identify them as a set of difference equations, which we call the \emph{fractional quantum T-Q equations}. These equations are the fractionalizations of the Baxter T-Q equation for the cluster integrable system constructed in Section.~\ref{sec:SD-glue} and can be used to construct the Lax matrices. Concatenating the Lax matrices gives the monodromy matrix, which is the generating function of all conserving Hamiltonians of the cluster integrable model. We prove the orbifold defect partition function is the stationary wave function shared by all conserving Hamiltonians. 


Finally we conclude our summary and point out some future direction in Section.~\ref{sec:Summary}.

\acknowledgments
The authors thank Saebyeok Jeong, Hee-Cheol Kim, Taro Kimura, Kimyeong Lee, Nikita Nekrasov, Yat-Hin Suen, and Xin Wang for discussion on related subjects. NL thanks the hospitality of Simons Center for Geometry and Physics where part of this work was done. 
The work of NL is supported by IBS project IBS-R003-D1. 

\section{Relativistic Toda lattice}\label{sec:RTL}

The relativistic Toda lattice describes $N$ particles on a circle. The $n$-th particle's position and moment are labeled by the hermitian operators $\hat\tq_n$ and $\hat\tp_n$ satisfying the commutation relation
\begin{align}
    \left[ \hat\tq_n, \hat\tp_m \right] =    \hbar_\text{RT} \delta_{n,m}, \ n,m=1,\dots,N . 
\end{align}
The model depends on a real parameter $R$, $\hbar = R \hbar_\text{RT}$. 
The Hamilton is 
\begin{align}\label{def:H-R-Toda}
    \hat{\rm H}_\text{RT} = \sum_{n=1}^N \left[ 1 + e^{\frac{-1}{2}\hbar} R^2 e^{\hat\tq_n - \hat\tq_{n+1}} \right] e^{R\hat\tp_n}.
\end{align}
The lattice is periodic, $\hat\tq_{N+n} = \hat\tq_n$. We can use the Baker–Campbell–Hausdorff formula
\begin{align}
    e^{\hat{A}} e^{\hat{B}} = e^{\hat{Z}}
\end{align}
with 
\begin{align}
    \hat{Z} = \hat{A} + \hat{B} + \frac{1}{2} [\hat{A},\hat{B}] + \frac{1}{12}[\hat{A},[\hat{A},\hat{B}]] - \frac{1}{12} [\hat{B},[\hat{A},\hat{B}]] + \cdots
\end{align}
to rewrite the Hamiltonian in the following form
\begin{align}
    \hat{\rm H}_\text{RT} = \sum_{n=1}^N e^{R\hat\tp_n} + R^2 e^{R \hat\tp_n + \hat\tq_n - \hat\tq_{n+1}}.
\end{align}
The relativistic Toda system is integrable. It has $N$ commuting Hamiltonians which can be obtained by constructing the Lax matrix
\begin{align}
    L_n (x,R) = \begin{pmatrix} x - \frac{e^{R\hat\tp_n}}{x} & Re^{-\hat\tq_n} \\ -Re^{\hat\tq_n+R\hat\tp_n} & 0 \end{pmatrix}.
\end{align}
The corresponding monodromy matrix is 
\begin{align}
    {\bf T}(x,R) = L_{N}(x,R) \cdots L_{1}(x,R),  
\end{align}
and the transfer matrix
\begin{align}
    T(x)= \Tr {\bf T}(x) = \sum_{n=0}^N (-1)^n x^{N-2n} \hat{\rm H}_n (\tq_1,\tp_1,\dots,\tq_N,\tp_N).
\end{align}
The $\hat{\rm H}_j$ are the commuting Hamiltonians. In particular
\begin{subequations}
\begin{align}
    & \hat{\rm H}_1 = \hat{\rm H}_\text{RT} \\
    & \hat{\rm H}_{N-1} = \sum_{n=1}^N \left[ 1 + e^{\frac{-1}{2}\hbar} R^2 e^{\hat\tq_{n-1} - \hat\tq_{n}} \right] e^{-R\hat\tp_n} \\
    & \hat{\rm H}_N = \exp \left( R \sum_{n=1}^N \hat\tp_n \right)
\end{align}
\end{subequations}

\paragraph{}
The non-relativistic limit can be obtained in the $R \to 0$ limit with $\hbar_\text{RT}$ fixed. The Hamiltonian \eqref{def:H-R-Toda} can be expand in power of $R$:
\begin{align}
\begin{split}
    \hat{\rm H}_\text{RT} =
    & \ N + R \sum_{n=1}^N \hat\tp_n + R^2 \left[ \sum_{n=1}^N \frac{\hat\tp_n^2}{2} + e^{\hat\tq_n-\hat\tq_{n+1}} \right]  \\
    & + \sum_{k=1}^{\infty} R^{k+2} \sum_{n=1}^N \left[ \frac{\hat\tp_n^{k+2}}{(k+2)!} + \frac{1}{k!} e^{\hat\tq_n-\hat\tq_{n+1}} \left(\hat\tp_n + \frac{\hbar_\text{RT}}{2i} \right)^k \right]
\end{split}
\end{align}
The coefficient of $R^1$ is the total momentum. The coefficient of $R^2$ is the non-relativistic Toda lattice Hamiltonian
\begin{align}
    {\rm H}_\text{Toda} = \sum_{n=1}^N \frac{\hat\tp^2_n}{2} + e^{\hat\tq_n-\hat\tq_{n+1}}.
\end{align}

\paragraph{}

The relativistic Toda lattices belong to a family of cluster integrable system proposed in \cite{goncharov2011dimers}. There are two ways to construct these integrable systems. The first one is through the dimer model, also known as the brane tiling. It is a study of perfect matching on a periodic bipartite graph embedded on a two torus, with the bipartite graph characterized by a Newton polygon $\mathscr{N}$. 
In \cite{goncharov2011dimers} it is proven that any dimer model on a bipartite graph on a torus give rise to a quantum integrable system. See \cite{ishii2016dimer} also for connection between the dimer model and Calabi-Yau three-fold. 

The dual of the dimer graph is a planar, periodic quiver. There is a one-to-one correspondence between the dimer graph and the periodic quiver based on the graph duality \cite{Franco:2005rj, Hanany:2005ve}. The dictionary can be summarized in the following table:

\begin{center}
    \begin{tabular}{|c|c|c|}
    \hline
    Dimer graph & Dual graph & Quiver gauge theory \\
    \hline \hline 
    Face & Vertex & Gauge group \\
    \hline
    Vertex & Face & Superpotential term \\
    \hline 
    Edge & Edge & Chiral superfield \\
    \hline
\end{tabular}
\end{center}

The quiver gauge theory described by the dimer model arises from the worldvolume of stack of D3 probing singular, toric Calabi-Yau (CY) 3-folds. The CY geometry, characterized by the same Newton polygon $\mathscr{N}$, emerges as the moduli space of vacua of the quiver gauge theory \cite{goncharov2011dimers,Hanany:2005ve,Franco:2005rj,Eager:2010ji}. 


This connects to the second way to construct the integrable model through the supersymmetric gauge theory associated to a bipartite torus graph. The key point of the construction is the identification between the spectral curve $\Sigma$ and the Seiberg-Witten curve $\Sigma$. This gauge theory can be construct by warpping an M5 brane on $\Sigma$. This M5 brane comes from the $(p,q)$-brane web in IIB. We compactify the $(p,q)$-brane on a circle of radius $R$ to pass through type IIA and then lift up to M-theory.  

\begin{figure}
    \centering
    \includegraphics[width=0.9\textwidth]{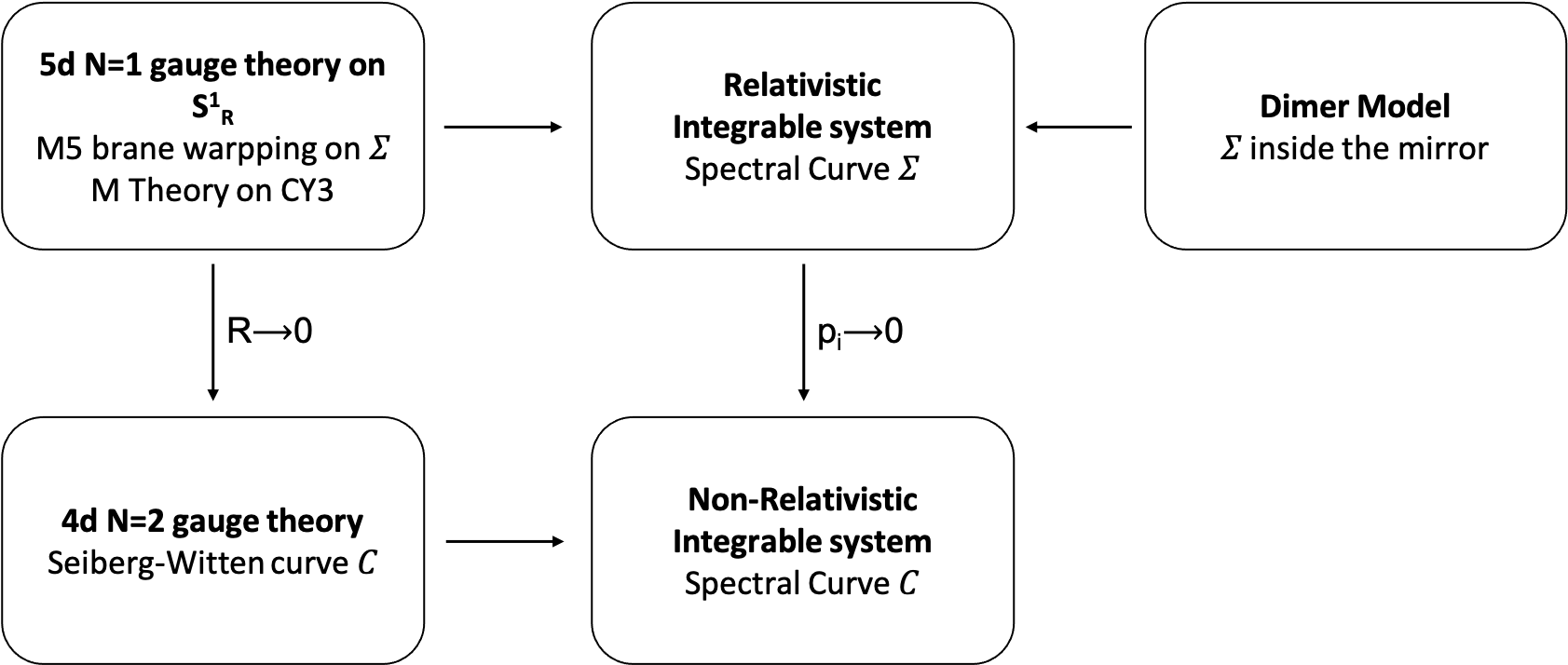}
    \caption{The relation between the dimer model, 5d $\CalN=1$ gauge theory compactified on a circle are based on different aspect to the spectral curve of $\Sigma$ the relativistic integrable model. In the $R\to0$ limit, the Seiberg-Witten curve $\CalC$ of the 4d $\CalN=2$ gauge theory is the spectral curve $\CalC$ of the non-relativistic integrable model.}
    \label{fig:theory-relation}
\end{figure}

The integrable model constructed from the dimer model (equivalently 5d gauge theory compactify on $S^1_R$) is relativistic by its nature. The radius $R$ of the compactified circle $S^1$ plays a role as inverse of speed of light. Taking the $R \to 0$ limit corresponds to taking the non-relativistic limit of the integrable model. In the gauge theory side, the limit $R\to 0$ gives 4d $\CalN=2$ supersymmetric gauge theory whose Seiberg-Witten curve $\CalC$ is the sepctral curve of the non-relativistic integrable theory. 
The relation between the theories are summerized in Figure.~\ref{fig:theory-relation}. 

\section{Dimer models and integrable systems}\label{sec:Dimer}

Dimer model is a study of the set of perfect matching on a graph $\Gamma$. For a bipartite graph, all the vertexes $\sv \in \Gamma$ are divided into two sets: the white vertexes $\sw_n$ and the black vertexes $\sb_n$. We choose the default orientation of edges such that each edge connects from one white vertex to a black vertex. The periodicity on the is realized by the unit of \emph{unit cell} on the torus.

\begin{figure}
    \centering
    \includegraphics[width=0.7\textwidth]{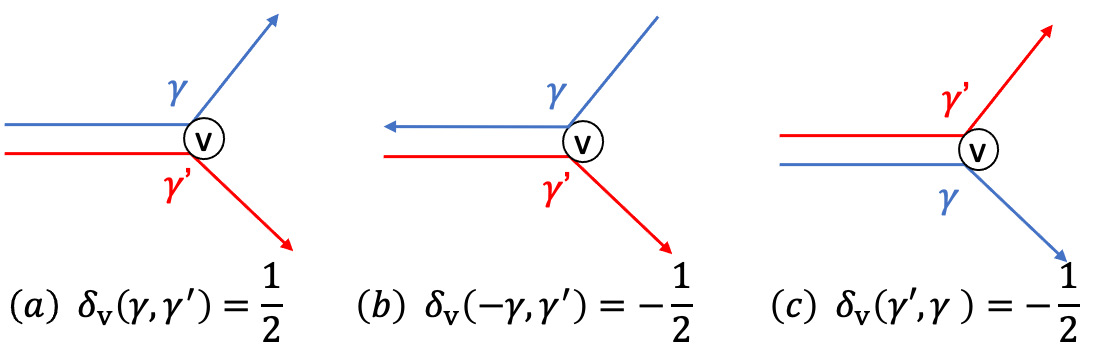}
    \caption{An illustration of $\delta_\sv(\g,\g')$. In $(a)$ the loops $\g$, $\g'$ are in the clockwise order with the same direction, $\d_\sv(\g,\g')=\frac{1}{2}$. Any change of the direction or clockwise order gives a sign. The arrow indicates the orientation of the loop.}
    \label{fig:delta-vertex}
\end{figure}

Let $\g$ and $\g'$ be an oriented loops on the dimer graph $\Gamma$, the Poisson commutator between the loops are defined as
\begin{align}\label{def:Poisson-loops}
    \{\g,\g'\} = \epsilon_{\g,\g'} \g \g'
\end{align}
where 
\begin{align}
    \epsilon_{\g,\g'} := \sum_{\sv} \text{sgn}(\sv) \delta_{\sv}(\g,\g'). 
\end{align}
The summation runs over the vertexes shared by the two loops $\g'$, $\g'$. 
Here $\text{sgn}(\sv=\sw)=1$ for the white vertexes, $\text{sgn}(\sv=\sb)=-1$ for the black vertexes. $\delta_\sv$ is a skew symmetric bilinear form with 
$$
    \d_\sv(w_n,w_m) = -\d_\sv(w_m,w_n) = - \d_{\sv}(-w_n,w_m) = \frac{1}{2} \BZ. 
$$
Here we define $\delta_\sv(\g,\g')=\frac{1}{2}$ if $\g$, $\g'$ are in the clockwise order and having the same direction, as illustrated in Figure.~\ref{fig:delta-vertex}. We will only encounter cubic vetexes in our study of dimer model. See \cite{goncharov2011dimers} for more general cases. 

As described in \cite{goncharov2011dimers}, the subtraction of perfect matching of the dimer model gives cycles. In order to construct the basis to all the loops. We will need to first select a perfect matching as the reference. Subtracting the reference perfect matching from another perfect matching gives the independent basis for the 1-loops. 

In this paper we will not discuss the choice of reference perfect matching in detail. Instead we will refer the interested readers to \cite{goncharov2011dimers}. 

\subsection{$Y^{N,0}$ model}\label{sec:YN0}

\begin{figure}
    \centering
    \includegraphics[width=0.5\textwidth]{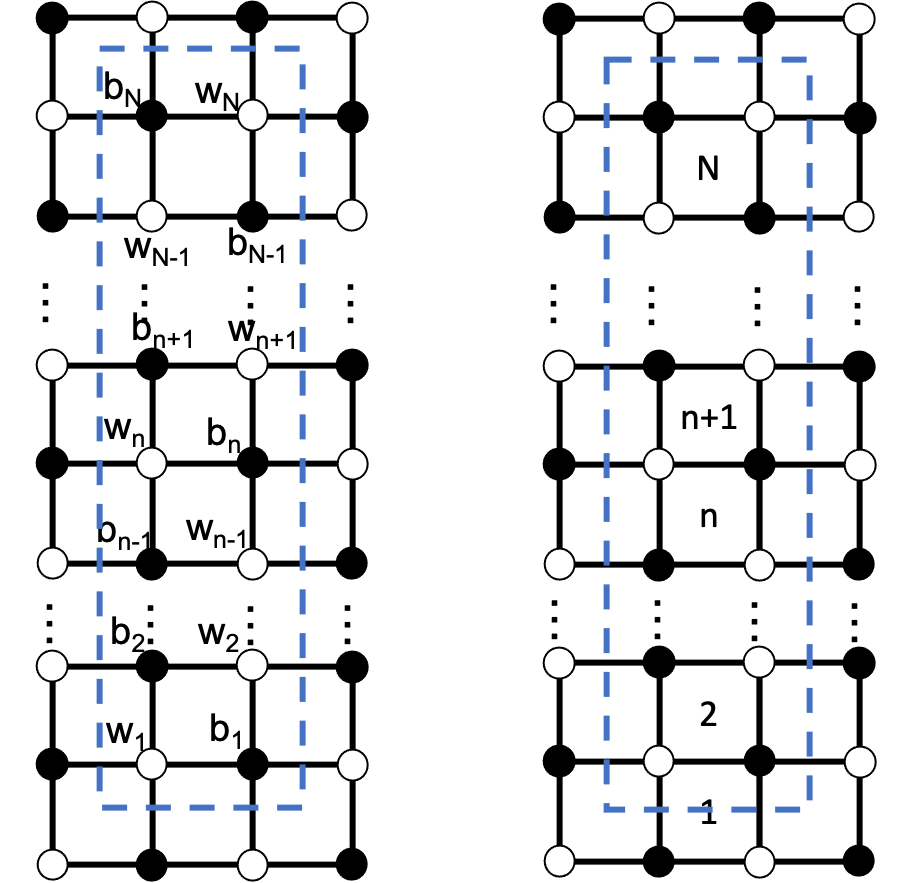}
    \caption{The brane tiling for $Y^{N,0}$ dimer model with $N$ even. A unit cell is encircled by the dashed blue line.}
    \label{fig:YN0}
\end{figure}

The $Y^{N,0}$ model is constructed by periodic square diagrams. The shape of the unit cell depends on whether $N$ is even (rectangular) or odd (rhombus). For concreteness, here we consider the case $N$ is even. Each unit cell consists two columns of $N$ squares along with $N$ white and black vertexes $\sw_n$, $\sb_n$, $n=1,\dots,N$. See Figure.~\ref{fig:YN0} for illustration.

\begin{figure}
    \centering
    \includegraphics[width=0.75\textwidth]{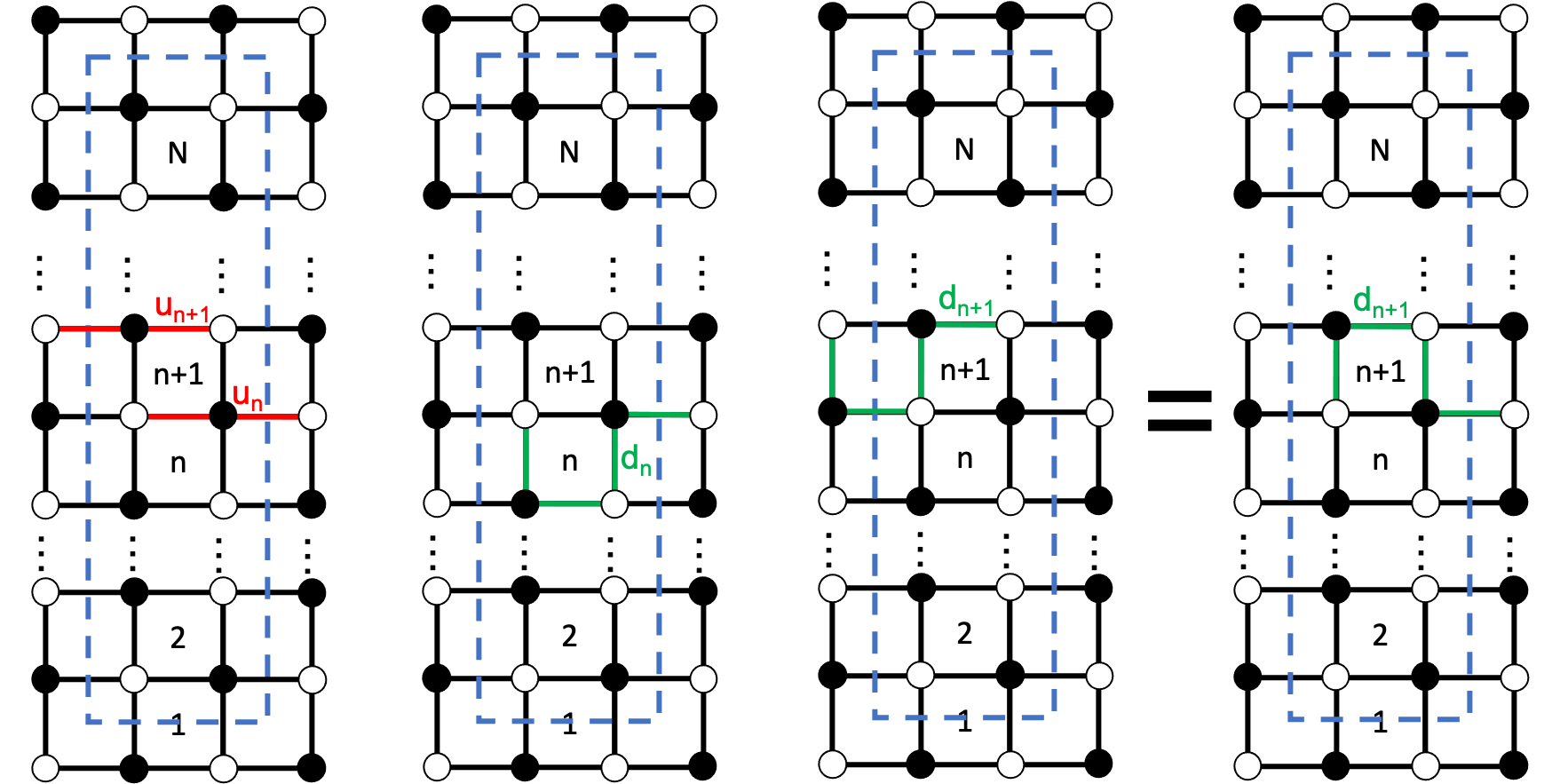}
    \caption{The 1-loops $u_n$ and $d_n$ on the square lattic for $Y^{N,0}$ dimer model with odd $n$ and $N$ even. }
    \label{fig:YN0-loop}
\end{figure}

A 1-loop on the squares diagram is a path connecting from a white vertex in a unit cell to the same white vertex in the next unit cell on the right (or left). To define the 1-loops, the reference perfect matching $M_f$ is chosen such that all dimers are placed horizontally, covering a black vertex on the left and white vertex on the right. 
Next we pick the two different perfect matching as follows
\begin{itemize}
    \item $M_1$: Dimers are placed horizontally with black vertex on the right and white vertex on the left. 
    \item $M_2$: Dimers are placed vertically in a unit cell, covering the vertical edges of the $n$-th square in the unit cell.
\end{itemize}
The $2N$ 1-loops are defined through the subtraction of reference perfect matching $M_f$ from the two perfect matching defined above: 
\begin{align}\label{def:bc}
    u_n, \  d_n, \ n=1,\dots,N.
\end{align}
The $u_n$ denotes the loop that goes straights right-forwardly based on $M_1-M_f$. $d_n$ are the loop that bends around the edge of the $n$-th squares, obtained from $M_2-M_f$. The bending always occurs at the squares whose top-left vertex is colored white. See Figure.~\ref{fig:YN0-loop} for illustration. 

The 1-loops obeys the Poisson commutation relation based on the overlapping of the shared edges based on \eqref{def:Poisson-loops}. Here we list out only the non-vanishing ones:
\begin{align}\label{eq:commu-bc}
    \{u_n,d_n\} = u_n d_n, \ \{d_{n+1},u_n\} = d_{n+1} u_n, \ \{d_{n+1},d_n\} = d_{n+1}d_n
\end{align}
with the periodicity $u_{n+N} = u_{n}$, $d_{n+N} = d_n$. 
The 1-loops can be expressed in terms of the canonical coordinates $\tp_n$, $\tq_n$ by
\begin{align}\label{def:ud-pq-N0}
    u_n = e^{-\tp_n}, \ d_n = R^2 e^{\tq_n - \tq_{n-1}} e^{-\frac{1}{2}(\tp_n+\tp_{n-1})}.
\end{align}
The canonical coordinates obey the Poisson commutation relation $\{\tq_n,\tp_m\}=\d_{n,m}$. 
The $n$-th Hamiltonian are defined as the sum of $n$-loops, where $n$-loops is the product of $n$ non-overlapping 1-loops. Here we list a few: 
\begin{subequations}
\begin{align}
    & H_1 = \sum_{n=1}^{N} u_n + d_n. \\
    & H_{N-1} = \sum_{n=1}^{N} u_{n+N-2} \cdots u_{n+1}(u_n+d_n). \\
    & H_N = u_1\dots u_N.
\end{align}
\end{subequations}

\subsubsection{Kastelyen matrix}
The dimer model can be connected to a toric diagram through the Kasteleyn matrix \cite{kenyon2003introduction}, which is a weighted adjacency matrix of the graph. The determinant of the Kasteleyn matrix happens to be the mirror curve of the Calabi-Yau three-fold \cite{Franco:2005rj,treumann2019kasteleyn}. 

Let us construct the Kasteleyn matrix for the dimer model in Figure.~\ref{fig:YN0}. Instead of assigning the explicit weight value to each of the edge as in \cite{kenyon2003introduction}, we will keep the assignment more general at this moment.  
It is convenient to label the edge according to whether they are horizontal ($\sH$ and $\tilde{\sH}$) or vertical ($\sV$ or $\tilde{\sV}$). $\sH$ edges are horizontal edges in the center of the unit cell. $\tilde{\sH}$ edges are horizontal edges go across the edge of the unit cell. $\sV$ edges are vertical edges that has black node on the top endpoints. $\tilde{\sV}$ edges are vertical edges that has white node on the top endpoint. 
A $0$ will be assigned to the Kasteleyn matrix component $K_{\sb_i,\sw_j}$ if there are no edge connecting between the white vertex $\sw_j$ and the black vertex $\sb_i$.  
We multiply the $\sV$ and $\tilde{\sV}$ edges a factor of $y$ if the edges cross through the top boundary of the unit cell, a factor of $y^{-1}$ if the edges cross through the bottom boundary of the unit cell. On the vertical direction we multiply a factor of $e^{-(-1)^ix}$ to $\tilde\sH$ edge to track which side of the unit cell the $\tilde{\sH}$ edge crosses. 
The Kasteleyn matrix is an $N \times N$ matrix whose components are 
\begin{align}\label{eq:K-matrix-N0}
    K_{\sb_i,\sw_j} = \begin{cases}
        (\sH_i+\tilde{\sH}_i e^{-(-1)^{i}x} ), & j=i; \\
        \tilde{\sV}_{\overline{j-1}} y^{-\d_{j,1}} , & i=j-1 \ \text{mod}(N); \\
        \sV_{j} y^{\d_{j,N}} & i=j+1, \ \text{mod}(N); \\
        0, & \text{otherwise}
    \end{cases}
\end{align}
The $(-1)^{i-1}$ appears on the diagonal of the Kasteleyn matrix $K_{\sb_i,\sw_i}$ keeps track of the orientation of the nodes. Remember that the default orientation is chosen from a white node to a black node. 

Kasteleyn matrix in \eqref{eq:K-matrix-N0} is tridiagonal. Its determinant can be rewritten in the spin-chain form using the following identity
\begin{align}\label{eq:KtoL}
    \det \begin{pmatrix} 
    A_1 & B_1 & & C_0 \\
    C_1 & \ddots & \ddots & \\
    & \ddots & \ddots & B_{n-1} \\
    B_0 & & C_{n-1} & A_n
    \end{pmatrix} 
    = (-1)^{n-1} \left[ \prod_j B_j + \prod_{j} C_j \right] + \Tr L_n \cdots L_{1} 
\end{align}
where
\begin{align}
    L_j = \begin{pmatrix}
        A_j & -B_{j-1}C_{j-1} \\ 1 & 0
    \end{pmatrix}.
\end{align}
The 1-loops given by the composition of the edges
\begin{align}
    u_n = \left( \sH_n \tilde\sH_{n}^{-1} \right)^{(-1)^{n-1}}, \ d_n =  \sH_n^{-1} \tilde\sV_{n-1} \tilde\sH_{n-1}^{-1} \sV_{n-1} 
\end{align}
for $n=1,\dots,N$. 
The Kasteleyn matrix elements $\sH$, $\tilde\sH$, $\sV$, and $\tilde\sV$ can be expressed in terms of $2N$ canonical coordinates $\tq_n$, $\tp_n$, $n=1,\dots,N$ by
\begin{align}
    \sH_n = e^{\frac{(-1)^n}{2} \tp_{n} }, \ \tilde\sH^{-1}_n = e^{\frac{(-1)^n}{2}\tp_{n}}, \ \tilde\sV_{n} \sV_{n} = R^2 e^{ (\tq_{n+1}-\tq_{n})}  
\end{align}
The determinant of the Kasteleyn matrix has the form 
\begin{align}
    (-1)^{n-1}\left[\prod_{n=1}^{N} \sV_n \right] y + (-1)^{n-1} \frac{\prod_{n=1}^N \tilde\sV_n }{y} + e^{\frac{N}{2}x} \prod_{i=1}^{\frac{N}{2}} \tilde\sH_{2i-1} \sH_{2i} \ T(e^{-x}) 
\end{align}
where 
\begin{align}
    T(e^{-x}) = e^{-Nx} + H_1 e^{-(N-1)x} + \cdots + H_N
\end{align}
is a degree $N$ polynomial in $e^{-x}$ whose coefficients are the conserving Hamiltonians. 

The reader may notice that the $d$-loop defined from the Kasteleyn matrix components are slightly different from what we wrote down in \eqref{def:ud-pq-N0}. But they still obeys the Poisson commutation relation in \eqref{eq:commu-bc}.

There is however a disadvantage of using the square lattice as shown in Fig.~\ref{fig:YN0} for the $Y^{N,0}$ model: $N$ must be even for the unit cell to be properly defined. Other wise the periodicity on the vertical direction is lost. 
There is an alternative way to define a $Y^{N,0}$ dimer model on a torus for all integer $N$. In order for the dimer graph to be constructed, we will need to start from the $Y^{N,N}$ dimer model, which will be discussed in the next section. 


\subsection{$Y^{N,N}$ model}\label{sec:YNN}
\begin{figure}
    \centering
    \includegraphics[width=0.5\textwidth]{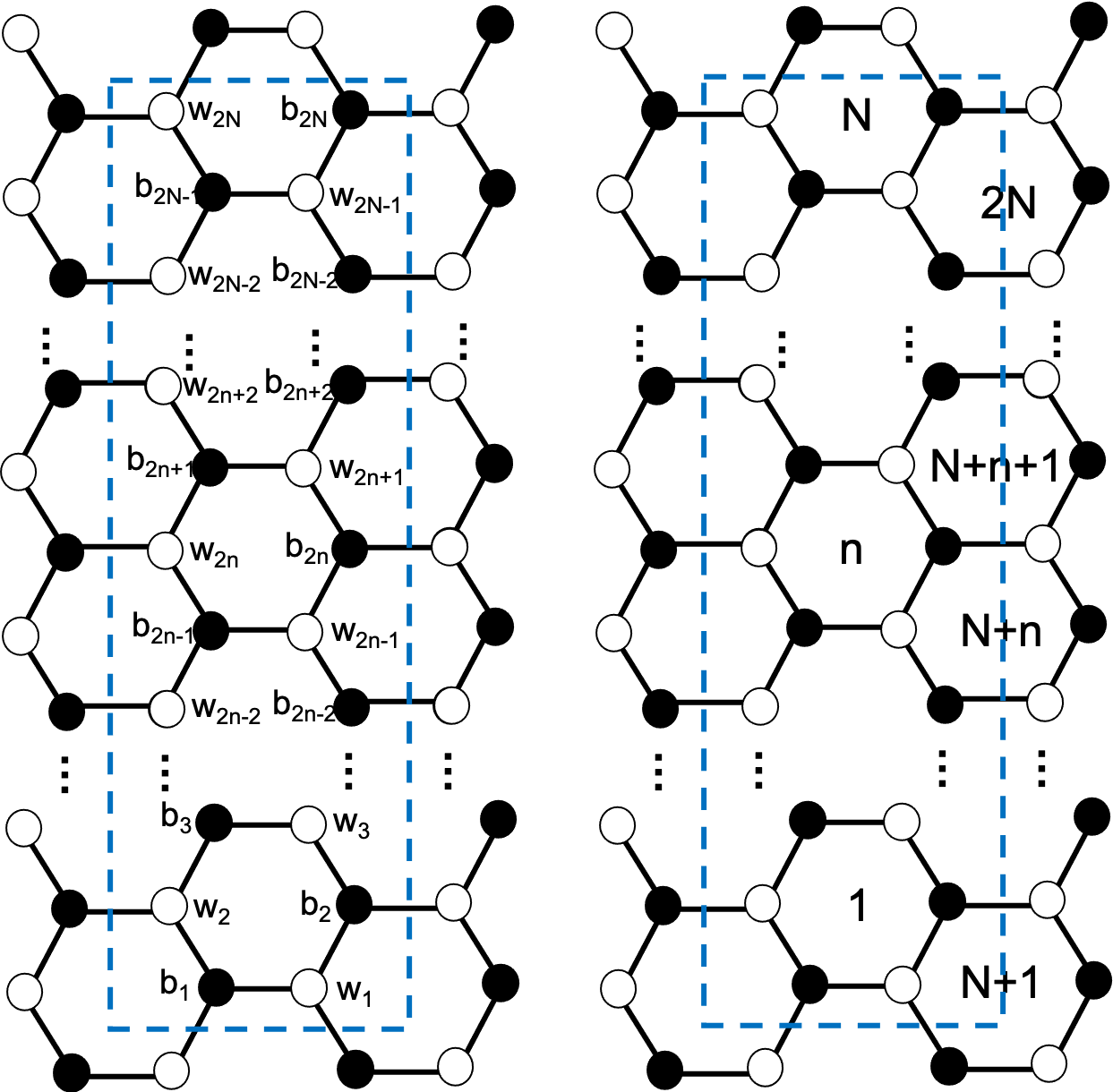}
    \caption{Brane tiling in $Y^{N,N}$ model realized in the hexagon graph.}
    \label{fig:YNN0}
\end{figure}

The dimer model for the $Y^{N,N}$ integrable system is well known to be the hexagon diagram \cite{Eager:2011dp}. Here $N$ can be either even or odd. 
Let us introduce notations for the $Y^{N,N}$ hexagon diagram (which consists left and right columns of $N$ hexagons in a beehive-stacking). Inside a unit cell there are $2N$ black nodes and $2N$ white nodes. We label the black nodes $\sb_i$ and white nodes $\sw_i$, $i=1,\dots,2N$, from bottom to the top as shown in Figure.~\ref{fig:YNN0}. 

The edges of hexagons in a unit cell can be divided into 4 groups
\begin{itemize}
    \item $\sH$: Horizontal edges in the center of the unit cell. 
    \item $\tilde{\sH}$: Horizontal edges across the boundary of the unit cell. 
    \item $\sU$: Top-right/bottom-left edges of a central hexagon. 
    \item $\sV$: Top-left/bottom-right edges of a central hexagon. 
\end{itemize}
The Kasteleyn matrix of $Y^{N,N}$ dimer model is  $2N \times 2N$ tri-diagonal matrix
\begin{align}
    K_{\sb_i,\sw_j} = \begin{cases}
        \sH_i & \text{, if } j=i, \ i\text{ is odd}; \\
        \tilde{\sH}_i e^{-x} & \text{, if } j=i, \ i\text{ is even}; \\ 
        \sU_{\overline{j-1}} \ y^{-\d_{j,1}} & \text{, if } i=\overline{j-1} \ \text{mod}(2N); \\
        \sV_{j} y^{\d_{j,N}} & \text{, if } i=\overline{j+1} \ \text{mod}(2N); \\
        0 & \text{otherwise}
    \end{cases}
\end{align}
with $i,j=1,\dots,2N$. 

\begin{figure}
    \centering
    \includegraphics[width=0.5\textwidth]{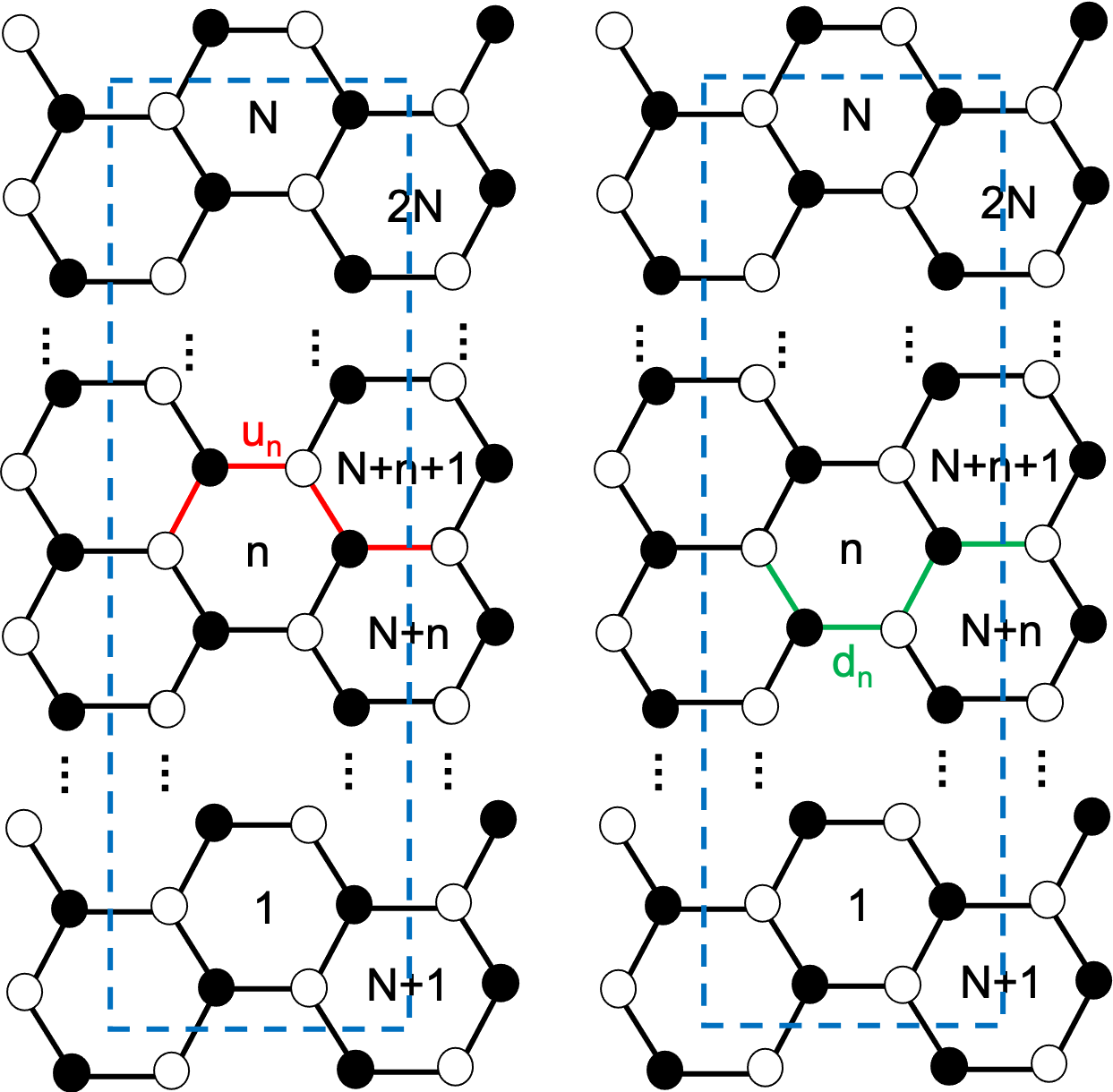}
    \caption{The $u_n$ and $d_n$ loops in a $Y^{N,N}$ brane tiling. }
    \label{fig:YNN}
\end{figure}

The 1-loops $u_n$ and $d_n$, connecting from the $2n$-th white vertex $\sw_{2n}$ to the same white vertex $\sw_{2n}$, $n=1,\dots,N$, in the next unit cell on the right, are
\begin{itemize}
    \item $u_n = \sw_{2n} \to \sb_{2n+1} \to \sw_{2n+1} \to \sb_{2n} \to \sw_{2n} = \tilde{\sH}_{2n}^{-1}\sU_{2n}\sH_{2n+1}^{-1}\sV_{2n}$, 
    \item $d_n = \sw_{2n} \to \sb_{2n-1} \to \sw_{2n-1} \to \sb_{2n} \to \sw_{2n} = \bar{\sH}_{2n}^{-1}\sV_{2n-1} \sH_{2n-1}^{-1}\sU_{2n-1}$.
\end{itemize}
Here $u_n$ denotes the loop that goes the top of the hexagon, the $d_n$ denotes the loop that goes to the bottom of the hexagon. See Figure.~\ref{fig:YNN} for illustration. 

The 1-loops have non-vanishing Poisson bracket if they overlap following \eqref{def:Poisson-loops}. The non-vanishing Poisson relations are 
\begin{align}
    \{u_{n},d_{n}\} = u_{n}d_{n}, \ \{d_{\overline{n+1}},u_n\} =  d_{\overline{n+1}} u_n, \ n=1,\dots,N. 
\end{align}
The $n$-th Hamiltonian is the sum over all $n$-loops, where the $n$-loops are the product of $n$ non-overlapping 1-loops. In particular
\begin{subequations}
\begin{align}
    H_1 & = \sum_{n=1}^{N} (u_n+d_n) \\
    H_2 & = \sum_{n=1}^{N-1} \left[ u_{n} \left( \sum_{m=n+1}^{N} u_{m} + d_{m} \right) + d_{n} \left( d_{n+1} + \sum_{m=n+2}^{N} u_{m} + d_{m} \right) \right] - u_{1}d_{N}  \\
    & \vdots \nonumber \\
    H_{N-1} & = \sum_{n=1}^{N} \sum_{m=1}^N u_{\overline{n+N-2}}^{\theta_{\overline{n+N-2}\geq m}} d_{\overline{n+N-2}}^{\theta_{\overline{n+N-2}<m}} \cdots u_{n}^{\theta_{n\geq m}} d_{n}^{\theta_{n<m}} \\
    H_N & = u_1\cdots u_N + d_1 \cdots d_N
\end{align}
\end{subequations}
The 1-loops can be expressed in terms of the canonical coordinates $\tq_n$, $\tp_n$ by
\begin{align}
    u_n = e^{-\tp_n}, \ d_n = R^2 e^{\tq_n-\tq_{n-1}} e^{\frac{1}{2}(\tp_{n-1}-\tp_n)}.
\end{align}



Let us explain the $H_{N-1}$ Hamiltonian: It is the sum over all $N-1$ loops by its definition. Due to periodicity, we can first focus on the case that the last white vertex $\sw_{2N}$ is the one ignored. 
We will start with the first $(N-1)$ loop $u_{N-1}\cdots u_1$. Then starting from the first hexagon, we flip the loop from a $u$-loop to a $d$-loop at white vertex $\sw_{2n}$. Notice that a $u$-loop at white vertex $\sw_{2n}$ can only be flipped if all the $u_m$ loops $m<n$ are already flipped to a $d$-loop (otherwise one will get $d_{n} u_{n-1}$ which is not Poisson commute). It is easy to see that the flipping generates a total $N$ different $N-1$ loops (from $u_{N-1}\cdots u_1$ to $d_{N-1}\cdots d_1$).  

We can now choose to neglect any one of the $N$ white vertexes $\sw_{2n}$, $n=1,\dots,N$ and repeat the procedure. Due to the periodicity of the lattice structure, the argument is all the same and all the $N-1$ loops can be easily written down. The total number of $N-1$ loops is hence $N \times N = N^2$.

\subsection{$Y^{N,\sk}$ model}\label{sec:YNk}

\begin{figure}
    \centering
    \includegraphics[width=0.5\textwidth]{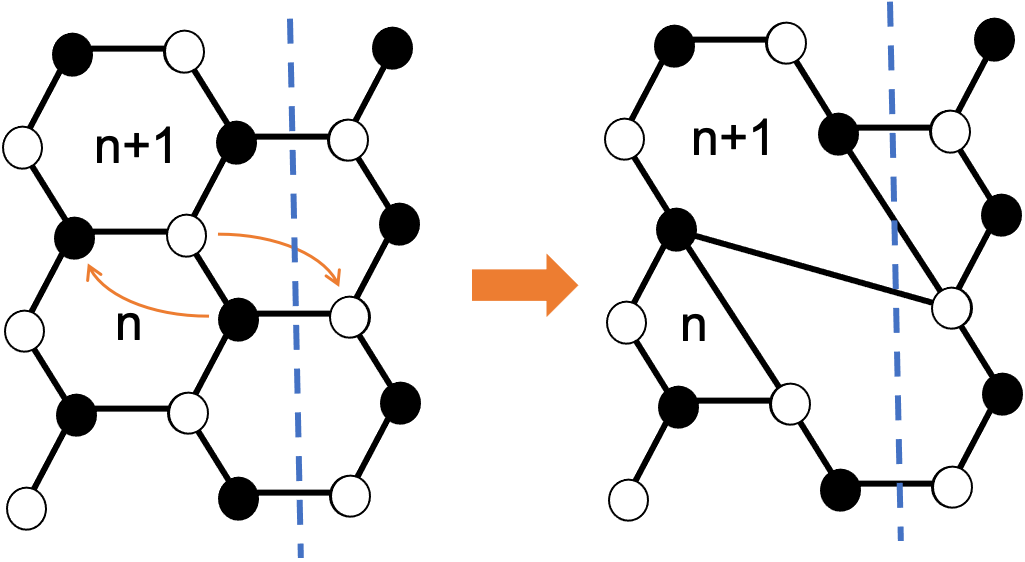}
    \caption{Gluing of vertices in the brane tiling.}
    \label{fig:gluing}
\end{figure}

The dimer model for $Y^{N,\sk}$ integrable system can be obtained by introducing impurity in the $Y^{N,N}$ diagram. This can be done solely on the level of the dimer diagram rathr than performed in $Y^{N,N}$ quiver diagram then taking the dual graph for dimer.
In the context of dimer diagram, the impurity is introduced by gluing vertexes: Suppose we choose the $n$-th hexagon in the unit cell as our starting point. 
We glue the black vertex $\sb_{2n}$ to $\sb_{2n+1}$ and white vertex $\sw_{2n+1}$ to $\sw_{2n}$ in the next unit cell on the right hand side. The gluing modifies 5 edges
\begin{itemize}
    \item $\tilde\sU_{2n+1}$ (originally $\sV_{2n+1}$): connects between $\sb_{2n+2}$ and $\sw_{2n}$ in the next unit cell. 
    \item $\tilde\sV_{2n}$ (originally $\sH_{2n+1}\cup\sU_{2n}\cup\tilde\sH_{2n}$) connects between $\sb_{2n+1}$ and $\sw_{2n}$ in the next unit cell. 
    \item $\sV_{2n-1}$ (originally $\sV_{2n-1}$) connects between $\sb_{2n+1}$ and $\sw_{2n-1}$. 
\end{itemize}
The reader might find the notation confusing (In particular $\tilde\sV_{2n}$ is used for the glued edge that was originally nothing to do with $\sV$). The notation is chosen such that both $\sV$ and $\tilde\sV$ are edges connecting $\sw_{2n}$ to $\sb_{2n+1}$. The tilde notation is reserved for the edges that cross the boundary of a unit cell. 
See Figure.~\ref{fig:gluing} for illustration.


\begin{figure}
    \centering
    \includegraphics[width=0.7\textwidth]{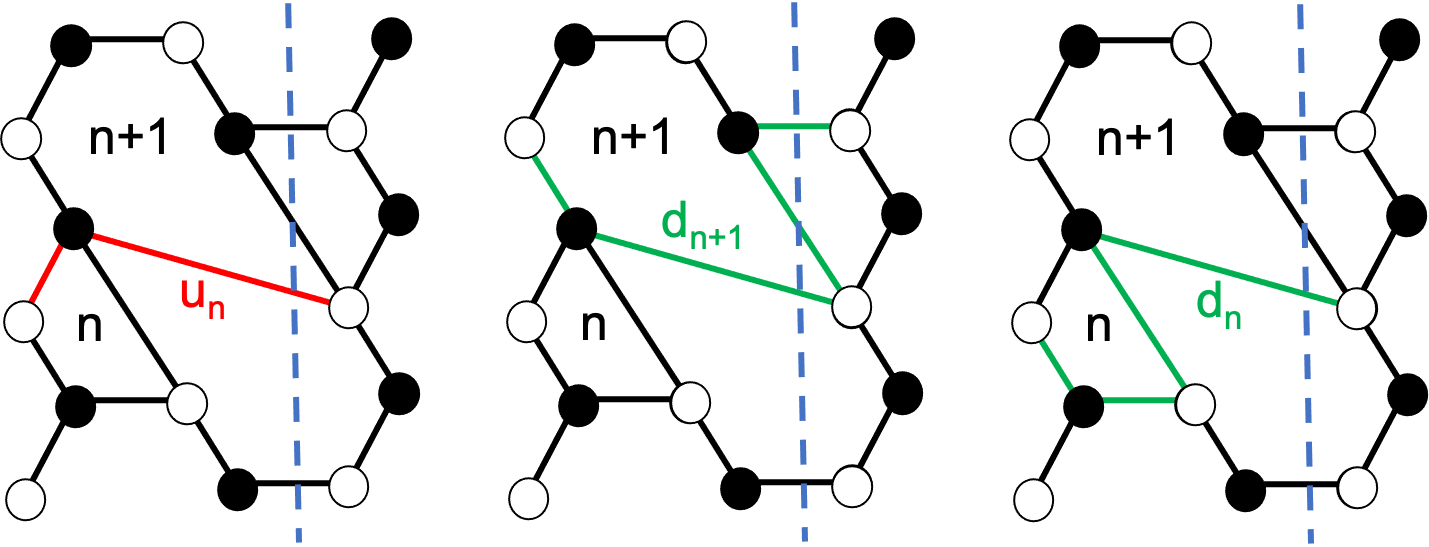}
    \caption{The 1-loops affected by the gluing.}
    \label{fig:newloop}
\end{figure}

The 1 loops affected by the gluing are the ones consisting the modified edges: $d_{n-1}$, $u_n$, and $d_n$. 
\begin{itemize}
    \item $d_{n+1} = \sw_{2n+2} \to \sb_{2n+1} \to \sw_{2n} \to \sb_{2n+2} \to \sw_{2n+2} = \tilde\sH_{2n+2}^{-1}\tilde\sU_{2n+1} \tilde\sV_{2n}^{-1} \sU_{2n+1}$ 
    \item $u_n = \sw_{2n} \to \sb_{2n+1} \to \sw_{2n} = \tilde\sV_{2n}^{-1}\sV_{2n}$
    \item $d_{n} = \sw_{2n} \to \sb_{2n-1} \to \sw_{2n-1} \to \sb_{2n+1} \to \sw_{2n} = \tilde\sV_{2n}^{-1} \sV_{2n-1} \sH_{2n-1}^{-1}\sU_{2n-1}  $
\end{itemize}
See Figure \ref{fig:newloop} for illustration.

The 1 loops ${d}_{n+1}$ and ${d}_{n}$ now share a common edge $\tilde\sV_{2n}$ (thus two vertexes $\sb_{2n+1}$ and $\sw_{2n}|_\text{next cell}$) after the gluing. The two will no longer Poisson commute based on \eqref{def:Poisson-loops}:
\begin{align}
    \{ {d}_{n+1}, {d}_{n} \} = {d}_{n} {d}_{n+1}. 
\end{align}

\begin{figure}
    \centering
    \includegraphics[width=0.5\textwidth]{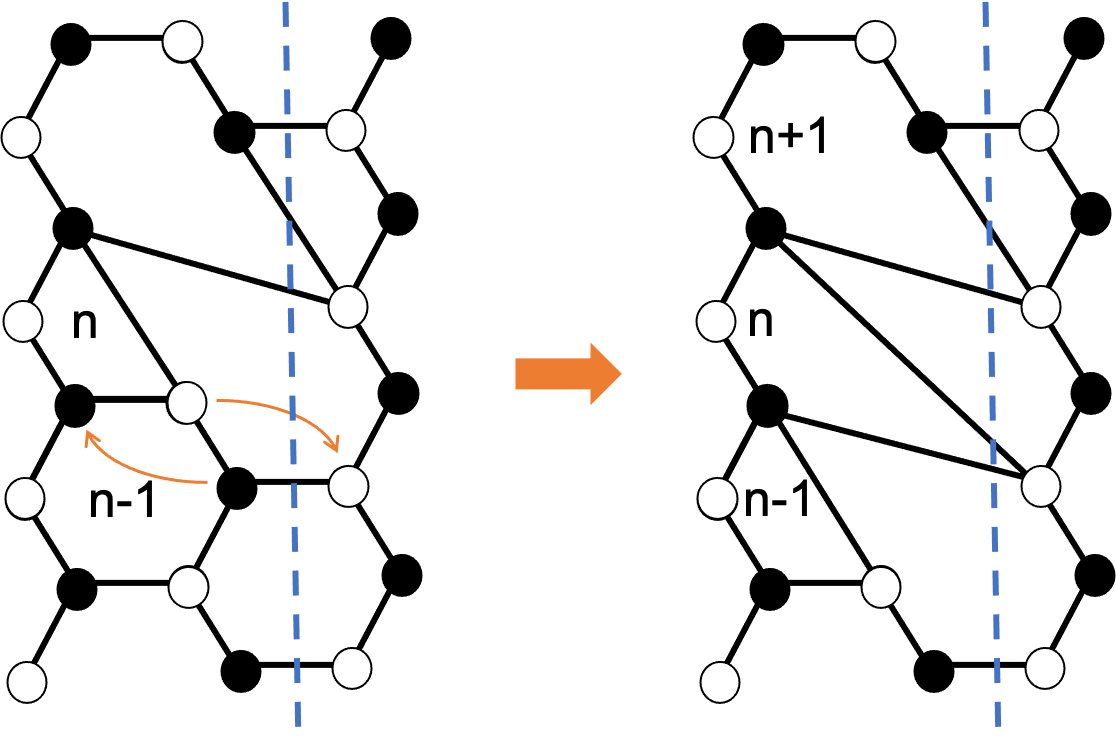}
    \caption{Second gluing of vertices.}
    \label{fig:gluing2}
\end{figure}

\paragraph{}
Next we move to the next hexagon below and glue the black vertex $\sb_{2n-2}$ to $\sb_{2n-1}$ and white vertex $\sw_{2n-1}$ to $\sw_{2n-2}$ in the next unit cell on the right. The gluing again modifies 5 edges: $\tilde\sV_{2n-1}$, $\sV_{2n-3}$, $\sU_{2n-2}$, $\sH_{2n-1}$, $\tilde\sH_{2n-2}$:
\begin{itemize}
    \item $\tilde \sU_{2n-1}$ (originally $\sV_{2n-1}$) connects between $\sb_{2n+1}$ and $\sw_{2n-2}$ in the next unit cell on the right. 
    \item $\tilde\sV_{2n-2}$ (originally $\sU_{2n-1}$, $\sH_{2n-1}$, and $\tilde\sH_{2n-2}$) connects between $\sb_{2n-1}$ and $\sw_{2n-2}$ in the next unit cell on the right. 
    \item $\sV_{2n-3}$ (originally $\sV_{2n-3}$) connects between $\sb_{2n-1}$ and $\sw_{2n-3}$. 
\end{itemize}
See Figure \ref{fig:gluing2} for illustration.

\begin{figure}
    \centering
    \includegraphics[width=0.7\textwidth]{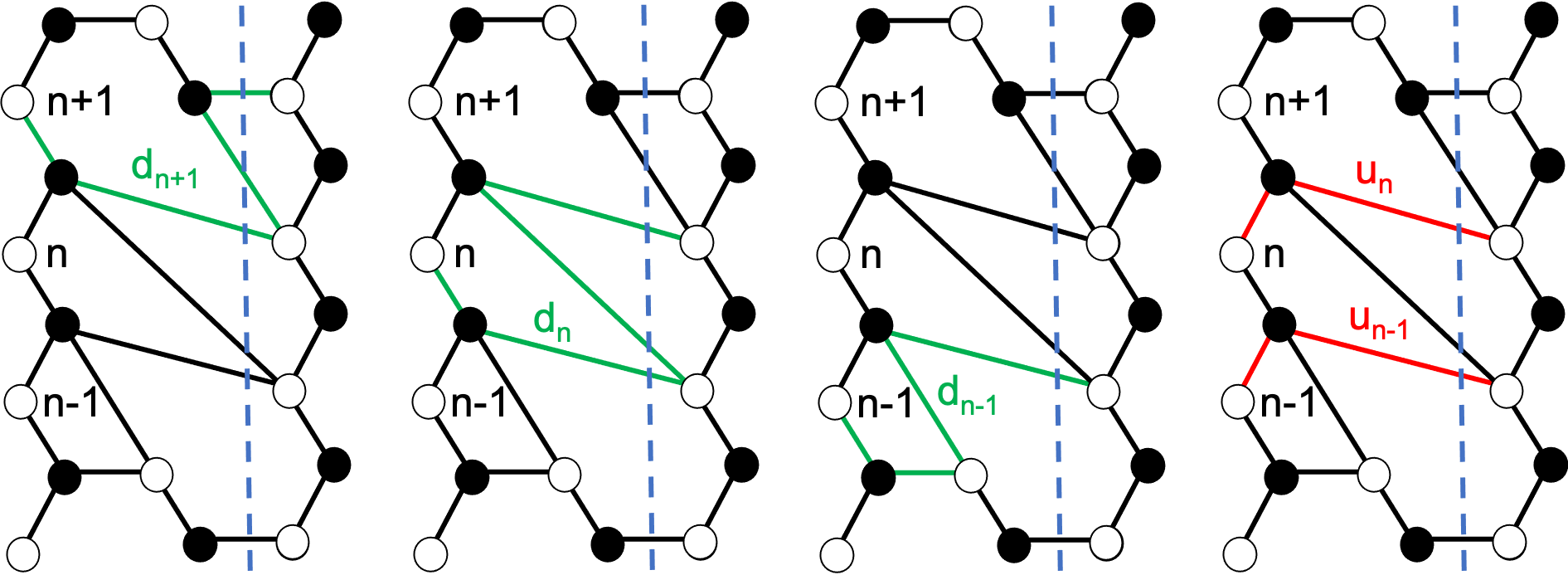}
    \caption{The 1-loops affected after performing gluing twice.}
    \label{fig:newloop2}
\end{figure}

The affected loops from the gluing are ${d}_{n}$, $u_{n+1}$, and $d_{n+1}$: 
\begin{itemize}
    \item ${d}_n = \sw_{2n} \to \sb_{2n-1} \to \sw_{2n-2} \to \sb_{2n+1} \to \sw_{2n} = \tilde\sV^{-1}_{2n}\tilde\sU_{2n-1}\tilde\sV^{-1}_{2n-2}\sU_{2n-1}$
    \item $u_{n-1} = \sw_{2n-2} \to \sb_{2n-1} \to \sw_{2n-2} = \tilde\sV^{-1}_{2n-2}\sV_{2n-2}$
    \item $d_{n-1} = \sw_{2n-2} \to \sb_{2n-3} \to \sw_{2n-3} \to \sb_{2n-1} \to \sw_{2n-2} = \tilde\sV_{2n-2}^{-1}\sV_{2n-3} \sH_{2n-3}^{-1}\sU_{2n-3}$
\end{itemize}
See Figure \ref{fig:newloop2} for illustration.

We now construct the dimer diagram for $Y^{N,\sk}$ theory by gluing $2(N-\sk)$ nodes $\sw_{2n+1}$, $\sb_{2n}$, $n=\sk,\dots,N-1$. The collection of 1-loops are
\begin{itemize}
    \item $u_n = \sw_{2n} \to \sb_{2n+1} \to \sw_{2n+1} \to \sb_{2n} \to \sw_{2n} =\tilde\sH_{2n}^{-1} \sU_{2n} \sH_{2n+1}^{-1} \sV_{2n}$, $n=1,\dots,\sk-1,N$; 
    \item $u_n = \sw_{2n} \to \sb_{2n+1} \to \sw_{2n} = \tilde\sV_{2n}^{-1} \sV_{2n}$, $n=\sk,\dots,N-1$;
    \item $d_n = \sw_{2n} \to \sb_{2n-1} \to \sw_{2n-1} \to \sb_{2n} \to \sw_{2n} = \tilde \sH_{2n}\sV_{2n-1}\sH_{2n-1}\sU_{2n-1}$, $n=1,\dots,\sk$;
    \item ${d}_N= \sw_{2N} \to \sb_{2N-1} \to \sw_{2N-2} \to \sb_{2N} \to \sw_{2N} =\tilde\sH_{2N}\tilde\sU_{2N-1}\tilde\sH_{2N-1}^{-1}\sU_{2N-1}$;
    \item ${d}_\sk = \sw_{2\sk} \to \sb_{2\sk-1} \to \sw_{2\sk-1} \to \sb_{2n+1} \to \sw_{2\sk} = \tilde\sV_{2\sk}\sV_{2\sk-1}\sH_{2\sk-1}^{-1}\sU_{2\sk-1}$;
    \item $d_n = \sw_{2n} \to \sb_{2n-1} \to \sw_{2n-2} \to \sb_{2n+1} \to \sw_{2n} = \tilde\sV_{2n}^{-1} \tilde\sU_{2n-1}\tilde\sV_{2n-2}^{-1}\sU_{2n-1}$, $n=\sk+1,\dots,N$. 
\end{itemize}
The Poisson commutation relation between the 1-loops are defined by their shared common edges. The only non-vanishing Poisson brackets are: 
\begin{align}\label{eq:commu}
\begin{split}
    & \{u_n,d_n\} = u_n d_n, \ \{d_{n+1},u_n\} = d_{n+1}u_n, n=1,\dots,N; \\
    & \{d_{n+1},d_{n}\} = d_{n+1} d_{n}, \ n=\sk,\dots,N-1.
\end{split}
\end{align}
The 1-loops can be expressed in terms of the canonical coordinates $\tp_n$, $\tq_n$, $n=1,\dots,N$ by
\begin{align}\label{eq:ud-qp standard}
    u_n = e^{-\tp_n}, \ d_n = R^2 e^{\tq_n - \tq_{n-1}} e^{-\frac{1}{2}(\tp_n+\tp_{n-1}) + \theta_{n\leq\sk}\tp_{n-1} }
\end{align}

The $n$-th Hamiltonian is the sum over the collection of $n$-loops, where $n$-loop is a product over $n$ non-overlapping 1-loops. It is straightforward to see that
\begin{subequations}
\begin{align}
    H_1 & = \sum_{n=1}^N u_n + d_n \\
    H_2 & = \sum_{n>m, \ n,m=1}^N (u_n+d_n)(u_m+d_m) - \sum_{n=1}^N u_n d_{\overline{n-1}} + d_{\overline{n+1}} u_n - \sum_{n=\sk+1}^N d_n d_{n-1} \\
    H_N & = u_N \cdots u_1
\end{align}
\end{subequations}
The number of terms in $H_2$ is 
\begin{align}
    \begin{pmatrix} 2N \\ 2 \end{pmatrix} - 2N - (N-\sk) = 2N (N-2) + \sk
\end{align}





In addition, we will now write down the exact form of $H_{N-1}$.
It is obvious that there are $N$ $(N-1)$-loops consists only $u$-loops. We label these $(N-1)$-loops by 
\begin{align}\label{def:All u (N-1)-loop}
    \bL_n = u_{n+N-2} \cdots u_{n}, \ n=1,\dots,N
\end{align}
with the periodicity $u_{n+N} = u_n$. 
We can always flip the last $u$-loop to a $d$-loop. This gives a total $2N$ $(N-1)$-loops. 
We then ask if one can further flip the second last $u$-loop $u_{n+1}$ in $\bL_n$. By the commutation relation between the 1-loops, it is achievable if $d_{n+1}$ Poisson commutes with $d_{n}$. That is if $n=\sk,\dots,N-1$ by \eqref{eq:commu}. Only $\bL_n$ with $n=1,\dots,\sk-1,N$ are allowed to have their second 1-loop $u_{n+1}$ to be flipped to $d_{n+1}$. 

Let us start with the following $N-1$ loop
\begin{align}
    u_{N-2} \cdots u_{1} {d}_{N}
\end{align}
obtained from flipping the last 1-loop of $\bL_N = u_{N-2} \cdots u_{1} u_{N}$. We can flip the second last loop $u_1 \to d_1$. Further more, we can keep flipping until we reach the following:
\begin{align}
    u_{N-2} \cdots u_{\sk+1} {d}_{\sk} d_{\sk-1} \cdots d_{1} {d}_N
\end{align}
with a total amount of flipping (hence the number of valid $(N-1)$-loops) equals to $\sk$.  
Next we move on to the following $N-1$ loop
\begin{align}
    u_{N-1}u_{N-2} \cdots u_{2} d_{1} 
\end{align}
and continue flipping starting from the second last $u$-loop until we reach
\begin{align}
    u_{N-1} \cdots u_{\sk+1} {d}_{\sk} d_{\sk-1} \cdots d_{1}
\end{align}
with a total number of $\sk-1$ $(N-1)$-loops. By repeating this process until we reach 
\begin{align}
    u_{N+\sk-2} \cdots u_{\sk+1} d_{\sk}
\end{align}
where we can not flip the second last loop $u_{\sk+1} \to d_{\sk+1}$. In total we obtain 
\begin{align}
    1+\cdots + \sk = \frac{\sk(\sk+1)}{2}
\end{align}
$(N-1)$-loops that has more than 1 flipping. 

There is a special case for $\sk=N-1$ with the $N-1$ loop $u_{N-2} \cdots u_{1} {d}_{N}$. We can only flip $\sk-1=N-2$ times to $d_{N-2} d_{N-3} \cdots d_{1} {d}_{N}$ instead of $\sk=N-1$ times. Hence the total number of terms in $H_{N-1}$ for $Y^{N,k}$, $\sk=0,\dots,N-1$ is exactly
\begin{align}
    2N + \frac{\sk(\sk+1)}{2} - \d_{\sk,N-1}. 
\end{align}

We define 
\begin{align}
\begin{split}
    & {\bC}_n = 1 + \frac{d_n}{u_n} + \frac{d_{n+1}d_n}{u_{n+1}u_n} + \cdots + \frac{d_\sk\cdots d_n}{u_\sk \cdots u_n}, \ n=1,\dots,\sk, \\
    & {\bC}_n = 1 + \frac{d_n}{u_n}, \ n=\sk+1,\dots,N-1, \\
    & {\bC}_N = 1 + \frac{d_N}{u_N} + \frac{d_1d_N}{u_1u_N} + \cdots + \frac{d_\sk\cdots d_1d_N}{u_\sk\cdots u_1u_N}.
\end{split}
\end{align}
The Hamiltonian $H_{N-1}$ takes the form 
\begin{align}\label{eq:YNK-HN-1}
    H_{N-1} = \sum_{n=1}^N \bL_n \bC_n 
\end{align}
with $\bL_n$ is defined in \eqref{def:All u (N-1)-loop}. 


\subsection{New dimers from non-standard gluing} \label{sec:SD-glue}

The construction for the $Y^{N,\sk}$ dimer diagram introduced in the last section is often considered as "standard" \cite{Eager:2011dp,Huang:2020neq}. However it is not the only way to glue the vertexes in dimer diagram $Y^{N,N}$. 

We will construct the dimer diagram $Y^{N,\sk}[S]$ characterized by a set $S\subset\{1,2,\dots,N\}$ (called gluing set) with $|S| = N-\sk$ in the following way: We start with the brane tiling hexagon diagram for $Y^{N,N}$ in Figure.~\ref{fig:YNN0} and glue the black vertexes $\sb_{2n}$ to $\sb_{2n+1}$ in the same unit cell and glue white vertexes $\sw_{2n+1}$ to $\sw_{2n}$ in the next unit cell for all $n \in S$.
The standard dimer graph for $Y^{N,\sk}$ introduced in the last section is a special case with $S=\{\sk,\dots,N-1\}$. 
Without losing generality, we can always choose $N\notin S$ and $ N-1 \in S$ when $0<\sk<N$ due to the periodicity of the dimer diagram. The Poisson commutation relation of the one loops $\{u_n,d_n\}$, $n=1,\dots,N$ can be calculated by \eqref{def:Poisson-loops}. The non-vanishing ones are: 
\begin{align}\label{eq:commu-ud-S}
\begin{split}
    & \{u_n,d_n\} = u_n d_n, \ \{d_{{n+1}},u_n\} = u_nd_{{n+1}}, \ n=1,\dots,N; \\
    & \{d_{{n+1}},d_{{n}}\} = d_{{n+1}} d_n  \text{ if } n \in S.    
\end{split}
\end{align}
with the periodicity $u_{N+n}=u_n$, $d_{N+n}=d_n$. 
We define $G_S(n) :\{1,\dots,N\} \to \{0,1\}$ to track the gluing by
\begin{align}\label{def:func-GS}
    G_S(n) = 
    \begin{cases}
        1 & n \notin S \\
        0 & n \in S
    \end{cases}.
\end{align}
The 1-loops can be expressed in terms of canonical coordinates $\tp_n$, $\tq_n$:
\begin{align}\label{def:ud-pq general}
    u_n = e^{-\tp_n}, \ d_n = R^2 e^{\tq_n - \tq_{n-1}} e^{-\frac{1}{2}(\tp_n+\tp_{n-1}) + G_S(n-1) \tp_{n-1} }. 
\end{align}

The Kasteleyn matrix of the dimer model $Y^{N,\sk}[S]$
\begin{align}\label{def:Kas-general}
    K_{\sb_i,\sw_j}(x,y) = 
    \begin{cases}
        \sH_j & \text{, if } i=j, \ j=2n+1, \ n \notin S; \\
        \tilde\sH_j e^{-x} & \text{, if } i=j, \ j=2n, \ n \notin S; \\ 
        -\sV_{j}+\tilde\sV_j e^{-x} & \text{, if } i=j+1, \ j=2n, \ n \in S; \\ 
        \sU_{\overline{j-1}} \ y^{-\delta_{j,1}} & \text{, if } i=\overline{j-1}, \\
        \sV_j \ y^{\delta_{j,N}} & \text{, if } i=\overline{j+1}, \ j=2n-1,2n, \ n \notin S;  \\
        \tilde\sU_j e^{-x} & \text{, if } i=j+2, j=2n, \  n\in S; \\
        0 & \text{, otherwise.}
    \end{cases}
\end{align}
is a tri-diagonal $(N+\sk) \times (N+\sk)$ matrix.
We would like to point out that when $n \in S$, $K_{\sb_{2n+1},\sw_{2n}}$ is on diagonal term and $K_{\sb_{2n+2},\sw_{2n}}$ sits next to diagonal since $\sb_{2n}$ has been glued to $\sb_{2n+1}$. 
One can construct $N$ $2\times 2$ Lax matrices $\tilde{L}_n(x)|_\text{dimer}$from the Kasteleyn matrix using \eqref{eq:KtoL}: 
\begin{align}
\begin{split}
    \det K_{\sb_i,\sw_j}(e^{-x},y) = & \ (-1)^{N+\sk-1} \left[ y \prod_{n\in S} \tilde\sU_{2n} \prod_{n\notin S} \sV_{2n} \sV_{2n+1} + \frac{1}{y} \prod_{n\in S} \sU_{2n-1} \prod_{n \notin S} \sU_{2n}\sU_{2n-1} \right] \\
    & + \Tr \tilde{L}_N(x)|_\text{dimer} \cdots \tilde{L}_{1}(x)|_\text{dimer}.
\end{split}
\end{align}
The Lax matrices are given by
\begin{align}\label{eq:Lax-dimer}
    \tilde{L}_n(x)|_\text{dimer} = 
    \begin{cases}
        \begin{pmatrix}
            \sH_{2n+1} & -\sU_{2n} \sV_{2n} \\ 1 & 0
        \end{pmatrix}
        \begin{pmatrix}
            \tilde\sH_{2n} e^{-x} & -\sU_{2n-1} \sV_{2n-1}^{\sk_S(n-1)} (\tilde\sU_{2n-1}e^{-x})^{1-\sk_S(n-1) } \\ 1 & 0 
        \end{pmatrix} & \text{, if } n \notin S; \\
        \begin{pmatrix}
            -\sV_{2n}+\tilde\sV_{2n}e^{-x} & \sU_{2n-1} \sV_{2n-1}^{\sk_S(n-1)} (\tilde\sU_{2n-1}e^{-x})^{1-\sk_S(n-1) } \\ 1 & 0
        \end{pmatrix} & \text{, if } n \in S. 
    \end{cases}
\end{align}
 
The $N$ conserving Hamiltonians $H_n$, $n=1,\dots,N$, are obtained from the coefficients of the polynomial in $e^{-x}$:
\begin{align}
\begin{split}
    & \Tr \tilde{L}_N(X)|_\text{dimer} \cdots \tilde{L}_{1}(X)|_\text{dimer} \\
    & = \prod_{n\in S} \tilde\sV_{2n} \prod_{n \notin S} \sH_{2n+1} \tilde\sH_{2n} \left[ e^{-Nx} - H_1 e^{-(N-1)x} + \cdots + (-1)^{N} H_{N} \right]. 
\end{split}
\end{align}
which $H_n$ is a collection of $n$-loops, where an $n$-loop is a product over $n$ non-overlapping 1-loops.

\subsubsection{Example: $Y^{4,2}[S]$}

\begin{figure}
    \centering
    \includegraphics[width=0.5\textwidth]{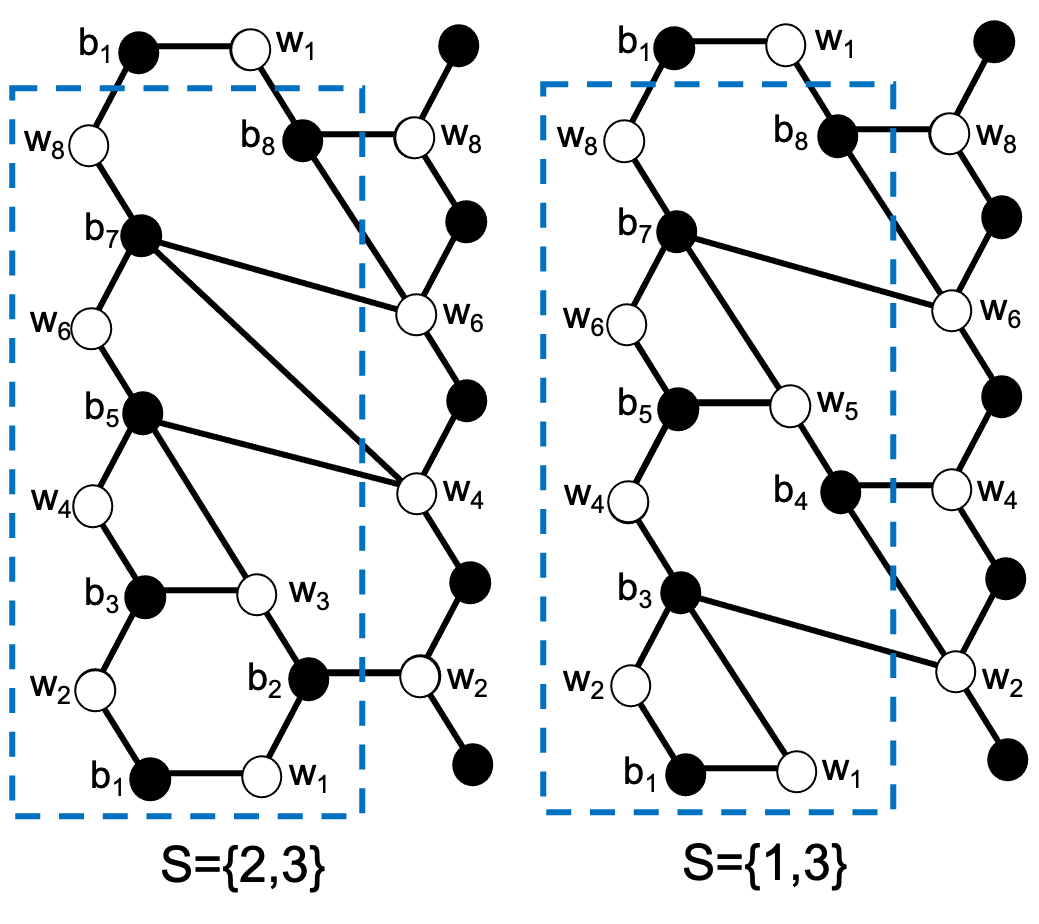}
    \caption{Two non-equivalent dimer graphs for $Y^{4,2}[S]$: The standard graph characterized by the gluing set $S=\{2,3\}$ on the left. the alternative one characterized by the gluing set $S=\{1,3\}$ on the right.}
    \label{fig:Alt-glue}
\end{figure}

There are two non-equivalent dimer diagrams for $Y^{4,2}$ that we can constructed from the hexagon diagram. They are characterized by the gluing set $S=\{2,3\}$ (standard) and $S=\{1,3\}$ (non-standard). See Figure.~\ref{fig:Alt-glue} for illustration. 

We choose the 1-loops in the dimer graph for $Y^{4,2}[S=\{2,3\}]$ in Figure.~\ref{fig:Alt-glue}:
\begin{align}
\begin{split}
    & u_1 = \sw_2 \to \sb_3 \to \sw_3 \to \sb_2 \to \sw_2, \ d_1 = \sw_2 \to \sb_1 \to \sw_1 \to \sb_2 \to \sw_2, \\
    & u_2 = \sw_4 \to \sb_5 \to \sw_4, \ d_2 = \sw_4 \to \sb_3 \to \sw_3 \to \sb_5 \to \sw_4, \\
    & u_3 = \sw_6 \to \sb_7 \to \sw_6, \ d_3 = \sw_6 \to \sb_5 \to \sw_4 \to \sb_7 \to \sw_6, \\
    & u_4 = \sw_8 \to \sb_1 \to \sw_1 \to \sb_8 \to \sw_8, \ d_4 = \sw_8 \to \sb_7 \to \sw_6 \to \sb_8 \to \sw_8.
\end{split}
\end{align}
The Kasteleyn matrix is 
\begin{equation}
    K_{\sb_i,\sw_j} = \left( 
    \begin{array}{c|c c c c c c c c}
     & \sw_1 & \sw_2 & \sw_3 & \sw_4 & \sw_6 & \sw_8 \\
     \hline 
    \sb_1 & \sH_1 & \sU_1 & & & & \sV_8y \\
    \sb_2 & \sV_1 & \tilde\sH_2 e^{-x} & \sU_2 & & & \\
    \sb_3 & & \sV_2 & \sH_3 & \sU_3 & & \\
    \sb_5 & & &  \sV_3 & \tilde\sV_4 e^{-x}-\sV_4 & \sU_5 & \\
    \sb_7 & & & & \tilde\sU_5 e^{-x} & \tilde\sV_6 e^{-x}-\sV_6 & \sU_7 \\
    \sb_8 & \frac{\sU_8}{y} & & & & \tilde\sU_7 e^{-x}  & \tilde\sH_8 e^{-x}
    \end{array}
    \right). 
\end{equation}
The Hamiltonians for the dimer model $Y^{4,2}[S=\{2,3\}]$ can be read off from the graph directly as a sum of one, two, three and four loops: 
\begin{subequations}
\begin{align}
    H_1 = & \ u_1+u_2+u_3+u_4+d_1+d_2+d_3+d_4 \\
    H_2 = & \ u_1u_2+u_1u_3+u_1u_4+u_2u_3+u_2u_4+u_3u_4 \nonumber \\
    & + u_1d_3+u_1d_4+u_2d_1+u_2d_4+u_3d_1+u_3d_2+u_4d_2+u_4d_3 \\
    & + d_1d_2+d_1d_3+d_1d_4+d_2d_4 \nonumber \\
    H_3 = & \ u_3u_2u_1+u_2u_1u_4+u_1u_4u_3+u_4u_3u_2 \nonumber \\
    & + u_3u_2d_1+u_2u_1d_4+u_1u_4d_3+u_4u_3d_2\\
    & + u_3d_2d_1+u_2d_1d_4+d_2d_1d_4 \nonumber \\ 
    H_4 = & \ u_1 u_2 u_3 u_4
\end{align}
\end{subequations}

\paragraph{}
Alternatively, we have a dimer graph characterized by $S=\{1,3\}$ as shown in Figure.~\ref{fig:Alt-glue}. We choose the 1-loops in the dimer graph $Y^{4,2}[S=\{1,3\}]$ as
\begin{align}
\begin{split}
    & u_1 = \sw_2 \to \sb_3 \to \sw_2, \ d_1 = \sw_2 \to \sb_1 \to \sw_1 \to \sb_3 \to \sw_2, \\
    & u_2 = \sw_4 \to \sb_5 \to \sw_5 \to \sb_4 \to \sw_4, \ d_2 = \sw_4 \to \sb_3 \to \sw_2 \to \sb_4 \to \sw_4, \\
    & u_3 = \sw_6 \to \sb_7 \to \sw_6, \ d_3 = \sw_6 \to \sb_5 \to \sw_5 \to \sb_7 \to \sw_6, \\
    & u_4 = \sw_8 \to \sb_1 \to \sw_1 \to \sb_8 \to \sw_8, \ d_4 = \sw_8 \to \sb_7 \to \sw_6 \to \sb_8 \to \sw_8. \\
\end{split}
\end{align}
The Kasteleyn matrix of $Y^{4,2}[S=\{1,3\}]$ 
\begin{equation}
    K_{\sb_i,\sw_j} = \left( 
    \begin{array}{c|c c c c c c c c}
     & \sw_1 & \sw_2 & \sw_4 & \sw_5 & \sw_6 & \sw_8 \\
     \hline 
    \sb_1 & \sH_1 & \sU_1 & & & & \sV_8y \\
    \sb_3 & \sV_1 & \tilde\sV_2 e^{-x}-\sV_2 & \sU_3 & & & \\
    \sb_4 & & \tilde\sU_3 e^{-x} & \tilde\sH_4 e^{-x} & \sU_4 & & \\
    \sb_5 & & &  \sV_4 & \sH_5 & \sU_5 & \\
    \sb_7 & & & & \sV_5 & \tilde\sV_6 e^{-x}-\sV_6 & \sU_7 \\
    \sb_8 & \frac{\sU_8}{y} & & & & \tilde\sU_7 e^{-x} & \tilde\sH_8 e^{-x}
    \end{array}
    \right). 
\end{equation}
The Hamiltonians for the dimer model $Y^{4,2}[S=\{1,3\}]$ can be read off from the graph directly as a sum of one, two, three and four loops: 
\begin{subequations}
\begin{align}
    H_1 = & \ u_1+u_2+u_3+u_4+d_1+d_2+d_3+d_4,\\
    H_2 = & \ u_1u_2+u_1u_3+u_1u_4+u_2u_3+u_2u_4+u_3u_4 \nonumber \\
    & + u_1d_3+u_1d_4+u_2d_4+u_2d_1+u_3d_1+u_3d_2+u_4d_2+u_4d_3 \\
    & + d_1d_3+d_1d_4+d_2d_3+d_2d_4, \nonumber\\
    H_3 = & \ u_3u_2u_1+u_2u_1u_4+u_1u_4u_3+u_4u_3u_2 \\
    & + u_3u_2d_1+u_2u_1d_4+u_1u_4d_3+u_4u_3d_2 + u_2d_1d_4+u_4d_3d_2 \nonumber\\
    H_4 = & \ u_1u_2u_3u_4 
\end{align}
\end{subequations}

Here we see the third Hamiltonian $H_3$ of $Y^{4,2}[S=\{1,3\}]$ dimer model has one less term comparing to $H_3$ of $Y^{4,2}[S=\{2,3\}]$. As it is impossible to have a three loop as a product over three $d$-loops in $Y^{4,2}[S=\{2,3\}]$.

\section{Defects in five dimensional gauge theory}\label{sec:gauge}

The four dimensional $\CalN=2$ supersymmetric gauge theories has an intrinsic connection \cite{DW1,Gorsky:1994dj,Nikita-Shatashvili} to the algebraic integrable system. 
The matching can be further extended in several directions. In \cite{Nekrasov:1996cz} Nekrasov proposed a non-perturbative solution to the 5d Seiberg-Witten gauge theory compactified on a circle using relativistic integrable model.  
Furthermore, Nekrasov and Shatashivilli extended the connection to the the quantum integrable model by introducing $\Omega$-deformation to the Seiberg-Witten theories \cite{Nikita-Shatashvili}. 

\subsection{The bulk theory}
 
We consider $\CalN=1$ SYM in 5d with gauge group $U(N)$ on the manifold $S^1_R \times \BR^4$, where $R$ is the radius of the circle $S^1_R$. The theory has $SO(1,4)$ Lorentz symmetry and $SU(2)$ R-symmetry. 
The Lagrangian of the theory is characterized by the complexified gauge coupling 
\begin{align}
    \tau = \frac{4\pi i}{g^2} + \frac{\vartheta}{2\pi}, \quad \kq = e^{2\pi i \tau},
\end{align}
and the Chern-Simons level $\sk$. 
We give non-zero vev to the adjoint scalar in the vector multiplet $\langle \Phi \rangle \neq 0$. This forces the theory to go on Coulomb branch and breaks the gauge group to its maximal torus. The vacuum is characterized by the Coulomb moduli parameters $\ba = (a_1,\dots,a_N)$, $\sum_{\alpha=1}^N a_\alpha = 0$. 

If we view the $\BR^4 = \BC_{1} \times \BC_2$, we can denote the coordinate on the first complex plane $\BC_1$ as $\bz_1$, and the second coordinate on the $\BC_2$ plane as $\bz_2$. We introduce an $\Omega$-background by viewing the $\BC_1 \times \BC_2$ as a bundle over the $S^1_R$, where as we go around the circle, we make an identification 
\begin{align}
    (\bz_1,\bz_2) \sim (e^{R\ve_1} \bz_1,e^{R\ve_2}\bz_2)
\end{align}
with $\ve_1,\ve_2\in \BC$ are the complex $\Omega$-deformation parameters \cite{Nekrasov:2002qd,Pestun:2016zxk,nekrasov2006seiberg}. 
The localization on the $\Omega$-deformed theory reduces the path integral in the partition function to an equivalent statistical system \cite{Nekrasov:1998ss,Nikita:I,Haouzi:2020yxy}
\begin{align}
    \CalZ_{\BC_{12}^2} = \CalZ^{\rm pert} (\ba,\boldsymbol{\ve}) \times \sum_{k=0}^\infty \kq^k \CalZ_k (\ba,\boldsymbol\ve)
\end{align}
where the $k$ instanton measure has the following integration form
\begin{align}\label{eq:Z-int}
    \CalZ_k (\ba,\boldsymbol\ve) 
    & = \frac{1}{k!} \int \prod_{j=1}^k \frac{d\phi_j}{2\pi i} \BE \left[\bN \bK^* + q_{12} \bN^* \bK - P_{12} \bK\bK^* \right] \det(\bK)^\sk. 
\end{align}
The perturbative contribution is given by
\[
    \CalZ^\text{pert}(\ba,\boldsymbol{\ve}) = \BE \left[ - \frac{\bN\bN^*}{P_{12}^*} \right]. 
\] 
The Chern characters are defined by
\begin{align}
    \bN = \sum_{\alpha=1}^N e^{a_\alpha}, \quad \bK = \sum_{j=1}^k e^{\phi_j}, \quad \bS = \bN - P_{12} \bK. 
\end{align}
$q_1=e^{\ve_1}$, $q_2=e^{\ve_2}$ are the exponentiated complex $\Omega$-deformation parameters, and
\begin{align}
    P_1 = 1 - q_1, \ P_2 = 1-q_2, \ P_{12} = (1-q_1)(1-q_2). 
\end{align}
Given a character $\bX = \sum_{\ti} \tn_\ti e^{\tx_\ti}$ we denote its dual character $\bX^* = \sum_{\ti} \tn_\ti e^{-\tx_\ti}$. $\BE$ is the index functor that converts the additive Chern class characters to multiplicative class
\begin{align}\label{def:E-operator}
    \BE \left[ \sum_{\ti}\tn_\ti e^{\tx_\ti} \right] = \prod_\ti \left( 2 \sinh \frac{\tx_i}{2} \right)^{-\tn_\ti} := \prod_\ti \left( \sh \tx_i \right)^{-\tn_\ti}.
\end{align}

The instanton partition function can be calculated by the supersymmetric localization in the presence of $\Omega$-deformation, whose parameters are $\boldsymbol{\ve} = (\ve_1,\ve_2)$. When $\sk\leq N$, the integration in \eqref{eq:Z-int} can be evaluated as follows \cite{NO1,Nikita:I}: First we rewrite the integral as an ordering
\begin{align}
    \int \frac{1}{k!} \prod_{j=1}^k \frac{d\phi_j}{2\pi i} = \oint \frac{d\phi_k}{2\pi i} \cdots \oint \frac{d\phi_1}{2\pi i} 
\end{align}
The first variable picks up pole at $\phi_1=a_\alpha$. The following ones can picks up poles recursively at one of the following
\begin{align}
    a_{\beta(\neq \alpha)}, \ \phi_{l(<j)}+\ve_1, \ \phi_{l(<j)}+\ve_2. 
\end{align}
As the result the instanton configuration is labeled by $N$ Young diagrams $\boldsymbol\lambda = (\lambda^{(1)},\dots,\lambda^{(N)})$, $\lambda^{(\alpha)} = (\lambda^{(\alpha)}_1,\lambda^{(\alpha)}_2,\dots,\lambda^{(\alpha)}_{\ell_\alpha})$ satisfying 
\begin{align}
    \lambda^{(\alpha)}_1 \geq \lambda^{(\alpha)}_{2} \geq \cdots \geq \lambda^{(\alpha)}_{\ell_\alpha} > \lambda^{(\alpha)}_{\ell_\alpha+1} = 0, \ 
\end{align}
The value $\lambda^{(\alpha)}_\ri$ labels the number of boxes in the $\ri$-th row in the Young diagram $\lambda^{(\alpha)}$. We will define 
\begin{align}
    \bK = \sum_{\alpha=1}^{N} \sum_{(\ri,\rj)\in \lambda^{(\alpha)}} e^{a_\alpha+(\ri-1)\ve_1+(\rj-1)\ve_2}. 
\end{align}
The supersymmetric localization finds the instanton partition function is a statistical ensemble over the instanton configurations
\begin{align}\label{def:inst-part}
    \CalZ_{{\BC}_{12}^2}(\ba,\boldsymbol\ve;\kq) = \sum_{\boldsymbol\lambda} \kq^{|\boldsymbol\lambda|} \CalZ_\text{bulk}(\ba,\boldsymbol\ve)[\boldsymbol\lambda] 
\end{align}
where we combine the perturbative contribution and the instanton contribution into
\begin{align}
    \CalZ_\text{bulk}(\ba,\boldsymbol\ve)[\boldsymbol\lambda] = \BE \left[ -\frac{\bS[\boldsymbol\lambda]\bS^*[\boldsymbol\lambda]}{P_{12}^*} \right] \prod_{\alpha=1}^N \prod_{(\ri,\rj)\in \lambda^{(\alpha)} } \left[ e^{a_\alpha+(\ri-1)\ve_1+(\rj-1)\ve_2} \right]^\sk.
\end{align}

\subsection{Line defect as $qq$-character}
The observables in the supersymmetric gauge theory that are $\Omega$-background protected (perserves the same supersymmetry as the $\Omega$-deformation) can be evaluated by the effective statistical mechanics system. 
We couple a 1d fermionic degrees of freedom to the bulk 5d gauge theory in the way that half of the supersymmetries are preserved. It becomes a half-BPS line defect from the 5d theory point of view. The action of the 1d fermion field $\chi$ couplied to the 5d gauge theory is 
\begin{align}
    S_\text{1d} = \int dt \ \chi^\dagger (\p_t -i A_t + \Phi + x) \chi. 
\end{align}
Here $A_t$ and $\Phi$ are the pull back of the vector and adjoint scalar field in the 5d vector multiplet. $x$ is the real mass of the fermions, or the background gauge field for the global $U(1)$ symmetry acting only on $\chi$. We focus on the case where $\chi$ is in the fundamental representation of the gauge group $U(N)$. 

The coupling of 1d fermionic degree of freedom to a gauge group is a classical way to introduce half-BPS Wilson loop. The Fock space of the 1d fermions contains the BPS Wilson loops in representations. Let us introduce the coupled 1d fermion to the path integral
\begin{align}\label{eq:Z-1d-5d-PI}
    \CalZ_\text{1d/5d}(x) = \int \ED\Psi_\text{5d}  \ED \chi \ e^{i S_\text{5d}[\Psi_\text{5d}] + i S_\text{1d}[\Psi_\text{5d},\chi,x]}.
\end{align}
Here $\Psi_\text{5d}$ denotes the 5d fields. In the five-dimensional context the partition function contains the generating functions in $X=e^x$ of the expectation values of the Wilson loop \cite{Kim:2016qqs,Tong:2014cha}:
\begin{align}\label{eq:Wilson-loop}
    \frac{\CalZ_\text{1d/5d}(x)}{\CalZ_{\BC^2_{12}}} = X^{-\frac{N}{2}} \sum_{n=0}^N (-X)^n \EW_{\wedge^n}
\end{align}
where $\wedge^n$ is the $n$-th anti-symmetric representation and the corresponding Wilson loop is defined by
\begin{align}
    \EW_{\wedge^n} = \Tr_{\wedge^n} \text{P} \exp \left[ i \int \ dt (A_t+i\Phi) \right].
\end{align}

We then turn on the $\Omega$-deformation on $\BC_{12}^2$ for localization computation. The 1d fermionic defect which warps on the $S^1_R$ preserves the same supersymmetric as the $\Omega$-deformation. The partition functon of the 1d-5d coupled system can be evaluated by the effective statistical mechanics system \cite{Kim:2016qqs,Haouzi:2020yxy}
\begin{align}\label{eq:Z-1d-5d}
    \CalZ_\text{1d/5d}(\ba,x,\boldsymbol\ve,\kq) = \sum_{\boldsymbol\lambda} \kq^{|\boldsymbol\lambda|} \CalZ_\text{bulk}(\ba,\boldsymbol\ve_1)[\boldsymbol\lambda] \left[ \EY(x)[\boldsymbol\lambda] + \cdots \right].
\end{align}
The $\EY(x)$-observable whose evaluation at state $\boldsymbol\lambda$ is computed as
\begin{align}
    \EY(x)[\boldsymbol\lambda] & = \BE \left[ -e^x \bS^*[\boldsymbol\lambda] \right] 
    = \prod_{\alpha=1}^N \sh(x-a_\alpha-\ve_1 \ell_\alpha ) \prod_{\ri=1}^{\ell_\alpha} \frac{\sh(x-a_\alpha-(\ri-1)\ve_1-\lambda^{(\alpha)}_\ri \ve_2 )}{\sh(x-a_\alpha-\ri\ve_1-\lambda^{(\alpha)}_\ri \ve_2 )}
\end{align}
whose poles and zeros are located at the boundary of the Young diagram.  

The 1d-5d coupled partition function $\CalZ_{1d/5d}(x)$ in \eqref{eq:Z-1d-5d} is analytic function in $x$ \cite{Kim:2016qqs,Haouzi:2020yxy,Nikita:II,NaveenNikita}. In particular, it defines the \emph{$qq$-character observables} \cite{Nikita:I}. The $qq$-character can be obtained by explicit evaluation of the contour integration arise from the localization computation \cite{Hwang:2014uwa,Kim:2012qf,Haouzi:2020yxy}. In \cite{Nikita:I}, it is shown that when the instanton figurations are labeled by $N$ Young diagrams $\boldsymbol\lambda$, we can obtain the exact form of the $qq$-character by studying the behavior of the instanton measure with adding/removing a point like instanton from an instanton configuration $\boldsymbol\lambda$. By adding an instanton at $X=e^x=e^{a_\alpha+ (\ri-1)\ve_1 + \lambda^{(\alpha)}_\ri \ve_2 }$ to $\bK$, the measures modifies by
\begin{align}
\begin{split}
    \kq \frac{\CalZ_\text{bulk}[\bK+X]  }{\CalZ_\text{bulk}[\bK]}
    & = \kq X^\sk \BE \left[ - \frac{(\bS-P_{12}X)(\bS-P_{12}X)^*}{P_{12}^*} + \frac{\bS\bS^*}{P_{12}^*} \right] \\
    & = \kq X^\sk \BE \left[ X (\bS-P_{12}X)^* + \bS X^* \right] \\
    & = \frac{\kq X^\sk}{\EY(x+\ve_+)[\bK+X]\EY(x)[\bK]}
\end{split}
\end{align}
The fundamental $qq$-character of the pure SYM in 5d with Chern-Simons level $\sk$ is 
\begin{align}\label{def:qq}
    \EX_\sk(x)[\boldsymbol\lambda] = \EY(x+\ve_+)[\boldsymbol\lambda] + \frac{\kq X^{\sk}}{\EY(x)[\boldsymbol\lambda]}
\end{align}
with $\ve_+ = \ve_1+\ve_2$. The $X^\sk$ in the second term comes from the contribution of Chern-Simons level when the instanton number is increased by one. 

The main statement in \cite{Nikita:II} proves that the expectation value of the $qq$-character 
\begin{align}
\begin{split}
    \langle \EX_\sk(x) \rangle \CalZ_{\BC_{12}^2}(\ba,\boldsymbol\ve;\kq) & = \sum_{\boldsymbol\lambda} \kq^{|\boldsymbol\lambda|} \CalZ_\text{bulk}(\ba,\boldsymbol\ve)[\boldsymbol\lambda] \EX_\sk(x)[\boldsymbol\lambda] \\
    & = T_\sk(x) \CalZ_{\BC_{12}^2}(\ba,\boldsymbol\ve,\kq) = \CalZ_\text{1d/5d}(\ba,x,\boldsymbol\ve,\kq)
\end{split}
\end{align}
is an analytic function in $x$ \cite{Nikita:I}. 

\paragraph{}
Let us consider the Nekrasov-Shatashivilli limit (NS-limit for short) $\ve_2 \to 0$ with $\ve_1$ fixed. The effective twisted superpotential of the three dimensional theory on $S^1\times \BC_2$ corresponding to the five dimensional theory subject to two dimensional $\Omega$-background \cite{Nikita-Shatashvili}:
\begin{align}\label{def:super_potential}
    \widetilde\CalW(\ba,\ve_1,\kq) = \ve_2 \lim_{\ve_2 \to 0} \log \CalZ_{\BC_{12}^2}.
\end{align}
The properties of partition function $\CalZ_{\BC^2_{12}}$ of the $A_1$ gauge theory along with the twisted super potential $\widetilde\CalW$ are well studied in various papers \cite{Nekrasov:1995nq,Nikita-Pestun-Shatashvili,Nikita-Shatashvili}. See also \cite{Nekrasov:2002qd, Nekrasov:2009uh,Nekrasov:2009ui,Nekrasov:1998ss}. 

We define the limit shape configuration $\boldsymbol\Lambda$ based on the superpotential $\widetilde\CalW$
\begin{align}\label{def:limitshape}
    \kq^{|\boldsymbol\Lambda|} \CalZ_\text{bulk}(\ba,\boldsymbol\ve) [\boldsymbol\Lambda] = e^{\frac{\widetilde\CalW(\ba,\ve_1,\kq)}{\ve_2}}. 
\end{align}
In other words, the limit shape configuration $\boldsymbol\Lambda$ dominates in the infinite sum in \eqref{def:inst-part}.
Let us define the following parameters 
\begin{align}\label{def:v_ai}
    v_{\alpha \ri} = a_\alpha + (\ri-1)\ve_1 + \Lambda^{(\alpha)}_\ri \ve_2, \ \mathring{v}_{\alpha \ri} = a_\alpha + (\ri-1)\ve_1
\end{align}
such that
\begin{align}\label{def:bV}
    \bS[\boldsymbol\Lambda] = \bN - P_1 P_2 \bK[\boldsymbol{\Lambda}] 
    = \bN - P_1 \mathring\bV + P_1 \bV, \ \bV := \sum_{\alpha=1}^{N} \sum_{\ri=1}^\infty e^{v_{\alpha \ri}}, \mathring\bV := \sum_{\alpha=1}^{N} \sum_{\ri=1}^\infty e^{\mathring{v}_{\alpha \ri}}
\end{align}

\subsection{Co-dimensional two defect: $Q$-observable}

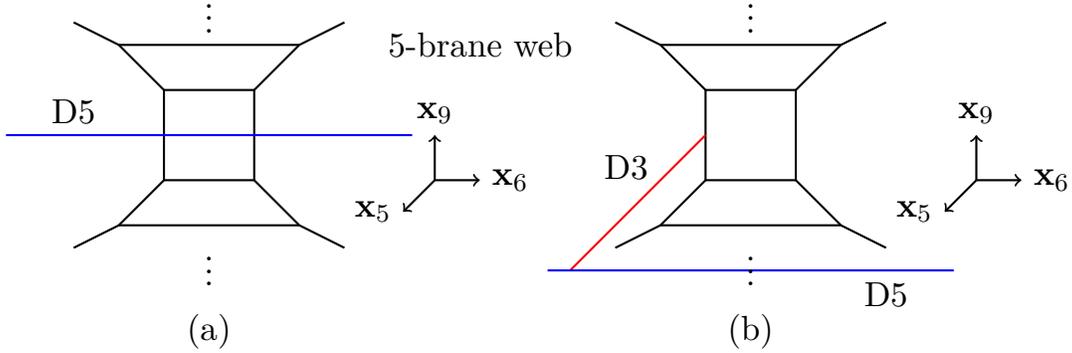
\begin{figure}[h!]
    \centering
    \begin{tikzpicture}[scale=0.6, every node/.style={scale=1.2}]
    \begin{scope}
    \draw[thick] (1,1) -- (-1,1);
    \draw[thick] (1,-1) -- (-1,-1);
    \draw[thick] (2,2) -- (-2,2);
    \draw[thick] (2,-2) -- (-2,-2);
    \draw[thick] (3,2.5) -- (2,2) -- (1,1) -- (1,-1) -- (2,-2) -- (3,-2.5);
    \draw[thick] (-3,2.5) -- (-2,2) -- (-1,1) -- (-1,-1) -- (-2,-2) -- (-3,-2.5);
    \draw[blue,thick] (-4.5,0) -- (4.5,0);

    \node[above] at (0,2) {$\vdots$};
    \node[below] at (0,-2) {$\vdots$}; 
    \node[right] at (3.7,2) {5-brane web};
    \node[above] at (-3,0) {D5};

     \draw[->,thick] (5,-1) -- (5,-0);
    \draw[->,thick] (5,-1) -- (6,-1);
    \draw[->,thick] (5,-1) -- (4.3,-1.7);

    \node[above] at (5,0) {$\bx_9$};
    \node[right] at (6,-1) {$\bx_6$};
    \node[left] at (4.3,-1.7) {$\bx_5$};
    
    \node[above] at (0,-5) {(a)};
    
    \end{scope}
    \begin{scope}[xshift = 12cm]
    \draw[thick] (1,1) -- (-1,1);
    \draw[thick] (1,-1) -- (-1,-1);
    \draw[thick] (2,2) -- (-2,2);
    \draw[thick] (2,-2) -- (-2,-2);
    \draw[thick] (3,2.5) -- (2,2) -- (1,1) -- (1,-1) -- (2,-2) -- (3,-2.5);
    \draw[thick] (-3,2.5) -- (-2,2) -- (-1,1) -- (-1,-1) -- (-2,-2) -- (-3,-2.5);
    \draw[red,thick] (-1,0) -- (-4,-3);
    \draw[blue,thick] (-4.5,-3) -- (4.5,-3);

    \node[above] at (0,2) {$\vdots$};
    \node[left] at (-2,-0.7) {D3};
    \node[below] at (0,-2) {$\vdots$}; 
    \node[below] at (3,-3) {D5};

    \draw[->,thick] (5,-1) -- (5,-0);
    \draw[->,thick] (5,-1) -- (6,-1);
    \draw[->,thick] (5,-1) -- (4.3,-1.7);

    \node[above] at (5,0) {$\bx_9$};
    \node[right] at (6,-1) {$\bx_6$};
    \node[left] at (4.3,-1.7) {$\bx_5$};

    \node[above] at (0,-5) {(b)};
    \end{scope}
    \end{tikzpicture}
    \caption{D3-brane creation and higgsing construction of $Q$-observable. (a) One of the D5-branes is aligned across the two NS5-branes, enabling the D5-brane to move along the $x^5$-direction. (b) A D3-brane ending on one of the NS5-branes is created as a result of the transition.}
    \label{fig:Q-operator}
\end{figure}

We consider a second type of defect: co-dimensional two defect in the 5d $\CalN=1$ supersymmetric gauge theory. It can be constructed by partially higgsing from higher rank gauge theory \cite{SNN}. 
Let us consider $N+1$ D5 branes stretching between two NS5 branes, creating 5-brane web, with two additional D5 branes stretching from one of the 5-brane web to infinity.  
The low energy effective theory is a 5d $SU(N+1)$ gauge theory with a one fundamental hypermultiplets and one anti-fundamental hypermultiplets. 
Now consider aligning the position (in the $\rx^9$ direction) of one of the D5 brane across the NS5. Such alignment allows the D5 to move along in the $\rx^5$ direction, creating a D3-brane stretched between the D5 and one of the NS5 brane. 
Thus, after transition, we are left with the original $N$ D5 branes in between two NS5 (and the 5-brane web that created by the junction between D5 and NS5) and a D3-brane ending on the 5-brane web. See Table.~\ref{tab:D-branes-Q} for the D-brane construction and Figure.~\ref{fig:Q-operator} for the illustration.

\begin{table}[h]
    \centering
    \begin{tabular}{|c||c|c|c|c|c|c|c|c|c|c|}
    \hline
     & 0 & 1 & 2 & 3 & 4 & 5 & 6 & 7 & 8 & 9 \\
    \hline \hline 
    $(1,0)$ D5 & x & x & x & x & x & & x & & & \\
    \hline 
    $(0,1)$ NS5 & x & x & x & x & x &  &  & &  & x \\
    \hline
    D3 & x & x & x & & & x & & & &  \\
    \hline
    \end{tabular}
    \caption{The $Q$-observable construction in a $(p,q)$ 5-brane web for 5d $\CalN=1$ pure SYM.}
    \label{tab:D-branes-Q}
\end{table} 

In the 5d theory point of view, this alignment corresponds to tune the masses for the fundamental hypermultiplet and anti-fundamental hypermultiplet to zero (up to $\ve_1$ and $\ve_2$).
Then we higgs the gauge group from $SU(N+1)$ to $SU(N)$ by giving vacuum expectation value to the fundamental and anti-fundamental hypermultiplet. The relevant $U(1)$ part of the gauge group is squeezed into the three-dimensional space, giving a three dimensional $\CalN=4$ sigma model in the Coulomb phase coupled to a remnant five-dimensional $\CalN=1$ gauge theory with gauge group $SU(N)$. 

We now implement the computation inspired by the type IIB brane construction following the exact procedure. We start with $U(N+1)_\sk$ gauge theory with one fundamental and one anti-fundamental matter, whose masses $m^+=a_{N+1}$, $m^- = a_{N+1}-\ve_1-2\ve_2$ are fine tuned w.r.t one of the Coulomb moduli $a_{N+1}=x+\ve_1+\ve_2$: 
\begin{align} \label{eq:higgs2}
\begin{split}
    &\bN = \sum_{\a=1} ^{N} e^{a_\a} + e^{x+\ve_1+\ve_2}.
\end{split}
\end{align}
The higgsing condition $a_{N+1}=m^+=m^-+\ve_1+2\ve_2$ restricts the instanton configuration for the last Coulomb moduli parameter $\lambda^{(N+1)}=\emptyset$. The instanton partition function of the $U(N+1)_\sk$ theory under higgising constrain gives
\begin{align}
\begin{split}
    & \CalZ_{U(N+1)}(\ba,a_{N+1}-\ve_1-\ve_2=m^+-\ve_1-\ve_2=m^-+\ve_2=x;\kq) \\
    & = \sum_{\boldsymbol\lambda} \kq^{|\boldsymbol\lambda|} \BE \left[ - \frac{(\bS+q_{12}e^x)(\bS+q_{12}e^x)^* - q_{12}(\bS+q_{12}e^x)^* - q_{1}^*e^{-x}(\bS+q_{12}e^x) }{P_{12}^*} \right] \det(\bK)^\sk \\
    & = \sum_{\boldsymbol\lambda} \kq^{|\boldsymbol\lambda|} \BE\left[ - \frac{\bS\bS^*}{P_{12}^*}\right] \det(\bK)^\sk \times \BE \left[ -\frac{e^x\bS^*}{P_{1}^*} \right] \\
    & = \sum_{\boldsymbol\lambda} \kq^{|\boldsymbol\lambda|} \CalZ_\text{bulk}[\boldsymbol\lambda] Q(x)[\boldsymbol\lambda] \\
    & = \langle Q(x) \rangle \CalZ_{\BC^2_{12}}(\ba,\boldsymbol\ve;\kq)
\end{split}
\end{align}
Here we define $Q$-observable:
\begin{align}\label{def:Q-func}
    Q(x)[\boldsymbol\lambda]= \BE \left[ - \frac{e^x \bS^*}{P_1^*} \right], \quad \frac{Q(x)}{Q(x-\ve_1)} = \EY(x). 
\end{align}
In the NS-limit where the bulk instanton configuration is locked to limit shape $\boldsymbol\Lambda$, the vacuum expectation value of the $Q$-observable, called $Q$-function $Q_\ba(x)$, has zeros at all $v_{\a \ri}$ \eqref{def:v_ai}:
\begin{align}
\begin{split}
    Q_\ba(x) & = 
    \lim_{\ve_2\to0}\langle Q(x) \rangle = Q(x)[\boldsymbol\Lambda] = \BE \left[ - \frac{e^x\bS^*[\boldsymbol\Lambda]}{P_1^*} \right] = \BE \left[ - \frac{e^xN^*}{P_1^*} + e^x \mathring\bV^* - e^x \bV^* \right] \\
    & = \prod_{\alpha=1}^N e^{\frac{(x-a_\alpha)(x-a_\alpha+\ve_1)}{4\ve_1}} (e^{-(x-a_\alpha)},q_1)_\infty \prod_{i=1}^N \frac{\sh(x-v_{\alpha i})}{\sh(x-\mathring{v}_{\alpha i})}.
\end{split}
\end{align}
It is easy to check the $Q$-function satisfies the Baxter T-Q equation for $Y^{N,\sk}$ model:
\begin{align}\label{eq:Baxter-TQ}
\begin{split}
    & T_\sk(x) = \langle \EX_\sk(x) \rangle = \EX_\sk(x)[\boldsymbol\Lambda] = \EY(x+\ve_1)[\boldsymbol\Lambda] + \frac{\kq e^{\sk x}}{\EY(x)[\boldsymbol\Lambda]} = \frac{Q_\ba(x+\ve_1)}{Q_\ba(x)} + \kq e^{\sk x} \frac{Q_\ba(x-\ve_1)}{Q_\ba(x)}. \\
    & \implies T_\sk(x) Q_\ba(x)  = Q_\ba(x+\ve_1) + \kq e^{\sk x} Q_\ba(x-\ve_1) 
\end{split}
\end{align}
The poles from the two terms on the RHS of the equation above must cancel each other as on the left $T_\sk(x)$ is analytic in $x$. This gives us the Bethe equation: 
\begin{align}
    1 = - \kq e^{\sk v_{\alpha \ri}} \frac{Q_\ba(v_{\alpha \ri}-\ve_1)}{Q_\ba(v_{\alpha \ri}+\ve_1)} = \kq e^{\sk v_{\alpha \ri}} \prod_{(\beta \rj) \neq (\alpha \ri)} \frac{\sh(v_{\alpha \ri}-v_{\beta \rj}-\ve_1)}{\sh(v_{\alpha \ri}-v_{\beta \rj}+\ve_1)}.
\end{align}
for all $v_{\alpha \ri}$. We identify $\{v_{\a\ri}\}$ as the Bethe roots of the $Y^{N,\sk}$ integrable model.

\subsection{Monodromy defect}

The monodromy defect is introduced in the form of an $\BZ_l$ oribifolding \cite{Nikita:IV} acting on $\BR^4=\BC_1 \times \BC_2$ by $(\bz_1,\bz_2) \to (\bz_1,\eta \bz_2)$ with $\eta^l = 1$. 
Here and below $\fR_\omega$ denotes the one-dimensional complex irreducible representation of $\BZ_l$, where the generator $\eta$ is represented by the multiplication of $\exp\left( \frac{2\pi i \omega}{l}\right)$ for $\omega \equiv \omega + l$. In general one can consider $\BZ_l$ orbifold of any integer $l$.
The orbifold generates a chainsaw quiver structure \cite{Nakajima:2011yq, Kanno:2011fw,JB2019}.
Such defect is characterized by a coloring function $c:[N] \to \BZ_l$ that assigns the $\BZ_l$ charge to the Coulomb moduli parameters. In addition when the bulk has non-zero Chern-Simons level $\sk\neq 0$, We will also need to assign fractional Chern-Simons levels $\sk_\o$ to the $\fR_\o$ representation of the orbifold, which must obeys
\begin{align}
    \sum_{\o\in \BZ_l} \sk_\o = \sk. 
\end{align}
A surface defect is called \emph{full-type/regular surface defect} if $l=N$ and the coloring function $c$ bijective. More detail discussion about the full-type surface defect can be found in \cite{Kanno:2011fw, Nikita:IV,Finkelberg:2010JEMS,Feigin:2011SM,Bruzzo:2010fk}

For later convenience, we will first fix the convention for our notation. We will denote $\hat\ve_1=\ve_1$, $\hat\ve_2$ for the $\Omega$-parameters on the orbifold. We denote
\begin{align}
    \hat{a}_{\o=c(\alpha)}
\end{align}
as the Coulomb moduli parameter that is assigned to the $\fR_\o$ representation of the regular surface defect.  
In order for the contour integration formula (where the Young diagram combintorial formula originate) to converge, the coloring function assigning $\sk_\o$ needs to obey
\begin{align}
    \sk_\o \leq 1
\end{align}
in the presence of regular surface defect.

There are $N$ couplings $(\hat\kq_\o)_{\o=0}^{N-1}$ for gauge defined on the space $\hat\BC^2_{12} = \BC_1 \times \BC_2/\BZ_N$. In the language of string theory, the bulk instantons fractionalize into $N$ types. The coupling $\hat\kq_\o$ counts the number of fractional instantons of type $\o$, so that $\hat\kq_{\o+N} = \hat\kq_\o$. The bulk coupling can be recovered by
\begin{align}
    \kq = \hat\kq_0 \hat\kq_1 \cdots \hat\kq_{N-1}. 
\end{align}
The coupling $\hat\kq_\omega$ is assigned to the representation $\fR_\omega$ of the $\BZ_l$. We define the fractional variables:
\begin{align} \label{def:z}
    \hat\kq_{\o} = \frac{z_{\o+1}}{z_\o}, \ z_{\o+N} = \kq z_{\o}, \o=0,\dots,N-1.
\end{align}
From string theory point of view, these variables $\{z_\omega\}_{\omega = 0, \dots, N-1}$ are interpreted as the (exponentiated) brane positions, whereas the couplings $\{\hat\kq_\omega\}_{\omega = 0,\dots,N-1}$ are (exponentiated) distances between the branes. 



For the convenience of later calculation, we scale $\ve_2 \to \hat\ve_2 = \frac{\ve_2}{N}$ and define the shifted moduli
\begin{align}
    \hat{a}_\o - {\hat\ve_2} \o = {a}_\o
\end{align}
which are neutral under the orbifolding. 
All the ADHM data can now be written in terms of the shifted moduli
\begin{align}
\begin{split}
    & \hat{\bN} = \sum_{\o=0}^{N-1} e^{a_\o} \hat{q}_2^{{\o}} \fR_\o, \quad \bN = \sum_{\o=0}^{N-1} \bN_\o; \\
    & \hat{\bK} = \sum_{\o=0}^{N-1} \bK_{\o} \hat{q}_2^{{\o}} \fR_\o, \quad \bK_{\omega} = \sum_{\alpha=1} e^{{a}_\alpha} \sum_{J=0}^\infty \sum_{\underset{c(\alpha)+j-1=\omega+NJ}{{(i,j)\in \lambda^{(\alpha)}}}} q_1^iq_2^J = \sum_{\Box \in \EK_\o} e^{c_\Box}; \\
    & \hat\bS = \hat\bN - \hat{P}_1 \hat{P}_2 \hat\bK = \sum_{\o=0}^{N-1} \bS_\o \hat{q}_2^{\o} \fR_\o. 
\end{split}
\end{align}
Here $c_\Box = a_{c(\alpha)}+(\ri-1)\ve_1 + J\ve_2$ with
\begin{align}\label{def:k_w}
     & k_\o = \# \EK_\o, \ \EK_\o = \left \{ \Box=(\alpha,(\ri,\rj)) \mid \alpha \in [N], \ (\ri,\rj)\in \lambda^{(\alpha)}, \ c(\alpha)+\rj-1 = \omega \text{ mod }(N) \right\}
\end{align}
denoting the number of squares in a colored Young diagram of the $\omega$-th type, that is in the $\fR_\omega$ representation of the $\BZ_N$ orbifold. The $\Omega$-deformation parameters on the orbifold 
\begin{align}
    \hat{q}_1 = q_1 \fR_0, \ \hat{q}_2 = q_2^{\frac{1}{N}} \fR_1, \ \hat{P}_1 = P_1 \fR_0, \ \hat{P}_2 = \fR_0 - q_2^{\frac{1}{N}} \fR_1.
\end{align}
We have defined the fractional characters
\begin{align}
\begin{split}
    & \bS_\o = \bN_\o - P_1 \bK_\o + P_1 \bK_{\o-1}, \ \o=1,\dots,N-1; \\
    & \bS_0 = \bN_0 - P_1 \bK_{0} + q_2 P_1 \bK_{N-1}. 
\end{split}
\end{align}
Then the character of the universal sheaf is obtained by summing over all the fractional characters $\bS_\o$:
\begin{align}\label{def:bulk-S}
    \bS = \sum_{\o=0}^{N-1} \bS_\o = \bN - P_{1}P_2 \bK_{N-1}. 
\end{align}
The defect instanton partition function on the oribifolded space $\hat{\BC}_{12}^2$ is an ensemble over all defect instanton configuration $\hat{\boldsymbol\lambda}$ 
\begin{align}\label{def:defect-part}
    & \hat\CalZ_{\hat{\BC}_{12}^2}(\ba,\boldsymbol\ve,{z}_\o;\kq) = \sum_{\hat{\boldsymbol\lambda}} \prod_{\omega=0}^{N-1} \hat\kq_\omega ^{k_\omega} \hat\CalZ_\text{defect}[\hat{\boldsymbol\lambda}] 
\end{align}
where the pseudo-measure is a product over the $\BZ_N$ invariant contribution from the vectormultiplet and the fractional Chern-Simons term:
\begin{align}\label{eq:defect-inst-part}
\begin{split}
    \hat\CalZ_\text{defect}[\hat{\boldsymbol\lambda}] 
    & = \BE \left[ -\frac{\hat\bS \hat\bS^*}{\hat{P}_{12}^*} \right]^{\BZ_N} \prod_{\o=0}^{N-1} (\det \bK_\o)^{\sk_\o} \\
    & = \BE \left[ - \frac{\bS\bS^*}{P_{12}^*} + \sum_{\o_1<\o_2} \frac{\bS_{\o_1}\bS_{\o_2}^*}{P_1^*} \right]
    \prod_{\o=0}^{N-1} \left[ \prod_{\Box \in \EK_\o} e^{c_\Box} \right]^{\sk_\o}. 
\end{split}
\end{align} 

We define the fractional $\EY_\o(x)$ function:
\begin{align}
    \EY_\o(x) = \BE \left[ - e^x \bS_\o^* \right]
\end{align}
so that the bulk $\EY(x)$ is the product over all fractional $\EY_\o(x)$ by the virtue of \eqref{def:bulk-S}: 
\begin{align}
    \EY(x) = \prod_{\o=0}^{N-1} \EY_\o(x). 
\end{align}

\paragraph{}
Here we will show that the defect partition function \eqref{def:defect-part} can be viewed as the expectation value of surface defect observable in the bulk theory. As we have seen the bulk instanon configuration is the colored instantons of the $(N-1)$-type, in other words assigned to the $\fR_{N-1}$ representation of the $\BZ_N$ group. We define the projection $\rho(\hat{\boldsymbol\lambda})=\boldsymbol\lambda$ from the colored partition $\hat{\boldsymbol\lambda}$ to the bulk partition $\boldsymbol\lambda$ \cite{Lee:2020hfu,Jeong:2018qpc} where
\begin{align}
    \lambda^{(\alpha)}_\ri = \left\lfloor \frac{\hat{\lambda}_\ri^{(\alpha)}+c(\alpha)}{N} \right \rfloor, \ 1 \leq \ri \leq \ell_\alpha , \ \alpha=1,\dots,N. 
\end{align}
$\ell_\alpha$ is the height of the colored partition $\hat\lambda^{(\alpha)}$.
Using the shifted moduli, the defect partition function can be reorganized as
\begin{align}\label{def:defect-part-bs}
    \hat\CalZ_{\hat{\BC}_{12}^2} = \sum_{\boldsymbol\lambda} \kq^{|\boldsymbol\lambda|} \CalZ_\text{bulk}[\boldsymbol\lambda] \sum_{\hat{\boldsymbol\lambda}\in \rho^{-1}(\boldsymbol\lambda)} \prod_{\o=0}^{N-1} z_\o^{k_{\o-1}-k_\o} \hat\CalZ_\text{surface}[\hat{\boldsymbol\lambda}]
\end{align}
The bulk and regular surface defect contribution to the canonical ensemble are 
\begin{align}
\begin{split}
    & \CalZ_\text{bulk} [\boldsymbol\lambda] = \BE \left[ -\frac{\bS\bS^*}{P_{12}^*} \right] \times \left[ \prod_{\Box\in \EK_{N-1}} e^{c_\Box} \right]^{\sk} \\
    & \hat\CalZ_\text{surface} [\hat{\boldsymbol\lambda}] = \BE \left[ \sum_{0\leq\o_1<\o_2\leq N-1} \frac{ \bS_{\o_1} \bS_{\o_2}^* }{P_1^*}\right] \prod_{\o=0}^{N-2} \left[ \frac{\det(\bK_\o) }{\det(\bK_{N-1})} \right]^{\sk_{\o}}
\end{split}
\end{align}
We can treat the defect term as an observable in the bulk theory
\begin{align}
    \hat\CalZ_{\hat{\BC}_{12}^2}(\ba,\boldsymbol\ve,\kq) = \sum_{\boldsymbol\lambda} \kq^{|\boldsymbol\lambda|} \CalZ_\text{bulk} [\boldsymbol\lambda] \ES_{c,\sk_\o}[\boldsymbol\lambda] = \langle \ES_{c,\sk_\o} \rangle \CalZ_{\BC_{12}^2}(\ba,\boldsymbol\ve,\kq). 
\end{align}
where
\begin{align}
    \ES_{c,\sk_\o}[\boldsymbol\lambda] = \sum_{\hat{\boldsymbol\lambda}\in \rho^{-1}(\boldsymbol\lambda)} \prod_{\o=0}^{N-1} z_\o^{k_{\o-1}-k_\o} \hat\CalZ_\text{surface}[\hat{\boldsymbol\lambda}]. 
\end{align}

By multiplying the perturbative factor, we define the full vacuum expectation value of the surface defect observable as
\begin{align}\label{def:Psi}
    \Psi(\ba,\boldsymbol\ve,\kq,z_\o) = \prod_{\o=0}^{N-1} z_\o^{-\frac{a_\o}{\ve_1}} \hat\CalZ_{\hat\BC_{12}^2}. 
\end{align}

In the NS-limit $\ve_2 \to 0$, the defect partition function $\Psi$ has the following asymptotics:
\begin{align}
    \lim_{\ve_2\to 0} \Psi = e^{\frac{\widetilde\CalW(\ba,\ve_1,\kq)}{\ve_2}} \psi(\ba,\kq,z_\o,\ve_1)
\end{align}
The singular part is the twisted superpotential identical to that of the bulk instanton in \eqref{def:super_potential}. 
We denote $\psi$ the \emph{normalized vev of the monodromy surface defect} by
\begin{align}\label{def:psi2}
    \psi(\ba,\ve_1,z_\o) = \lim_{\ve_2\to 0} \frac{\Psi(\ba,\boldsymbol\ve,\kq,z_\o)}{\CalZ_{\BC^2_{12}} (\ba,\boldsymbol\ve,\kq)} = \prod_{\o=0}^{N-1} z_\o^{-\frac{a_\o}{\ve_1}} \ES_{c,\sk_\o}[\boldsymbol\Lambda].
\end{align}
The normalized vev of the monodromy surface defect in comparison to the twisted superpotential in the bulk has been less studied. However, as shown in many recent studies \cite{Chen:2019vvt,Kimura:2022zsx,Lee:2020hfu,Jeong:2023qdr}, the normalized surface defect partition function is identified as the wave function of the scattering state in the quantum integrable system, as we will see in the next chapter.

\section{Bethe/gauge correspondence}\label{sec:Baxter}

In this section we formally establish the Bethe/gauge correspondence between the new dimer integrable system and 5d $\CalN=1$ SYM with defects by reconstructing all commuting quantum Hamiltonians of the new dimer integrable system, proving its quantum integrability. 

\subsection{Quantum Hamiltonian}

Let us start by showing the normalized vev of the monodromy surface defect of $SU(N)_\sk$ SYM satisfies the Schr\"{o}dinger equation of the $Y^{N,\sk}$ model. 
The differential equation acting on the monodromy defect can be derived from the regularity property of the $qq$-character \cite{Nikita:V}.

The fractional $qq$-character can be obtained by study the change of measure under the introduction of a single instanton of the $\omega$-th type $\bK_\o \to \bK_\o + X$, $X=e^x$ in the defect instanton pseudo measure \eqref{eq:defect-inst-part}:
\begin{align}
\begin{split}
    \hat\kq_\o\frac{\hat\CalZ_\text{defect}[\bK_\o+X]}{\hat\CalZ_\text{defect}[\bK_\o]} 
    & = \hat\kq_\o X^{\sk_{\o}} \BE \left[ e^x q_1 (\bS_{\o+1}+P_1X)^* + \bS_\o X^* \right] \\
    & = \frac{\hat\kq_\o e^{\sk_\o x}}{\EY_{\o+1}(x+\ve_1+\ve_2\d_{\o,N-1})[\bK_{\o}+X] \EY_\o(x)[\bK_\o]}. 
\end{split}
\end{align}
We construct the fractional $qq$-character:
\begin{align}\label{def:qq-frac}
\begin{split}
    \EX_\o (x)[\hat{\boldsymbol\lambda}] & = \EY_{\o+1}(x+\ve_1)[\hat{\boldsymbol\lambda}] + \frac{\hat\kq_\o e^{\sk_\o x}}{\EY_\o(x)[\hat{\boldsymbol\lambda}]}, \\
\end{split}
\end{align}
with $\EY_N(x)[\hat{\boldsymbol\lambda}] = \EY_0(x+\ve_2)[\hat{\boldsymbol\lambda}]$. The vacuum expectation value of fractional $qq$-character, which is defined through an ensemble over all defect instanton configuration, 
\begin{align}
\begin{split}
    \langle \EX_\o(x) \rangle_{\BZ_N} & = \frac{1}{\hat\CalZ_{\hat\BC_{12}^2} } \sum_{\hat{\boldsymbol\lambda}} \prod_{\omega} \hat\kq_\omega ^{k_\omega} \hat\CalZ_\text{defect} [\hat{\boldsymbol\lambda}] \EX_\o(x)[\hat{\boldsymbol\lambda}] \\
    & = \langle \EY_{\o+1}(x+\ve_1) \rangle_{\BZ_N} + \left \langle \frac{\hat\kq_\o e^{\sk_\o x}}{\EY_\o(x)} \right \rangle_{\BZ_N}
\end{split}
\end{align}
is an analytic function in $x$ \cite{Nikita:IV}.

A function $f(X=e^x)$ analytic in $x$ means that it can only have pole at $X=0$ and $X=\infty$, which corresponds to $x \to -\infty$ and $x\to \infty$ respectively. $f(X)$ is analytic in $X$ once the pole at $X=0$ is removed. In the case of fractional $qq$-character $\EX_\o(x)$, we consider
\begin{align}
\begin{split}
    f(X) = & \left \langle \frac{\EX_\o(X)}{\sqrt{X}}  \right \rangle_{\BZ_N} .
\end{split}
\end{align}
It's expectation value is analytic in $X$. The coefficient of $X^{-1}$ order in the Laurent expansion of the $qq$-character in $X$ should vanish
\begin{align}
    \left[ X^{-1} \right] \left \langle \frac{\EX_\o(X)}{\sqrt{X}} - \frac{1}{X} \underset{X=0}{\text{Res}} \frac{\EX_\o(X)}{\sqrt{X}} \right \rangle_{\BZ_N} = 0.
\end{align}
Despite the equation seems to be trivial by its definition. 
We can explicitly calculate the right hand side of the equation above by considering the large $X$ expansion of the $qq$-character. The building block $\EY_\o(x)$ in the large $X=e^x$ behaves
\begin{align}\label{eq:Y-large-X}
    \EY_{\o}(x) = \sqrt{X} e^{\frac{a_\o}{2}} q_1^{\frac{k_\o-k_{\o-1} }{2}} \left( 1-\frac{e^{a_\o}}{X}\right) \exp \left[ \frac{1-q_1}{X} \left( \sum_{\Box\in \EK_{\o}} e^{c_\Box} - \sum_{\Box\in \EK_{\o-1}} e^{c_\Box}q_2^{\d_{\o,0}} \right) + \cdots \right]
\end{align}

Let us first consider the case where the bulk Chern Simons level $\sk=0$ ($\sk_\o = 0$ for all $\omega=0,\dots,N-1$). Take the large $X$ expansion on $f(X)$ using \eqref{eq:Y-large-X} and consider the coefficient of $X^{-1}$ in the Laurant series in $X$:
\begin{align}
\begin{split}
    & \left[ X^{-1} \right] \left[ \frac{\EX_\o(X)}{\sqrt{X}} - \frac{1}{X} \underset{X=0}{\text{Res}} \frac{\EX_\o(X)}{\sqrt{X}} \right] \\
    & = e^{\frac{-a_{\o+1}+\ve_1}{2}} e^{\frac{\ve_1}{2}(k_{\o}-k_{\o+1})} \left[ - e^{a_{\o+1}-\ve_1} + \sum_{\Box \in \EK_{\o+1}} e^{c_\Box}(q_1^{-1}-1) + \sum_{\Box\in \EK_{\o}} e^{c_\Box} (1-q_1^{-1}) q_2^{\delta_{N-1,\o}} \right] \\
    & \quad + \hat\kq_\o e^{\frac{a_\o}{2}} e^{\frac{\ve_1}{2} (k_{\o}- k_{\o-1}) } + e^{\frac{a_{\o+1}-\ve_1}{2}} e^{\frac{\ve_1}{2}(k_{\o+1}-k_\o) }.
\end{split}
\end{align}

We consider a linear combination 
\begin{align}\label{eq:liner-combo-C}
    \sum_{\o=0}^{N-1} \hat{C}_\o \left \langle \left[ X^{-1} \right] \left[ \frac{\EX_\o(X)}{\sqrt{X}} - \frac{1}{X} \underset{X=0}{\text{Res}} \frac{\EX_\o(X)}{\sqrt{X}} \right] \right\rangle_{\BZ_N} \Psi
\end{align}
with the coefficient $\hat{C}_\o$ chosen to cancel the terms related to $e^{c_\Box}$:
\begin{align}
\begin{split}
    0 
    &= \hat{C}_{\o} \left\langle e^{\frac{-a_{\o+1}+\ve_1}{2}} q_1^{\frac{k_{\o}-k_{\o+1}}{2}} \sum_{\Box \in \EK_{\o}} e^{c_\Box}  \right\rangle_{\BZ_N}
    - \hat{C}_{\o-1} \left \langle e^{\frac{-a_\o+\ve_1}{2}} q_1^{\frac{k_{\o-1}-k_\o}{2}} \sum_{\Box\in \EK_\o} e^{c_\Box} \right \rangle_{\BZ_N} \Psi \\
    & = \left[ \hat{C}_\o q_1^{\frac{1}{2}\nabla^z_{\o+1} + \frac{1}{2} } - \hat{C}_{\o-1} q_1^{\frac{1}{2}\nabla^z_{\o}+\frac{1}{2}} \right] \left \langle \sum_{\Box \in \EK_{\o}} e^{c_\Box} \right \rangle_{\BZ_N} \Psi.
\end{split}
\end{align}
By choosing
\begin{align}
    \hat{C}_\o = q_1^{-\frac{1}{2}\nabla^z_{\o+1}-\frac{1}{2}}
\end{align}
the linear combination gives
\begin{align}\label{eq:k=0,linear}
\begin{split}
    0= & q_1 \sum_{\o=0}^{N-1} \hat{C}_\o \left \langle \left[ X^{-1} \right] \left[ \frac{\EX_\o(X)}{\sqrt{X}} - \frac{1}{X} \underset{X=0}{\text{Res}} \frac{\EX_\o(X)}{\sqrt{X}} \right] \right\rangle_{\BZ_N} \Psi \\
    & = \left[ -  \langle \bS \rangle_{\BZ_N} + \sum_{\o=0}^{N-1} \hat\kq_\o q_1^{-\frac{1}{2} \nabla^z_{\o}-\frac{1}{2}\nabla^z_{\o+1}} + q_1^{-\nabla^z_{\o+1}}  \right] \Psi. 
\end{split}
\end{align}
$\bS$ is the universal sheaf defined in \eqref{def:bulk-S}
\begin{align}
    \bS = \sum_{\o=0}^{N-1} e^{a_\o} - (1-q_1)(1-q_2) \sum_{\Box \in \EK_{N-1}} e^{c_\Box}
\end{align}
which only depends on the bulk instanton configuration $\boldsymbol\lambda$. 

In addition, the expectation value over the defect partition function ensemble for any bulk observable $\CalO_\text{bulk}$ that is finite in the $\ve_2 \to 0$ limit factorized by
\begin{align}
    \lim_{\ve_2\to 0} \langle \CalO_\text{bulk} \rangle_{\BZ_N} =  \CalO_\text{bulk}[\boldsymbol\Lambda]. 
\end{align}

In the $\ve_2\to 0$ limit, \eqref{eq:k=0,linear} becomes time-independent Schr\"{o}dinger equation for $\psi(\ba,\kq,z_\o,\ve_1)$:
\begin{align}
    \hat{\rm H}|_{\sk=0} \psi(\ba,\kq,z_\o,\ve_1) = E \psi(\ba,\kq,z_\o,\ve_1), \ E = \bS[\boldsymbol\Lambda]
\end{align}
with the Hamiltonian
\begin{align}\label{eq:H-N-0}
    \hat{\rm H}|_{N,\sk=0} = \sum_{\o=0}^{N-1} q_1^{-\nabla^z_{\o+1}} + \hat\kq_\o q_1^{-\frac{1}{2}(\nabla^z_{\o+1} + \nabla^z_\o) }. 
\end{align}

\paragraph{}

When the bulk theory has non-zero Chern-Simons level $\sk \neq 0$. We will need to assign the fractional Chern-Simons levels $\sk_\o$ to each of the representation $\fR_\o$ of the $\BZ_N$ orbifold. Up to a permutation, we can always assign $\sk_\o=1$
for $\o=0,\dots,\sk-1$ and $\sk_\o=0$ for $\o=\sk,\dots,N-1$. 
We take the large $X$ expansion on the $N$ fractional $qq$-characters in \eqref{def:qq-frac} and consider the linear combination as in \eqref{eq:liner-combo-C}. 
Let us denote
\begin{align}\label{def:hatD}
    \hat{D}_\o = \hat\kq_\o q_1^{-\frac{1}{2}\nabla^z_\o - \frac{1}{2}\nabla^z_{\o+1}} = \hat{D}_{\o+N}.
\end{align}
When $0<\sk<N$, we choose the coefficients $\hat{C}_\o$ as
\begin{subequations}
\begin{align}
    & \hat{C}_{{\o-1}} = \left[ 1 + \hat{D}_\o + \hat{D}_{\o+1} \hat{D}_{\o} + \cdots + \hat{D}_{\sk-1} \cdots \hat{D}_\o \right] q_1^{-\frac{1}{2}\nabla^z_{\o}-\frac{1}{2}} , \ \o=1,\dots,\sk-1; \\
    & \hat{C}_{\o-1} = q_1^{-\frac{1}{2}\nabla^z_{\o}-\frac{1}{2}}, \ \o=\sk,\dots,N-1; \\
    & \hat{C}_{N-1} = \left[ 1 + \hat{D}_0 + \hat{D}_1 \hat{D}_0 + \cdots + \hat{D}_{\sk-1} \cdots \hat{D}_0 \right] q_1^{-\frac{1}{2}\nabla^z_{0}-\frac{1}{2}}. 
\end{align}
\end{subequations}
The linear combination in \eqref{eq:liner-combo-C} with $\hat{C}$ chosen as above gives
\begin{align}
\begin{split}
    0 = & q_1 \sum_{\o=0}^{N-1} \hat{C}_\o \left \langle \left[ X^{-1} \right] \left[ \frac{\EX_\o(X)}{\sqrt{X}} - \frac{1}{X} \underset{X=0}{\text{Res}} \frac{\EX_\o(X)}{\sqrt{X}} \right] \right \rangle_{\BZ_N} \Psi \\
    = & - \langle \bS \rangle_{\BZ_N} \Psi + \hat{\rm H} \Psi.
\end{split}
\end{align}
The Hamiltonian is
\begin{align}\label{eq:H-N-k}
\begin{split}
    \hat{\rm H}|_{N,\sk} 
    = &  \sum_{\o=0}^{\sk-1} \left( 1 + \hat{D}_{\o} + \cdots + \hat{D}_{\sk-1}\cdots \hat{D}_{\o} \right) q_1^{-\nabla^z_{\o}} + \sum_{\o=\sk}^{N-2} q_1^{-\nabla^z_{\o}} + \hat{D}_{\o} \\
    & + q_1^{-\nabla^z_{N-1}} + \hat{D}_{N-1} + \hat{D}_0 \hat{D}_{N-1} + \cdots + \hat{D}_{\sk-1} \cdots \hat{D}_0 \hat{D}_{N-1}.
\end{split}
\end{align}
The total number of terms in the Hamiltonian is 
\begin{align}
    \sum_{\o=0}^{\sk-1} \sk-\o + \sum_{\o=\sk}^{N-2} 2 + 2 \times (\sk+1) - \d_{\sk,N-1} = 2N + \frac{\sk(\sk+1)}{2} - \d_{\sk,N-1}.  
\end{align}
In particular when $\sk=N-1$ one of them is $\hat{D}_{N-2} \cdots \hat{D}_{0} \hat{D}_{N-1} q_1^{-1} = \kq e^{a}$. This term can be moved to the other side of the Schr\"{o}dinger equation. 
See Appendix.~\ref{sec:compu-sec4} for detail of calculation. 




\paragraph{}
When the bulk Chern-Simons level equals to the rank of the gauge group $\sk=N$, i.e. $\sk_{\o}=1$ for all $\o=0,\dots,N-1$, the coefficients $\hat{C}_\o$ of the linear combination is chosen as
\begin{align}
\begin{split}
    & \hat{C}_{\o} = \left[ {1 + \hat{D}_{\o+1} + \hat{D}_{\o+2} \hat{D}_{\o+1} + \cdots + \hat{D}_{\o+N-1} \cdots \hat{D}_{\o+1}} \right] q_1^{-\frac{1}{2} \nabla^z_{\o} - \frac{1}{2} }, \\
\end{split}
\end{align}
The linear combination \eqref{eq:liner-combo-C} with the coefficient $\hat{C}_\o$ chosen gives
\begin{align}
\begin{split}
    0 = & q_1 \sum_{\o=0}^{N-1} \hat{C}_\o \left \langle \left[ X^{-1} \right] \left[ \frac{\EX_\o(X)}{\sqrt{X}} - \frac{1}{X} \underset{X=0}{\text{Res}} \frac{\EX_\o(X)}{\sqrt{X}} \right] \right \rangle_{\BZ_N} \Psi \\
    = & - \left( 1 - \kq e^{a} q_1^{\frac{1-N}{2}} \right) \langle \bS \rangle_{\BZ_N} \Psi + \hat{\rm H}|_{N,\sk=N} \Psi
\end{split}
\end{align}
The Hamiltonian  
\begin{align}\label{eq:H-N-N}
    \hat{\rm H}|_{N,\sk=N} = \sum_{\o=0}^{N-1} \left[ 1 + \hat{D}_{\o} + \cdots + \hat{D}_{\o+N-2} \cdots \hat{D}_\o \right] q_1^{-\nabla^z_{\o}}, \ \hat{D}_{\o} = \hat\kq_{\o} q_1^{-\frac{1}{2}\nabla^z_{\o+1}-\frac{1}{2}\nabla^z_\o}
\end{align}
consists $N^2$ terms. 

We find that the Hamiltonians $\hat{\rm H}$ in \eqref{eq:H-N-0}, \eqref{eq:H-N-k}, and \eqref{eq:H-N-N} are the quantization of the $H_{N-1}$ Hamiltonian of $Y^{N,\sk}$ model in \eqref{eq:YNK-HN-1} with the identification
\begin{align}
    \hat\kq_{n} = R^2 e^{\rx_{n+1}-\rx_{n}},\ \rx_{n+N} = \rx_{n}, \ n=1,\dots,N. 
\end{align}
The quantum uplift of the 1-loops $u_n,d_n$ in the $Y^{N,\sk}$ dimer model \eqref{eq:ud-qp standard} is
\begin{align}\label{eq:1-loop to pq}
\begin{split}
    \hat{u}_n = e^{\ve_1\p_{\rx_{n}}}, \hat{d}_n = R^2 e^{\rx_{n}-\rx_{n-1}} e^{ \frac{\ve_1}{2}(\p_{\rx_{n}}+\p_{\rx_{n-1}}) - \ve_1\theta_{n\leq \sk}\p_{\rx_{n-1}} }, \ n=1,\dots,N. 
\end{split}
\end{align}

\subsection{Quantum T-Q equation}
In the previous section, we prove that the wave function of $Y^{N,\sk}$ model with standard gluing is given by the normalized vacuum expectation value of the monodromy surface defect:
\begin{align}
    \psi(\ba,\ve_1,z_\o) = \lim_{\ve_2\to 0} \frac{\Psi(\ba,\boldsymbol\ve,\kq,z_\o)}{\CalZ_{\BC^2_{12}}(\ba,\boldsymbol\ve,\kq)} = \prod_{\o=0}^{N-1} z_\o^{-\frac{a_\o}{\ve_1}} \ES_{c,\sk_\o}[\boldsymbol\Lambda].
\end{align}
Our goal in this section is to explore the correspondence further. Specifically: 
We will prove normalized vacuum expectation value of the monodromy surface defect $\psi$ is the shared eigenfunction of the commuting Hamiltonians of $Y^{N,\sk}[S]$ model specified by gluing set $S$.

The main tool we use is \emph{gauge origami} introduced in \cite{Nikita:III}. A review can be found in the Appendix.~\ref{sec:GO}. 

We consider three stacks of branes: $n_{12,0}=N$ regular branes on $S^1\times \BC^2_{12}$, one $\CalR_1$-type brane warpping on $S^1 \times \BC^2_{13}$, and one $\CalR_0$-type brane warpping on $S^1 \times \BC^2_{34}$:
\begin{align}\label{eq:gosetup}
\begin{split}
    & {\bN}_{12} = \sum_{\alpha=1}^N e^{a_\alpha} \CalR_0 \\
    & {\bN}_{13} = e^{x'} q_1 q_3 \CalR_1 \\
    & {\bN}_{34} = e^x \CalR_0
\end{split}
\end{align}
There are three gauge coupling $\kq_{0,1,2}$ and three Chern-Simons levels $\sk_{0,1,2}$ for each of the representation $\CalR_{0,1,2}$ of the cyclic group $\Gamma_{34}=\BZ_3$. 
We take the freezing limit $\kq_{1} = \kq_2 = 0$ and denote $\kq = \kq_0$, $\sk=\sk_0$. The gauge origami partition function is the invariant part of the $\Gamma_{34}$ orbifold in \eqref{def:part-GO}.
Combining all ingredients, the partition function of the gauge origami becomes
\begin{align}
\begin{split}
    \CalZ_\text{GO} & = \sum_{\boldsymbol{\lambda}} \kq^{|\boldsymbol{\lambda}|} \CalZ_\text{bulk}[\boldsymbol\lambda] \left[ \sh(x-x') \EY(x+\ve_{12})Q(x') + \kq e^{\sk x} \sh(x-x'+\ve_2) \frac{Q(x')}{\EY(x)} \right] \\
    & = \langle T(x,x')Q(x') \rangle \CalZ_{\BC^2_{12}}.
\end{split}
\end{align}
The $qq$-character is
\begin{align}
    T(x,x'){Q}(x') = \sh(x-x') \EY(x+\ve_{12}) {Q}(x') + \kq e^{\sk x} \sh(x-x'+\ve_2) \frac{{Q}(x')}{\EY(x)}. 
\end{align}
By the compactness of the spiked instanton \cite{Nikita:II}, the expectation value of the left hand side
$\left \langle T(x,x') {Q}(x') \right \rangle$
is a regular function in $x$. In particular
\begin{subequations}
\begin{align}
    & T(x=x',x') {Q}(x') = \kq e^{\sk x'} \sh\ve_2 \frac{{Q}(x')}{\EY(x')} = \kq e^{\sk x'} \sh\ve_2 {Q}(x'-\ve_1) \\
    & T(x=x'-\ve_2,x') {Q}(x') = \sh(-\ve_2) \EY(x'+\ve_1) {Q}(x') = -\sh\ve_2 {Q}(x'+\ve_1)
\end{align}
\end{subequations}
Define 
\begin{align}
    \ET(x') = \frac{T(x=x',x') - T(x=x'-\ve_2,x') }{\sh\ve_2}
\end{align}
and 
\begin{align}\label{def:EQ}
    \EQ(x) = e^{-\frac{\sk x(x+\ve_1)}{4\ve_1}} Q(x). 
\end{align}
We obtain the elliptic T-Q equation:
\begin{align}\label{eq:TQ-bulk}
    \langle \ET(x) \EQ(x) \rangle = e^{\frac{\sk}{2}(x+\ve_1)} \langle \EQ(x+\ve_1) \rangle + \kq e^{\frac{\sk}{2}x} \langle \EQ(x-\ve_1) \rangle
\end{align}
where $e^{\frac{N}{2}x} \cdot \ET(x)$ is a degree $N$ polynomial in $e^x$: 
\begin{align}
    e^{\frac{N}{2}x} \cdot \ET(x) =  e^{Nx} + E_1 e^{(N-1)x} + \cdots + E_N.  
\end{align}
Eq.~\eqref{eq:TQ-bulk} can be consider a quantum uplift of the Baxter T-Q equation in \eqref{eq:Baxter-TQ} with both $\Omega$-deformation parameters $\ve_{1,2}$ non-vanishing. 

Recall that in the presence of $\Omega$-deformation, any local observable or line operator warpping on the $S^1$ must be located at the origin of the $\BC^2_{12}$ to preserve the spacetime isometry. By the same token, the co-dimension two defect $Q(x)$, which warps on the $S^1$ and the first complex plan $\BC_1$, must be placed at the origin of the second complex plane $\BC_2$. Therefore when both $Q$-bservable and the line defect $\ET(x)$ are inserted, $\ET(x)$ lies on top of $Q(x)$ becoming a line defect observable.  

\subsection{Fractional T-Q equation}
We introduce the second orbifold $\Gamma_{24} = \BZ_N$ acting on the space $\BC^2_{24}$ on $Z$. We will now follow the convention in Section.~\ref{sec:gauge}, for general coloring functions $c: [N] \to \{0,\dots,N-1\}$, we denote
\begin{align}
    \hat{a}_{\o=c(\alpha)}.
\end{align}
We also define the shifted parameters $a_{\o} = \hat{a}_\o - \o \hat\ve_2$ such that $a_\o$ are charged neutral with the $\BZ_N$ orbifold. 
We take the fractional Chern-Simons level $0 \leq \sk_{\o}=\sk_{\o+N}\leq 1$ general with 
$$
    \sum_{\o=0}^{N-1} \sk_\o = \sk.
$$
The $\Omega$-deformation parameters works similarly by: $\hat\ve_1=\ve_1$, $\hat\ve_3=\ve_3$, $\hat\ve_4=(-\ve_1-\ve_3-\hat\ve_2)$. 
The gauge origami data can be expressed in terms of the shifted parameters:
\begin{align}
\begin{split}
    \bN_{12} & = \sum_{\o'=0}^{N-1} e^{{a}_{\o'}} \hat{q}_2^{\o'} \CalR_0 \otimes \fR_{\o'}, \\
    {\bN}_{13} & = \sum_{\o'=0}^{N-1} e^{x'_{\o'}+\ve_1+\ve_3} \hat{q}_2^{\o'} \CalR_1 \otimes \fR_{\o'}, \\
    {\bN}_{34} & = e^x \hat{q}_2^{\o} \CalR_0 \otimes \fR_{\o}.
\end{split}
\end{align}
with  
\begin{align}\label{def:x'}
    {x}'_{\o'} = \begin{cases} x' \quad\quad\quad\quad \text{if} \quad 0 \leq \o' \leq \o \\ x' -\ve_1 \quad\;\;\; \text{if} \quad \o < \o' \leq N-1 \end{cases}. 
\end{align}
The Guage origami partition function is an ensemble over the $\Gamma=\BZ_3\times \BZ_N$ invariant contribution to the canonical ensemble in \eqref{def:part-GO}. 
With some decent but tedious computation, we find te gauge origami partition function can be organized in to the following form 
\begin{align}
\begin{split}
    \hat\CalZ_\text{GO} & = \sum_{\hat{\boldsymbol\lambda}} \prod_{\o=0}^{N-1} \hat{\kq}_\o^{k_\o} \hat\CalZ_\text{defect}[\hat{\boldsymbol\lambda}] T_\o (x,x')[\hat{\boldsymbol\lambda}] \hat{Q}_\o(x')[\hat{\boldsymbol\lambda}] \\
    & = \left \langle T_\o(x,x')\hat{Q}_\o(x') \right \rangle_{\BZ_N} \hat\CalZ_{\hat\BC^2_{12}}
\end{split}
\end{align}
The fractional $qq$-character, which is the correlation function of the gauge theory observables, consists both the fractional $Q$ and $\EY$ observables: 
\begin{align}
\begin{split}
    T_\o (x,x') \hat{Q}_\o(x') = & \ \sh(x-x') \EY_{\overline{\o+1}}(x+\ve_1+\ve_2 \d_{\o,N-1}) \hat{Q}_\o(x') \\
    & + \hat\kq_\o e^{\sk_\o x} \sh({x-x'+\ve_1+\ve_2\d_{\o,N-1} }) \frac{\hat{Q}_\o(x')}{\EY_\o(x)}
\end{split}
\end{align}
with 
\begin{align}
    \hat{Q}_\o(x) = \prod_{\o'=0}^{\o} Q_{\o'}(x') \prod_{\o'=\o+1}^{N-1} Q_\o(x'-\ve_1). 
\end{align}
We define
\begin{align}
\begin{split}
    & \EQ_{\o}(x') = \prod_{\o'=0}^{\o} e^{-\frac{\sk_{\o'}(x'+\ve_1)x'}{4\ve_1}} \prod_{\o'=\o+1}^{N-1} e^{-\frac{\sk_\o(x'-\ve_1)x'}{4\ve_1}} \times \hat{Q}_\o(x'). 
\end{split}
\end{align}
as a fractional version of $\EQ(x)$ in \eqref{def:EQ}. 

First for $\o=0,\dots,N-2$, 
We consider two special cases:
\begin{subequations}
\begin{align}
    T_\o(x=x',x'){\EQ}_\o(x') & = \hat\kq_\o e^{\sk_\o x'} \sh({\ve_1 }) \frac{{\EQ}_\o(x')}{\EY_{\o}(x'_\o)} \\
    & = \hat\kq_\o e^{\frac{\sk_\o}{2} x'} \sh({\ve_1 }) {\EQ}_{\o-1}(x') \nonumber\\
    T_\o(x=x'-\ve_1,x') {\EQ}_\o(x')
    & = \sh({-\ve_1}) \EY_{\overline{\o+1}}(x') {\EQ}_\o (x') \\
    & = -\sh({\ve_1 }) e^{\frac{\sk_{\o+1}}{2}x} {\EQ}_{\o+1} (x') \nonumber
\end{align}
\end{subequations}
and define
\begin{align}
    \langle \ET_\o(x') {\EQ}_\o(x') \rangle_{\BZ_N} = \left \langle \frac{T_\o(x=x',x'){\EQ}_\o(x') - T_\o(x=x'-\ve_1,x'){\EQ}_\o(x')}{\sh({\ve_1 })} \right \rangle_{\BZ_N}. 
\end{align}
For $\o=N-1$, we consider 
\begin{subequations}
\begin{align}
    T_{N-1}(x=x',x'){\EQ}_{N-1}(x') & = \kq_{N-1} e^{\sk_{N-1} x'} \sh({\ve_2 }) \frac{{\EQ}_{N-1}(x')}{\EY_{N-1}(x'_\o)} \\
    & = \kq_{N-1} e^{\frac{\sk_{N-1}}{2} x'} \sh({\ve_2 }) {\EQ}_{N-2}(x') \nonumber\\
    T_{N-1}(x=x'-\ve_2,x') {\EQ}_{N-1}(x')
    & = \sh({-\ve_2 }) \EY_{0}(x'+\ve_1) {\EQ}_{N-1} (x') \\
    & = -\sh({\ve_2 }) e^{\frac{\sk_{N}}{2}(x+\ve_1)} {\EQ}_{N} (x') . \nonumber
\end{align}
\end{subequations}
Note that the fractional Chern-Simons level obeys $\sk_{\o+N}=\sk_{\o}$. 

By comparing $\EQ_{-1}(x)$ and $\EQ_{N-1}(x)$, we define the periodicity of fractional $\EQ$-function is $\EQ_{\o+N}(x') = \EQ_{\o}(x'+\ve_1)$, $\o=0,\dots,N-1$. 
By multiplying the perturbative factor, we define the full vacuum expectation value of the regular surface defect observable as
\begin{align}
    \Psi = \prod_{\o=0}^{N-1} z_\o^{-\frac{a_\o}{\ve_1}} \hat\CalZ_{\hat\BC_{12}^2}
\end{align}
The fractional T-Q equation is now written as
\begin{align}\label{eq:frac-T-Q}
\begin{split}
    & \ET_{\o}(x') \langle \EQ_\o(x') \rangle_{\BZ_N} \Psi \\
    & = e^{\frac{\sk_{\o+1}}{2}(x'+\ve_1\d_{\o,N-1})} \langle \EQ_{\o+1}(x') \rangle_{\BZ_N} \Psi
    + R^2 e^{\rx_{\o+1}-\rx_{\o}} e^{\frac{\sk_\o}{2} x'} \langle \EQ_{\o-1}(x') \rangle_{\BZ_N} \Psi
\end{split}
\end{align}
where $\ET_\o(x')$ can be obtained from taking the large $X$ expansion on $T_\o(x,x')$: 
\begin{align}
    \ET_{\o}(x') = \sh (x'+\ve_1\d_{\o,N-1}+\ve_1\p_{\rx_{\o+1}}) + \sk_{\o} R^2 e^{\frac{x'}{2}} e^{\rx_{\o+1}-\rx_{\o}} e^{-\frac{\ve_1}{2}\p_{\rx_\o}}. 
\end{align}
See Appendix.~\ref{sec:comput-sec5} for detail of the calculation.

\subsection{Construction of Lax operators}

Let us define $2 \times 1$ vector
\begin{align}
    \Xi_{\o}(x) = \begin{pmatrix}
        \langle \EQ_{\o}(x) \rangle_{\BZ_N} \\ \langle \EQ_{\o-1}(x) \rangle_{\BZ_N}
    \end{pmatrix}
\end{align}
to translate the fractional T-Q equation \eqref{eq:frac-T-Q} to degree one matrix equation:
\begin{align}\label{eq:Lax-prime}
\begin{split}
    \Xi_{\o+1}(x) & = e^{-\frac{\sk_{\o+1}}{2}(x+\ve_1\d_{\o,N-1})} \begin{pmatrix}
        \ET_\o(x) & R^2 e^{\frac{\sk_\o}{2}x} e^{\rx_{\o+1}-\rx_{\o}}  \\ e^{\frac{\sk_{\o+1}}{2}(x+\ve_1\d_{\o,N-1})} & 0
    \end{pmatrix} \Xi_{\o}(x) \\
    & := e^{-\frac{\sk_{\o+1}}{2}(x+\ve_1\d_{\o,N-1})} \tilde{L}_{\o+1}(x) \Xi_{\o}(x).
\end{split}
\end{align}
The matrix $\tilde{L}_{\o}(x)$, $\o=0,\dots,N-1$, takes the form 
\begin{align}
    \tilde{L}_{\o+1}(x) = 
    \begin{pmatrix}
        \sh(x+\ve_1\d_{\o,N-1}+\ve_1\p_{\rx_{\o+1}}) + \sk_{\o} R^2 e^{\frac{x}{2}} e^{\rx_{\o+1}-\rx_{\o}} e^{-\frac{\ve_1}{2}\p_{\rx_\o}} & -R^2 e^{\frac{\sk_\o}{2}x} e^{\rx_{\o+1}-\rx_{\o}}  \\ e^{\frac{\sk_{\o+1}}{2}(x+\ve_1\d_{\o,N-1})} & 0
    \end{pmatrix}
\end{align}
with the periodicity $\tilde{L}_{\o+N}(x) = \tilde{L}_{\o}(x+\ve_1)$. 
We take the gauge transformation $\Theta_{\o}(x) = g_\o \Xi_\o (x)$ with
\begin{align}
\begin{split}
    g_\o(x) & = \begin{pmatrix}
        1 & 0 \\ 0 & R e^{\frac{\sk_\o}{2}(x+\ve_1\d_{\o,N})} e^{\rx_{\o}} e^{-\frac{\ve_1}{2}\p_{\rx_\o} }
    \end{pmatrix}
    \begin{pmatrix}
        1 & 0 \\ \sk_\o & -1
    \end{pmatrix}
    \begin{pmatrix}
        1 & 0 \\ 0 & e^{\frac{\ve_1}{2}\p_{\rx_\o}}
    \end{pmatrix} \\
    & = \begin{pmatrix}
        1 & 0 \\ R \sk_\o e^{\frac{(x+\ve_1\d_{\o,N})}{2}} e^{\rx_\o} e^{-\frac{\ve_1}{2}\p_{\rx_{\o}} } & -R e^{\frac{\sk_\o }{2}(x+\ve_1\d_{\o,N})} e^{\rx_\o}
    \end{pmatrix}
\end{split}
\end{align}
It should be noted that $g_{\o+N}(x) = g_{\o}(x+\ve_1)$ following the same periodicity as $\tilde{L}_{\o}(x)$.

In terms of $\Theta_\o(x)$, \eqref{eq:Lax-prime} becomes
\begin{align}
    \Theta_{\o+1}(x) = e^{-\frac{\sk_{\o+1}}{2}(x+\ve_1\d_{\o,N-1})} L_{\o+1}(x) \Theta_{\o}(x)
\end{align}
where the gauge transformed $L_\o(x)$ takes the form 
\begin{align}
    L_{\o+1}(x) = \begin{pmatrix}
        \sh(x+\ve_1\d_{\o,N-1}+\ve_1\p_{\rx_{\o+1}}) & R e^{\rx_{\o+1}} \\
        -R e^{-\rx_{\o+1}} e^{-\sk_{\o+1}\ve_1\p_{\rx_{\o+1}}} & \sk_{\o+1} R^2 e^{\frac{x+\ve_1\d_{\o,N-1}-\ve_1}{2}} e^{-\frac{\ve_1}{2}\p_{\rx_{\o+1}}}
    \end{pmatrix}, 
\end{align}
The Lax operator $L_\o(x)$ acts on the space $\EH_\o \otimes V_\text{aux}$. $\EH_\o$ is the Hilbert space for the $\o$-th particle. $V_\text{aux} = \BC^2$ is an auxiliary space. By construction the Lax operator satisfies the periodicity condition
\begin{align}
    L_{\o+N}(x) = L_{\o}(x+\ve_1).
\end{align}

We identify the $\Omega$-deformation parameter with the Planck constant
\begin{align}\label{def:ve-hbar}
    \ve_1 = - \hbar. 
\end{align}

The monodromy matrix can be defined as an ordered product over the Lax operator
\begin{align}
    \bT^{(n)}(x) = L_{n+N-1}(x) \cdots L_{n}(x)
\end{align}
with any integer $n=0,\dots,N-1$. 
When acting on the vector $\Theta_{n-1}(x)$, the monodromy matrix transforms
\begin{align}\label{eq:transport}
    \Theta_{n-1}(x+\ve_1) = e^{-\frac{1}{2} \sum_{\o=0}^{n-1} \sk_\o (x+\ve_1) -\frac{1}{2} \sum_{\o=n}^{N-1} \sk_\o x } \ \bT^{(n)}(x) \Theta_{n-1}(x).
\end{align}
Hence the monodromy matrix is an operator on the complete tensor product with of all the local Hilbert spaces $\EH_\o$ and the auxiliary space $V_\text{aux}$
\begin{align}
    \bT^{(n)}(x) \in \text{End} \left( \EH_0 \hat\otimes \EH_1 \hat\otimes \cdots \hat\otimes \EH_{N-1} \otimes V_\text{aux} \right).
\end{align}

The integrability of the $Y^{N,k}$ model is quantum integrable by the fact that the Lax operators $L_\o(x)$ satisfies the RLL-relation (train track relation) \eqref{eq:RLL}. The $R$-matrix, defined in $V_\text{aux} \otimes V_\text{aux} = \BC^2 \otimes \BC^2$ space, is given in \eqref{def:R-matrix}. See Appendix.~\ref{sec:integrability} for detail.

\subsection{Matching with new dimer Kasteleyn matrix}
Now we show that the integrable system we obtained is the quantum version of the $Y^{N,\sk}[S]$ dimer model defined in Section.~\ref{sec:SD-glue}. The gluing set $S$ (which is characterized by the function $G_S(n)$ in \eqref{def:func-GS}) chosen to match with the assignment of the Chern-Simons levels $\sk_n$ by
\begin{align}
    G_S(n) = \sk_{n}.
\end{align}
Recall that $\sk_N = \sk_0$ in the gauge theory. 

In $Y^{N,\sk}[S]$ dimer model, $N$ $2\times 2$ Lax matrices $\tilde{L}_n(x)|_\text{dimer}$ in \eqref{eq:Lax-dimer} can be built based on the Kasteleyn matrix $K_{\sb_i,\sw_j}$ in \eqref{def:Kas-general}: 
\begin{align}
    \tilde{L}_n(x)|_\text{dimer} = 
    \begin{cases}
        \begin{pmatrix}
            \sH_{2n+1} & -\sU_{2n} \sV_{2n} \\ 1 & 0
        \end{pmatrix}
        \begin{pmatrix}
            \tilde\sH_{2n} e^{-x} & -\sU_{2n-1} \sV_{2n-1}^{\sk_S(n-1)} (\tilde\sU_{2n-1}e^{-x})^{1-\sk_S(n-1) } \\ 1 & 0 
        \end{pmatrix} & \text{, if } n \notin S; \\
        \begin{pmatrix}
            -\sV_{2n}+\tilde\sV_{2n}e^{-x} & \sU_{2n-1} \sV_{2n-1}^{\sk_S(n-1)} (\tilde\sU_{2n-1}e^{-x})^{1-\sk_S(n-1) } \\ 1 & 0
        \end{pmatrix} & \text{, if } n \in S. 
    \end{cases}
\end{align}

To match with the dimer model, We gauge transformed the Lax operators and take the classical limit (with $\tq_n=\rx_n$, $\tp_n = -\ve_1 \p_{\rx_n}$)
\begin{align}\label{eq:gauge-trans-Lax}
\begin{split}
    \tilde{L}_n(x)|_\text{gauge} 
    & = \begin{pmatrix} 1 & 0 \\ 0 & R^{-1} e^{\tq_n} e^{-\frac{\sk_n}{2}(x+\tp_n)} \end{pmatrix} L_n(x) \begin{pmatrix} 1 & 0 \\ 0 & R^{-1} e^{-\tq_{n-1}} e^{\frac{\sk_{n-1}}{2}(x+\tp_{n-1})} \end{pmatrix} \\
    & = \begin{pmatrix}
        \sh(x-\tp_n) & R^2 e^{\tq_{n}-\tq_{n-1}} e^{\frac{\sk_{n-1}}{2}(x+\tp_{n-1}) } \\ -e^{-\frac{\sk_n}{2}(x-\tp_n)} & \sk_n R^2 e^{\frac{x}{2}} e^{\tq_{n}-\tq_{n-1}} e^{\frac{\sk_{n-1}}{2}(x+\tp_{n-1})}
    \end{pmatrix}, \\
\end{split}   
\end{align}
We identify the gauge transformed Lax matrix from gauge theory with the dimer model Lax matrix:
\begin{align}
    \tilde{L}_n(x)|_\text{gauge} = - \tilde{L}_n(x)|_\text{dimer}.
\end{align}
with the dictionary between the components of the Kasteleyn matrix and $2N$ canonical coordinate $\{\tq_n,\tp_n\}$:
\begin{align}
\begin{split}
    & \sU_{2n} = 1, \ \sV_{2n} = e^{\frac{x}{2}-\frac{\tp_n}{2}}, \ \sH_{2n+1} = 1, \ \tilde\sH_{2n} = e^{\frac{x}{2}+\frac{\tp_n}{2}}, \ n \notin S; \\
    & \sU_{2n-1}\sV_{2n-1} = R^2 e^{\frac{x}{2}} e^{\tq_n-\tq_{n-1}} e^{\frac{1}{2}\tp_{n-1}}, \ \sU_{2n-1} \tilde\sU_{2n-1} e^{-x} = R^2 e^x e^{\tq_n-\tq_{n-1}}, \ n=1,\dots,N; \\
    & \tilde\sV_{2n} = e^{\frac{x}{2}+\frac{\tp_n}{2}}, \ \sV_{2n} = e^{\frac{x}{2}-\frac{\tp_n}{2}}, \ n \in S. 
\end{split}
\end{align}
The 1-loops in the dimer graph $Y^{N,\sk}[S]$ can now be expressed in terms of the canonical coordinates by
\begin{align}
    u_n = e^{-\tp_n}, \ d_n = R^2 e^{\tq_n - \tq_{n-1}} e^{-\frac{1}{2}(\tp_n+\tp_{n-1}) + \sk_{n-1} \tp_{n-1}}.
\end{align}
This matches with \eqref{eq:1-loop to pq}.

Despite the individual matrix elements of the Kasteleyn matrix can be $x$-dependent, the 1-loops which the conserving Hamiltonians build upon are $x$-independent. Furthermore the coefficient of the highest order term $e^{-Nx}$
$$
    \prod_{n\in S} \tilde\sV_{2n} \prod_{n \notin S} \sH_{2n+1} \tilde\sH_{2n} = e^{\frac{k}{2} x + \sum_{n=1}^N \tp_n} = e^{\frac{\sk}{2}x}.
$$

It should be noted that the Lax matrices from the Dimer model does not satisfy RLL-relation if one simply promote its components to quantum operators. A gauge transformation is required as shown above. 
This degree of freedom comes from the fact that there are much more edges than the number of 1-loops in a given dimer graph. 
Recall that Poisson structure of a dimer integrable system is defined on oriented loops \eqref{def:Poisson-loops} instead of individual edges, it is allowed to modify the assignments to the edges as long as the 1-loops are fixed, which can be achieved by conjugating the Kasteleyn matrix $K_{\sb_i,\sw_j}$ with an arbitrary diagonal matrix $D$. \footnote{The conjugate matrix must be diagonal otherwise it modifies the dimer graph.}
The spectral curve, which consists the generating function of the integral of motion, is unchanged  
\[
    0 = \det (D K_{\sb_i,\sw_j} D^{-1}) = \det K_{\sb_i,\sw_j}. 
\]

\subsection{Spectral equations from factorization of defects}
We define the Fourier transform of the fractional $\EQ_\o(x)$
\begin{align}\label{def:Ups}
    \Upsilon_\o(y) = \sum_{x\in \mathsf{L}} \langle \EQ_\o(x) \rangle_{\BZ_N} y^{-\frac{x}{\ve_1}}
\end{align}
which satisfies the periodicity $\Upsilon_{\o+N}(y) = y \Upsilon_{\o}(y)$. 
$\mathsf{L}$ is a lattice chosen such that the summation converges. Eq.~\eqref{eq:transport} becomes an eigenvalue equation in the Fourier space:
\begin{align}\label{eq:Lax-F}
    \left[ e^{\frac{1}{2} \sum_{\o=0}^n \sk_\o (-\ve_1y\p_y+\ve_1) + \sum_{\o=n+1}^{N-1} -\sk_\o \ve_1y\p_y} y - \bT^{(n)}(-\ve_1y\p_y) \right] 
    \begin{pmatrix}
        \Upsilon_{n-1}(y) \\ \tilde\Theta_{n-1}(y). 
    \end{pmatrix}.
\end{align}
Here $\tilde{\Theta}_{n-1}(y)$ is the Fourier transform of the second component of the vector $\Theta_{n-1}(x)$. 

We will now consider $n=0$. The first vector component of $\Theta_{-1}(x)$ is the bulk $Q$-function 
\begin{align}
    \EQ_{-1}(x) = \prod_{\o'=0}^{N-1} e^{-\frac{\sk_{\o'}(x-\ve_1)x}{4\ve_1}} Q_{\o'}(x-\ve_1) = e^{-\frac{\sk(x-\ve_1)x}{4\ve_1}} Q(x-\ve_1) = \EQ(x-\ve_1).
\end{align}
Here we write down the four components of the monodromy matrix
\begin{align}
    \bT^{(0)}(x) = \begin{pmatrix}
        A(x) & B(x) \\ C(x) & D(x)
    \end{pmatrix}
\end{align}
to explicitly write down the the two equations in \eqref{eq:Lax-F} 
\begin{subequations}
\begin{align}
    & e^{-\frac{\sk}{2}\ve_1 y\p_y} y \Upsilon_{-1}(y) - A(-\ve_1 y\p_y) \Upsilon_{-1}(y) - B(-\ve_1y\p_y) \tilde\Theta_{-1}(y) = 0, \\
    & e^{-\frac{\sk}{2}\ve_1 y\p_y} y \tilde\Theta_{-1}(y) - C(-\ve_1y\p_y) \Upsilon_{-1}(y) - D(-\ve_1y\p_y) \tilde\Theta_{-1}(y) = 0. 
\end{align}
\end{subequations}
We multiply the first equation by $e^{-\frac{\sk}{2}\ve_1y\p_y}y-D(-\ve_1y\p_y+\ve_1)$ and the second line by $B(-\ve_1y\p_y+\ve_1)$ then take the sum of the two. It yields
\begin{align}
\begin{split}
    0 = & \left[ e^{-\frac{\sk}{2}\ve_1y\p_y} y - D(-\ve_1y\p_y+\ve_1) \right] \left[ e^{-\frac{\sk}{2}\ve_1y\p_y}y - A(-\ve_1y\p_y) \right] \Upsilon_{-1}(y) \\
    & - B(-\ve_1y\p_y+\ve_1) C(-\ve_1y\p_y) \Upsilon_{-1}(y) \\
    & + \left[ B(-\ve_1y\p_y+\ve_1)D(-\ve_1y\p_y) - D(-\ve_1y\p_y+\ve_1) B(-\ve_1y\p_y) \right] \tilde\Theta_{-1}(y).
\end{split}
\end{align}
The commutation relation between the components of the monodromy matrix is governed by the RTT-relation (train track relation). Notice that the RTT-relation does not require the argument of the monodromy matrix to be a scalar. Indeed it gives 
\begin{align}
    \left[ B(-\ve_1y\p_y+\ve_1)D(-\ve_1y\p_y) - D(-\ve_1y\p_y+\ve_1) B(-\ve_1y\p_y) \right] = 0.
\end{align}

The fractional $\Upsilon_{-1}(y)$  satisfies
\begin{align}\label{eq:Ups-deg-2}
\begin{split}
    & \Big[ e^{-\frac{\sk}{2}\ve_1y\p_y} y e^{-\frac{\sk}{2}\ve_1y\p_y} y - \left( A(-\ve_1y\p_y) + D(-\ve_1y\p_y) \right) e^{-\frac{\sk}{2}\ve_1y\p_y} y  \\
    & \ + D(-\ve_1y\p_y+\ve_1) A(-\ve_1y\p_y) - B(-\ve_1y\p_y+\ve_1) C(-\ve_1y\p_y) \Big] \Upsilon_{-1}(y) = 0.
\end{split}
\end{align}
The second term is propotional to the trace of the monodromy matrix, and the term on the second line is identified as the quantum determinant of the monodromy matrix, which is
\begin{align}
\begin{split}
    \text{q-det }\bT^{(0)}(x)
    & = D(x+\ve_1)A(x) - B(x+\ve_1) C(x) \\
    & = \prod_{\o=0}^{N-1} \text{q-det } L_{\o}(x) 
    = \prod_{\o=0}^{N-1} \hat\kq_\o e^{{\sk_\o}x} 
    = \kq e^{\sk x}.
\end{split}
\end{align}
We now perform inverse Fourier transform on \eqref{eq:Ups-deg-2} and divide a common factor $e^{\frac{\sk}{2}x}$: 
\begin{align}\label{eq:TQ-frac}
    e^{\frac{\sk}{2}(x+\ve_1)} \langle \EQ(x+\ve_1) \rangle_{\BZ_N} - \Tr \bT^{(0)}(x) \langle \EQ(x) \rangle_{\BZ_N} + \kq e^{\frac{\sk}{2}x} \langle \EQ(x-\ve_1) \rangle_{\BZ_N} = 0. 
\end{align}
After multiplying a factor of $e^{\frac{N}{2}x}$, the trace of the monodromy matrix is a degree $N$ polynomial in $X=e^x$ whose coefficients are the conserving Hamiltonian 
\begin{align}
     X^{\frac{N}{2}} \cdot \Tr \bT^{(0)}(x) = \hat{\rm H}_N X^N + \hat{\rm H}_{N-1} X^{N-1} + \hat{\rm H}_{N-2} X^{N-2} + \cdots + \hat{\rm H}_0 
\end{align}
where $\hat{\rm H}_n$ are operators acting on the $N$ particle Hilbert space $\EH_{1} \hat\otimes \cdots \hat\otimes \EH_{N}$.

If we turn off the $\Omega$-deformation parameter $\ve_2 \to 0$ for the complex plane $\BC_2$ transverse to the plane $\BC_1$ of the defect $\EQ(x)$, the bulk observables $\EQ(x)$ and $\ET(x)$ can be arbitrarily separated on $\BC_2$. The cluster decomposition implies fractionalization of the correlation function without any contact term. 
Thus \eqref{eq:TQ-bulk} leads to
\begin{align}\label{eq:TQ-bulk-limit}
    e^{\frac{\sk}{2}(x+\ve_1)} \EQ(x+\ve_1)[\boldsymbol\Lambda] + \kq e^{\frac{\sk}{2}x} \EQ(x-\ve_1)[\boldsymbol\Lambda] = \ET(x)[\boldsymbol\Lambda] \EQ(x)[\boldsymbol\Lambda].
\end{align}
In particular $e^{\frac{N}{2}} \cdot \ET(x)[\boldsymbol\Lambda]$ is a degree $N$ polynomial in $X=e^x$ whose coefficients are the expectation value of the Wilson loop in the bulk theory \eqref{eq:Wilson-loop}:
\begin{align}
\begin{split}
    X^{\frac{N}{2}} \ET(x)[\boldsymbol\Lambda] 
    & = \EX_\sk(x)[\boldsymbol\Lambda] = \lim_{\ve_2 \to 0} \frac{\CalZ_\text{1d/5d}(x)}{\CalZ_{\BC^2_{12}}} \\
    & = (1+\d_{N,\sk}) X^N - \EW_{\wedge^{N-1}} X^{N-1} + \EW_{\wedge^{N-2}} X^{N-2} \cdots + \EW_{\wedge^0} . 
\end{split}
\end{align}

Furthermore, the bulk canonical ensemble in \eqref{def:defect-part-bs} is locked to the limit shape configuration $\boldsymbol\Lambda$. The expectation value of bulk $\EQ_{N-1}(x)=\EQ(x)$ function fractionalizes to
\begin{align}
    \lim_{\ve_2\to 0}\langle \EQ(x) \rangle_{\BZ_N} \Psi = e^{\frac{\widetilde\CalW(\ba,\ve_1,\kq)}{\ve_2}} \EQ(x)[\boldsymbol\Lambda] \left[ \psi(\ba,\ve_1,\rx_\o,\kq) + \CalO(\ve_2) \right], \ 
\end{align}
\eqref{eq:TQ-frac} becomes
\begin{align}\label{eq:TQ-frac-limit}
\begin{split}
    & e^{\frac{\sk}{2}(x+\ve_1)} \EQ(x+\ve_1)[\boldsymbol\Lambda] \psi(\ba,\ve_1,\rx_\o,\kq) + \kq e^{\frac{\sk}{2}x} \EQ(x-\ve_1)[\boldsymbol\Lambda] \psi(\ba,\ve_1,\rx_\o,\kq) \\
    & = \Tr \bT^{(0)}(x) \EQ(x)[\boldsymbol\Lambda] \psi(\ba,\ve_1,\rx_\o,\kq) .
\end{split}
\end{align}
We multiply \eqref{eq:TQ-bulk-limit} by $\psi(\ba,\ve_1,\rx_\o,\kq)$ and compare it with \eqref{eq:TQ-frac-limit}. The function $\EQ(x)[\boldsymbol\lambda]$ is an over all constant since it has no dependence in $\rx_\o$ and can be canceled. We obtain
\begin{align}\label{eq:All-Schrodinger}
\begin{split}
    & \Tr \bT^{(0)}(x) \psi(\ba,\ve_1,\rx_\o,\kq) = \ET(x) \psi(\ba,\ve_1,\rx_\o,\kq), \\
    & \implies \hat{\rm H}_n \psi(\ba,\ve_1,\rx_\o,\kq) = \EW_{\wedge^{n}} \psi(\ba,\ve_1,\rx_\o,\kq). 
\end{split}
\end{align}
We thus prove the normalized vev of the monodromy surface defect $\psi$ \eqref{def:psi2} is the common eigenfunction of the $N$ commuting Hamiltonians of the quantum $Y^{N,\sk}[S]$ integrable model. And the eigenvalues are given by the expectation value of the BPS Wilson loop in the anti-symmetric representation of the gauge group. 


\section{Summary and future direction}\label{sec:Summary}

In this paper we have established the relation between the $\CalN=1$ 5d super Yang-Mills theory and the $Y^{N,\sk}[S]$ cluster integrable system, which are the generalization of the affine $A$-type relativistic Toda lattice. 
By considering non-standard gluing of the vertexes in the dimer model, we construct new varieties of $Y^{N,\sk}[S]$ cluster integrable system characterized by such a gluing. 

The gauge theories dual to these new $Y^{N,\sk}[S]$ cluster integrable systems are the five dimensional $\CalN=1$ $U(N)_\sk$ super Yang-Mills theories on $S^1\times \hat\BC_1 \times \hat\BC_2/\BZ_N$ with a regular surface defect warping on $\BC_1$. 
We derive the quantum T-Q equation and its fractional version from the gauge origami. 
Through them we prove that the normalized vev of the monodromy surface defect is the common eigenfunction of the commuting Hamiltonians of the quantum $Y^{N,\sk}[S]$ cluster integrable system, with the eigen value being the expectation value of the BPS Wilson loop in the anti-symmetric representation of the gauge group in the bulk \eqref{eq:All-Schrodinger}. 

Let us end up this paper by pointing out some potential furture research directions: 

\begin{enumerate}
    \item There are 5d suersymmetric gauge theories with Chern-Simons level higher than the gauge group rank. These theories are not toric and can not be constructed from M-theory warping on a Calabi-Yau threefold. On the level of the supersymmetric partition function, the integration form of the instanton measure in \eqref{eq:Z-int} diverges. Hence the $qq$-character computation is not valid when $\sk>N$. It will be interesting to explore the Bethe/gauge correspondence of such gauge theory. Studies of such non-Lagrangian theories have been conducted in \cite{Kim:2023qwh}. 
    \item We introduce non-standard gluing in Section.~\ref{sec:SD-glue} in the dimer model. One may notice that all the gluing happens at the right edges of the center hexagon in the unit cell. More general gluing choices can be considered in a dimer graph with the graph kept planar. It would be interesting to check of those dimer models also has a integrable systems dual, and if so what gauge theories are in the Bethe/gauge correspondence of those systems. 
    \item The Toda lattice can be defined on a root system of any given affine algebra \cite{bogoyavlensky1976perturbations}. In \cite{Martinec:1995by} and later \cite{Gorsky:1999gx} it has been shown that a four-dimensional $\CalN=2$ super Yang-Mills theory with gauge group $G$ of $BCD$-type is in Bethe/gauge correspondence to a Toda lattice defined on the affine Lie algebra of the dual group $G^\vee$. The correspondence is established on the level between Seiberg-Witten curve of the gauge theory and the spectral curve of the classical integrable model. The extension of such a correspondence to the quantum level turns out much harder comparing to the $A$-type. This is due the lack of Young diagram representations for the instanton configurations for the $BCD$-type gauge theories with $\Omega_{\ve_1,\ve_2}$-deformation introduced for localization. In addition the $qq$-characters for the $BCD$-type often consists infinite sums of $\EY$-observables \cite{Haouzi:2020yxy}. In the unrefined limit $\ve_1=-\ve_2$ the instanton configuration has Young diagram representation \cite{Rui-dong1} and the $qq$-character consists only finite terms \cite{Nawata:2023wnk}. It will be interesting to know if the Bethe/gauge correspondence of the $BCD$-type can be extended to the quantum level to the same degree as the $A$-type like in \eqref{eq:All-Schrodinger}. 
    \item The eigenvalue of $N$ Hamiltonians for $Y^{N,\sk}[S]$ model in \eqref{eq:All-Schrodinger} comes from the bulk elliptic T-Q equation \eqref{eq:TQ-bulk}. This value is nothing but the expectation value of the BPS Wilson loop in the anti-symmetric representation of the gauge group. As the bulk configuration knows nothing about the detail of the defect, it indicates the $Y^{N,\sk}[S]$ integrable model, despite potentially having non-identical of Hamiltonians for different gluing set $S$, will share the same energy spectrum. 
    This would indicate potential duality among the $Y^{N,\sk}[S]$ with different guling set $S$. 
\end{enumerate}

\newpage
\appendix



\section{Integrability of Lax operator} \label{sec:integrability}
Lax operator $L(x)$ defined on the space $\EH \otimes V_\text{aux}$, $V_\text{aux} = \BC^2$ takes the following form
\begin{align}
    L(x) = \begin{pmatrix} \sh(x-\hbar\p_\rx) & R e^{\rx} \\ -R e^{-\rx} e^{\sk\hbar\p_\rx} & \sk R^2 e^{\frac{x+\hbar\p_\rx}{2}} \end{pmatrix}
\end{align}
We define R-matrix acting on the space $V_\text{aux} \otimes V_\text{aux}$ by
\begin{align}\label{def:R-matrix}
    R_{a_1,a_2}(x-y) = 
    \begin{pmatrix} 
        \sh(x-y+\hbar) & 0 & 0 & 0 \\
        0 & \sh (x-y) & \sh\hbar & 0 \\
        0 & \sh\hbar & \sh(x-y) & 0 \\
        0 & 0 & 0 & \sh(x-y+\hbar)
    \end{pmatrix}
\end{align}
such that
\begin{align}
    L_{a_1}(x) = L(x) \otimes I = 
    \begin{pmatrix}
        \sh(x-\hbar\p_\rx) & 0 & R e^{\rx} & 0 \\
        0 & \sh(x-\hbar\p_\rx) & 0 & R e^{\rx} \\ 
        -R e^{-\rx} e^{\sk\hbar\p_\rx} & 0 & \sk R^2 e^{\frac{x+\hbar\p_\rx}{2}} & 0 \\
        0 & -R e^{-\rx} e^{\sk\hbar\p_\rx} & 0 & \sk R^2 e^{\frac{x+\hbar\p_\rx}{2}}
    \end{pmatrix}
\end{align}
and 
\begin{align}
    L_{a_2}(y) = I \otimes L(y) = 
    \begin{pmatrix} 
        \sh(y-\hbar\p_\rx) & R e^{\rx} & 0 & 0 \\
        -R e^{-\rx} e^{\sk\hbar\p_\rx} & \sk R^2 e^{\frac{y+\hbar\p_\rx}{2}} & 0 & 0 \\
        0 & 0 & \sh(y-\hbar\p_\rx) & R e^{\rx} \\
        0 & 0 & -R e^{-\rx} e^{\sk\hbar\p_\rx} & \sk R^2 e^{\frac{y+\hbar\p_\rx}{2}}
    \end{pmatrix}
\end{align}

The commutation relations between two elements in the Lax operator is governed by the RLL-relation (train track relation)
\begin{align}\label{eq:RLL}
    R_{a_1,a_2}(x-y)L_{a_1}(x) L_{a_2}(y) = L_{a_2}(y) L_{a_1}(x) R_{a_1,a_2}(x-y)
\end{align}
which can be verified true by direct computation.  

The monodromy matrix ${\bf T}^{(n)}_a(x)$ is defined as an ordered product over the Lax operators
\begin{align}
    {\bf T}^{(n)}_a(x) = L_{n+N-1,a}(x) \cdots L_{n,a}(x).
\end{align}
It is obvious that the monodromy matrix ${\bf T}^{(n)}_a(x)$ satisfies the same train track commutation relation as the Lax operators 
\begin{align}\label{eq:RTT}
    R_{a_1,a_2}(x-y){\bf T}^{(n)}_{a_1}(x) {\bf T}^{(n)}_{a_2}(y) = {\bf T}^{(n)}_{a_2}(y) {\bf T}^{(n)}_{a_1}(x) R_{a_1,a_2}(x-y)
\end{align}
for all $n$.

\section{Note on gauge origami} \label{sec:GO}

The 5d SYM theories has a string theory construction using type IIB $(p,q)$ 5-brane web \cite{Aharony:1997bh} with $N$ D5 brane ($(1,0)$-brane) extended in the 012346 direction, suspend between two NS5 branes ($(0,1)$-brane) that extend in the 012349 direction. Charge conservation results the creation of $(p,q)$ 5-brane web when a D5 brane ends on an NS5 brane. 

The 1d/5d coupled system introduced in \eqref{eq:Z-1d-5d-PI} in the previous section admits a different brane realization which was first proposed in \cite{Gomis:2006sb}. 
Our starting point is a ten-dimensional flat spacetime theory IIA string theory on $S^1 \times \BR^8 \times \BR$. We compactify the $x_0$ direction to a circle and introduce the following D-branes in Table.~\ref{tab:D-branes}.

\begin{table}[h]
    \centering
    \begin{tabular}{|c||c|c|c|c|c|c|c|c|c|c|}
    \hline
     & 0 & 1 & 2 & 3 & 4 & 5 & 6 & 7 & 8 & 9 \\
    \hline \hline
    $k$ D0 & x & & & & & & & & & \\
    \hline 
    $N$ D4 & x & x & x & x & x & & & & & \\
    \hline 
    1 D4' & x & & & & & x & x & x & x &  \\
    \hline
    F1 & x & & & & & & & & & x \\
    \hline
    \end{tabular}
    \caption{The direction of various branes.}
    \label{tab:D-branes}
\end{table}

The $N$ D4 branes are obtained by T-dualize the D5 brane in the $\bx_6$ direction. The NS5 branes are removed since the one D4' further breaks the supersymmetry to exactly 8 supercharges. 

The effective theory on the D4 branes is a 5d $U(N)$ gauge theory on $S^1 \times \BR^4$. Separating the D4 branes along the $\bx_9$ direction gives non-zero vev to the adjoint scalar in the vector multiplet $\langle \Phi \rangle \neq 0$. Pulling the D4' brane a distance $x$ away in the $\bx_9$-direction creates F1 strings with non-zero tension stretching between the D4 and D4' branes. This introduce a set of fermionic particles in the worldvolume of the 5d $U(N)$ gauge theory whose mass are proportional to the distance $x$. The fermions transforms in the fundamental representation of the $U(N)$ gauge group.  

This brane realization is a special case of the \emph{gauge origami} introduced in \cite{NaveenNikita,Nikita:III}. A type IIB realization is given there, which is one T-dual away from ours here. It is consistant by the fact that the theories of interests are the $\CalN=2$ 4d $U(N)$ supersymmetric gauge theories. Our presentation here is simply the $K$-theoretic version of the cohomological results obtained there. Indeed this can be seen after taking the 4d limit of our 5d instanton partition function with a Wilson loop warpping on the circle in $\bx_0$. 

The setting for the gauge origami is as follows: The ten-dimensional spacetime of the type IIA theory is denoted as $S^1 \times Z \times \BR$. $Z$ is a Calabi-Yau fourfold. 
We denote the coordinates on the 4 complex planes $\BC_{a}$ as $\bz_{a}$, $a\in \underline{\bf 4}=\{1,2,3,4\}$. We also use the notation
\begin{align}
    \underline{\bf 6}=\{12,13,14,23,24,34\}
\end{align}
and $A=ab\in \underline{\bf 6}$ to denote $\BC^2_{A} = \BC_a \times \BC_b \subset Z$ corresponding two complex planes.  In a toric origami there can be at most six stacks of D4 branes with the multiplicity $n_A \in \BZ_{\geq 0}$ and the world volume $S^1 \times \BC^2_A$. Let $a_{A,\alpha} \in \BR$, $A\in \underline{\bf 6}$, $\alpha=1,\dots,n_A$ denote their location on $\BR$ which we assume to be generic. The union
\begin{align}
    \bigcup_A n_A \BC^2_A \subset Z
\end{align}
is called the \emph{worldvolume of the gauge origami}. 
For each of the 6 stack we define the vector spaces $\bN_A = \BC^{n_A}$. Let us denote the Chan-Paton space of the six stacks of D-branes by $\bN_A$. They carry the representation of $\Gamma$ and therefore can be decomposed into
\begin{align}\label{eq:char}
    \bN_A = \bigoplus_{\substack{i\in\{0,\dots,p-1\} \\ \o \in \{0,\dots,l-1\} }} e^{a_{A,i,\o}} \CalR_i \otimes \fR_\o    
\end{align}

In this paper we shall only look at the cases $Z=\BC^4=\BC^4_{1234}$, a local K3 orbifold $Z=\BC^2_{12} \times \BC^2_{34}/\Gamma_{34}$ with the cyclic group $\BZ_p$, and the local CY3 orbifold with another cyclic group $Z=\BC_1 \times \BC^3_{234}/ (\Gamma_{24} \times \Gamma_{34})$ with another cyclic group $\BZ_l$. The cyclic group $\Gamma_{ab}$ acts on the two dimensional space $\BC^2_{ab}$ by 
\begin{align}\label{eq:Gamma-action}
    (\bz_a,\bz_b) \mapsto (\eta \bz_a,\eta^{-1} \bz_b)
\end{align}
where $\eta$ is the generator of the cyclic group $\Gamma_{ab}$ represented as the root of unity. The 5d gauge theory worldvolume warps on $S^1 \times \BC^2_{12} $. We use the notation 
\begin{align}
\begin{split}
    & \CalR_i \in \Gamma^\vee_{34}, \ i=0,1,\dots,p-1 \equiv -1 \\
    & \fR_{\o} \in \Gamma^\vee_{24}, \ \o=0,1,\dots,l-1 \equiv -1
\end{split}
\end{align}
for the irreducible representation of the cyclic group. We will denote $\Gamma = \Gamma_{24} \times \Gamma_{34}$. 

The global symmetry of the gauge origami is
$$
    \sG = \bigtimes_{A \in \underline{\bf 6} } \bigtimes_{i=0}^{p-1} \bigtimes_{\o=0}^{l-1} U(n_{A,i,\o}) \times SU(4). 
$$
The first one is the rotation on the D4 branes carrying the same representation of $\Gamma$. The second one is the isometry of $Z$. 

We may insert D0 branes warpping on $S^1$ dissolved into the worldvolume of the gauge origami times the $S^1$. In the T-dual picture where the D0 becomes a point-like BPS object D(-1) in type IIB, the D(-1) branes introduced to the intersecting gauge theories are called \emph{spiked instanton}, which we will also call the D0 branes in the type IIA picture here. Let $\bK$ be the Chan-Paton space of the D0-instantons. It is decomposed according to which D4-brane they are placed on 
$$
    \bK = \bigoplus_{A \in \underline{\bf 6}} \bigoplus_{\alpha=1}^{n_A} K_{A,\alpha}
$$
Let us denote $\bK_A = \bigoplus_{\alpha=1}^{n_A} K_{A,\alpha}$ for each $A \in \underline{\bf 6}$. 

We introduce an $\Omega$-background by viewing $Z$ as a bundle over the $S^1$, where as we go around the circle, we make an identification
\begin{align}
    (\bz_1,\bz_2,\bz_3,\bz_4) \sim (\bz_1e^{R\ve_1},\bz_2e^{R\ve_2},\bz_3e^{R\ve_3},\bz_4e^{R\ve_4})
\end{align}
with $\ve_{1,2,3,4} \in \BC$ are the complex $\Omega$-deformation parameters satisfying 
\begin{align}
    \ve_1+\ve_2+\ve_3+\ve_4=0.
\end{align}
The localization on the $\Omega$-deformed theory reduces the path integral in the partition function to an equivalent statistical system summing over the fixed points on the moduli space of the spiked instanton, which admits a ADHM-like realization in terms of the linear maps between $\bN_A$ and $\bK_A$. The discrete fix points on this moduli space under the action of the maximal torus $\sT_\sG$ of the symmetry group $\sG$ is labeled by a set of partitions $\boldsymbol\lambda=\left( \boldsymbol\lambda^{(A,\alpha)} \right)_{A \in \underline{\bf 6}, \ \alpha\in \{1,\dots,n_A\}}$ \cite{Nikita:II}.   

Since $\sT_\sG$ is an abelian group, the space $\bN_A$ and $\bK_A$ are decomposed into its one-dimensional irreducible representation. Let us denote the decomposition by denoting the Chern character of the spaces here 
\begin{align}\label{eq:character}
\begin{split}
    & \bN_{A} = \sum_{\alpha=1}^{n_{A}} e^{a_{A,\alpha} }  \\
    & \bK_{A} = \sum_{\alpha=1}^{n_{A}} \sum_{\Box\in \boldsymbol\lambda^{(A,\alpha)} } e^{c_{A,\Box}} 
\end{split}
\end{align}
where $c_{A,\Box} = a_{A,i,\alpha}+(\ri-1)\ve_a+(\rj-1)\ve_b$ with $A=ab\in \underline{\bf 6}$. The character $\bS_A$ for the universal bundle is
\begin{align}
    \bS_{A} = \bN_A - {P}_{ab} \bK_{A}
\end{align}
The exponentiated $\Omega$-deformation parameters are
\begin{align}
\begin{split}
    & q_a = e^{\ve_a}, \ P_a = 1-q_a, \ P_{ab}=(1-q_a)(1-q_b), \ \prod_{a=1}^4 q_a = 1.
\end{split}
\end{align}
Note that each one dimensional subspace represented by each term of the equivalent Euler character in \eqref{eq:character} carries a representation of $\Gamma$ according to the decomposition \eqref{eq:char}. 

In this note, we will only consider the case $p=3$ to generate $SU(N)$ gauge theory on $S^1 \times \BC^2_{12}$. 
The general formula for the gauge origami partition function is derived in \cite{Nikita:III}:
\begin{align}\label{def:part-GO}
\begin{split}
    \CalZ_\text{GO} = \sum_{\boldsymbol\lambda} & \boldsymbol\kq^{|\boldsymbol\lambda|} \det(\bK)^{\boldsymbol\sk} \BE \left[ - \sum_{A\in \underline{\bf 6}} \frac{{P}_{\min\bar{A}} \bS_A\bS_A^*}{P_A^*} - \sum_{A=12,13,14} q_A \bS_{\bar{A}} \bS_A^* - \sum_{\substack{A<B \\ |A\cap B|=1}} q_{\max B} {P}_{\overline{A\cup B}} \frac{\bS_A \bS_B^*}{{P}_{A\cap B}^*}  \right]^\Gamma
\end{split}
\end{align}
The $\BE$ operator is defined in \eqref{def:E-operator} and the notation $[\cdots]^\Gamma$ is to pick up only the $\Gamma$-invariant part. Also we use the notation $\bar{A} = ab$ where $a,b\notin A$, $a<b$. 
The summation is taken over all the partitions $\boldsymbol\lambda=\left( \boldsymbol\lambda^{(A,\alpha)} \right)_{A \in \underline{\bf 6}, \ \alpha\in \{1,\dots,n_A\}}$. 
The gauge coupling are collectively denoted as $\boldsymbol\kq = (\kq_{i,\o})_{\substack{i\in \{0,\dots,p-1\} \\ \o \in \{0,\dots,l-1\} }}$. The Chern-Simons levels are collectively denoted as $\boldsymbol\sk = (\sk_{i,\o})_{\substack{i\in \{0,\dots,p-1\} \\ \o \in \{0,\dots,l-1\} }}$. We also use the notation
\begin{align}
    \boldsymbol\kq^{|\boldsymbol\lambda|} = \prod_{\substack{i\in \{0,\dots,p-1\} \\ \o \in \{0,\dots,l-1\} }} \kq_{i,\o}^{|\bK_{i,\o}|}, \ \det(\bK)^{\boldsymbol\sk} = \prod_{\substack{i\in \{0,\dots,p-1\} \\ \o \in \{0,\dots,l-1\} }} \det(\bK_{i,\o})^{\sk_{i,\o}}
\end{align}
where $\bK_{i,\o}$ is the subspace of $\bK$ that carries the representation $\CalR_i \otimes \fR_\o$ of $\Gamma$. 


\subsection{The bulk gauge theory}

Let us now construct the bulk theory from the gauge origami. We consider a single stack of $D4$ branes on $S^1 \times \BC^2_{12}$. We assign the $\BZ_3$-charge as
\begin{align}
    \bN_{12} = \sum_{\alpha=1}^N e^{a_\alpha} \CalR_0. 
\end{align}
The gauge theory constructed in this way is a $\hat{A}_2$ theory with gauge group $U(N) \times {Id} \times {Id}$. Here $Id$ denotes the trivial group whose only element is the identity. The moduli space of the spiked instanton is classified, a priori, by a single $N$ tuples of Young diagrams $\boldsymbol\lambda$ from the only non-trivial gauge group. 

Throughout this note, we will constantly using the trick of taking $\kq_1=\kq_2=0$. There by killing all the instanton in node 1 and 2, $\boldsymbol\lambda_1=\boldsymbol\lambda_2 = \varnothing$ had they existed. We call this procedure \emph{freezing} $\hat{A}_2 \to A_1$, or simply freezing, for short. The remaining gauge coupling is denoted as $\kq = \kq_0$. The remaining Chern-Simons level is denoted as $\sk=\sk_0$. The gauge origami partition function reduced to 
\begin{align}
    \CalZ_\text{GO} = \sum_{\boldsymbol\lambda} \kq^{|\boldsymbol\lambda|} \BE \left[ - \frac{\bS\bS^*}{P_{12}^*} \right] \det(\bK_{12})^\sk = \CalZ_{\BC^2_{12}}
\end{align}
where we have defined 
\begin{align}
    \bN = \sum_{\alpha=1}^N e^{a_\alpha}, \ \bS = \bN - P_{12} \bK_{12}. 
\end{align}
This is the 5d $U(N)$ SYM theory with Chern-Simons level $\sk$ we studied in Section.~\ref{sec:gauge}. We emphasize here that $\Gamma_{34}$-orbifold plays an auxiliary role reducing the theory with adjoint fields to SYM. 

\subsection{\textit{qq}-character} \label{sec:qq-go}
Now we study two transversal stack of branes. Such a configuration defines a BPS operators on the interesction of the braneworld. Integrating out the degree of freedom from one of the stack of D-brane creates $qq$-character. 

Consider the gauge origami with two stacks of branes: one on $S^1 \times \BC^2_{12}$ and the other one on $S^1 \times \BC^{2}_{34}$. As a minimal modification of the $A_1$ bulk theory we start with a single brane on $S^1 \times \BC^2_{34}$. We still have the degree of freedom to assign the $\BZ_3$-representation to the brane. It turns out the only interesting assignment is to assign it to a singlet representation
\begin{align}
    {\bN}_{12} = \sum_{\alpha=1}^N e^{a_\alpha} \CalR_0, \ {\bN}_{34} = e^x \CalR_0. 
\end{align}
The corresponding gauge origami partition function is computed as
\begin{align}
    \CalZ_\text{GO} = \sum_{\boldsymbol\lambda} \kq_0^{|\boldsymbol\lambda|} \det(\bK_{12})^{\sk_0} \prod_{i=0,1,2}\kq_i^{|\bK_{34,i}|} \det(\bK_{34,i})^{\sk_i} \BE \left[ - \frac{{P}_3\bS_{12}\bS^*_{12} }{{P}_{12} } - \frac{{P}_1 \bS_{34} \bS_{34}^* }{{P}_{34}} - {q}_{12} \bS_{34} \bS_{12}^* \right]^{\BZ_3} .
\end{align}
In the freezing limit $\kq_1=\kq_2=0$. The only instanton configuration for $\boldsymbol\lambda_{34}$ is $\bK_{34,0}=0$ or $e^x$, $\bK_{34,1}=\bK_{34,2}=0$. Therefore the gauge origami partition function splits into the expectation value of two observables 
\begin{align}
    \CalZ_\text{GO} = \sum_{\boldsymbol\lambda} \kq^{|\boldsymbol\lambda|} \CalZ_\text{bulk}[\boldsymbol\lambda] \left( \EY(x+\ve_+) + \frac{\kq e^{\sk x}}{\EY(x)} \right) = \langle \EX_\sk (x) \rangle \CalZ_{\BC^2_{12}}
\end{align}
In the language of five-dimensional gauge theory, it is the expectation value of the $qq$-character \eqref{def:qq}. 

\subsection{\textit{Q}-observable}

We consider a class of co-dimensional two canonical defect called $Q$-observables, constructed by considering two stacks of branes, one on $S^1 \times \BC_{12}^2$ and the other one on $S^1 \times \BC^2_{13}$ in gauge origami. We assign the $\BZ_3$-charges as 
\begin{align}
    \bN_{12} = \sum_{\alpha=1}^N e^{a_\alpha} \CalR_0, \ \bN_{13} = e^x q_1q_3 \CalR_1. 
\end{align}
The gauge origami partition function is
\begin{align}
    \CalZ_\text{GO} = \sum_{\boldsymbol\lambda} \kq_0^{|\boldsymbol\lambda_{12}|} \prod_{i=0,1,2} \kq_i^{|\bK_{13,i}|} \BE \left[ - {P}_3\frac{\bS_{12}\bS_{12}^*}{{P}_{12}^*} - \frac{P_2\bS_{13}\bS_{13}^* }{{P}_{13}} + {q}_2 {P}_4 \frac{\bS_{13}\bS_{12}^* }{{P}_1^*} \right]^{\BZ_3} \det(\bK_{12})^{\sk_0} \prod_{i=0,1,2} \det(\bK_{13,i})^{\sk_i}.
\end{align}
In the freezing limit $\kq_1=\kq_2=0$, only $\bK_{13}=\varnothing$ has non-zero contribution to the ensemble. The partition function reduces to a sum over $N$-tuples of Young diagrams which we still denote by $\boldsymbol\lambda=\boldsymbol\lambda_{12}$. It gives
\begin{align}
    \CalZ_\text{GO} = \sum_{\boldsymbol\lambda} \kq^{|\boldsymbol\lambda|} \BE \left[ - \frac{\bS\bS^*}{P_{12}^*} - \frac{e^x\bS^*}{P_1^*} \right] \det(\bK_{12})^{\sk}.
\end{align}
The first term gives the usual measure of the $A_1$-quiver gauge theory. The second term is obtained by integrating out the field on the brane on $\BC_{13}^2$. Thus in the five-dimensional theory point of view it is a co-dimensional two defect on the $S^1 \times \BC_1$, which is the $Q$\emph{-observable} we defined in \eqref{def:Q-func}. The gauge origami partition function provides the expectation value for the $Q$-observable
\begin{align}
    \CalZ_\text{GO} = \langle Q(x) \rangle \CalZ_{\BC^2_{12}}. 
\end{align}


The $Q$-observable is a half-BPS co-dimensional two defect in the 5d gauge theory, obtained by coupling 3d $\CalN=4$ gauged linear sigma model (supported on $S^1 \times \BC_{\ve_1} \times \{0\} \subset S^1 \times \BC_{\ve_1} \times \BC_{\ve_2}$) in the Coulomb phase. 
In the 5-brane web picture, it is a D3 brane extending in the 0125 direction which ends on the NS5 brane on the left.

\section{Detail of Calculations}

\subsection{Detail calculation for section 4}\label{sec:compu-sec4}
The expectation value of the defect $qq$-character \eqref{def:qq} 
\begin{align}
    \langle \EX_\o(x) \rangle_{\BZ_N} \Psi = \langle \EY_{\o+1}(x+\ve_1) \rangle_{\BZ_N} \Psi + \left \langle \frac{\hat\kq_\o e^{\sk_\o x}}{\EY_\o(x)} \right \rangle_{\BZ_N} \Psi
\end{align}
is analytic in $x$. That means 
\begin{align}
\begin{split}
    & \left \langle \frac{\EX_\o(X)}{\sqrt{X}} - \frac{1}{X} \underset{X=0}{\text{Res}} \frac{\EX_\o(X)}{\sqrt{X}} \right \rangle_{\BZ_N} \Psi \\
    & = \left \langle \frac{\EY_{\o+1}(X)}{\sqrt{X}} + \frac{\hat\kq_\o X^{\sk_\o}}{\sqrt{X} \ \EY_\o(X)} - \frac{1}{X} \underset{X=0}{\text{Res}} \left( \frac{\EY_{\o+1}(X)}{\sqrt{X}} + \frac{\hat\kq_\o X^{\sk_\o}}{\sqrt{X} \ \EY_\o(X)} \right) \right \rangle_{\BZ_N} \Psi
\end{split}
\end{align}
is analytic in $X=e^x$. The right hand side of the equation above can be calculated by large $X$ extension of its building block $\EY_\o(x)$ using \eqref{eq:Y-large-X}:   
\begin{align}
\begin{split}
    & \frac{\EX_\o(X)}{\sqrt{X}} - \frac{1}{X} \underset{X=0}{\text{Res}} \frac{\EX_\o(X)}{\sqrt{X}} \\
    & = e^{\frac{-a_{\o+1}+\ve_1}{2}} \left( 1 - \frac{e^{a_{\o+1}-\ve_1}}{X} \right) \prod_{\Box \in \EK_{\o+1}} e^{-\frac{\ve_1}{2}} \frac{1-\frac{e^{c_\Box}}{X}}{ 1-\frac{e^{c_\Box-\ve_1}}{X} } \prod_{\Box \in \EK_{\o}} e^{\frac{\ve_1}{2}} \frac{1- \frac{e^{c_\Box-\ve_1}}{X} q_2^{\delta_{N-1,\o}} }{1-\frac{e^{c_\Box}}{X} q_2^{\delta_{N-1,\o}} } \\
    & \quad - \frac{\hat\kq_\o X^{\sk_\o} e^{\frac{-a_\o}{2}} }{1 - \frac{X}{e^{a_\o}}} \prod_{\Box\in \EK_{\o}} e^{\frac{\ve_1}{2}} \frac{{\frac{e^{c_\Box}}{X}} - 1}{ \frac{e^{c_\Box+\ve_1} }{X} - 1 } \prod_{\Box \in \EK_{\o-1}} e^{-\frac{\ve_1}{2}} \frac{\frac{e^{c_\Box+\ve_1}}{X}q_2^{\delta_{0,\o}}-1}{\frac{e^{c_\Box}}{X}q_2^{\delta_{0,\o}}-1} + \frac{e^{\frac{a_{\o+1}-\ve_1}{2}}}{X} e^{\frac{\ve_1}{2}(k_{\o+1}-k_{\o})}.
\end{split}
\end{align}
The large $X$ expansion with $\sk_\o=1$ gives
\begin{align}
\begin{split}
    & \left[ X^{-1} \right] \left[ \frac{\EX_\o(X)}{\sqrt{X}} - \frac{1}{X} \underset{X=0}{\text{Res}} \frac{\EX_\o(X)}{\sqrt{X}} \right] \\
    & = e^{-\frac{a_{\o+1}-\ve_1}{2}} e^{\frac{\ve_1}{2}(k_{\o}-k_{\o+1})} \left[ - e^{a_{\o+1}-\ve_1} + e^{a_{\o+1}-\ve_1} e^{k_{\o+1}-k_{\o}} + \sum_{\Box \in \EK_{\o+1}} e^{c_\Box} (q_1^{-1}-1) + \sum_{\Box \in \EK_{\o}} e^{c_\Box} q_2^{\delta_{N-1,\o}} (1-q_1^{-1}) \right] \\
    & \quad + \hat\kq_\o e^{\frac{a_\o}{2}} e^{\frac{\ve_1}{2}(k_\o-k_{\o-1})} \left[ e^{a_\o} + \sum_{\Box\in \EK_\o} e^{c_\Box} (q_1-1) + \sum_{\Box \in \EK_{\o-1}} e^{c_\Box}q_2^{\delta_{\o,0}} (1-q_1) \right]. 
\end{split}
\end{align}
In the case $\sk_\o=0$, the expansion reads 
\begin{align}
\begin{split}
    & \left[ X^{-1} \right] \left[ \frac{\EX_\o(X)}{\sqrt{X}} - \frac{1}{X} \underset{X=0}{\text{Res}} \frac{\EX_\o(X)}{\sqrt{X}} \right] \\
    & = e^{\frac{-a_{\o+1}+\ve_1}{2}} e^{\frac{\ve_1}{2}(k_{\o}-k_{\o+1})} \left[ - e^{a_{\o+1}-\ve_1} + \sum_{\Box \in \EK_{\o+1}} e^{c_\Box}(q_1^{-1}-1) + \sum_{\Box\in \EK_{\o}} e^{c_\Box} (1-q_1^{-1}) q_2^{\delta_{N-1,\o}} \right] \\
    & \quad + \hat\kq_\o e^{\frac{a_\o}{2}} e^{\frac{\ve_1}{2} (k_{\o}- k_{\o-1}) } + e^{\frac{a_{\o+1}-\ve_1}{2}} e^{\frac{\ve_1}{2}(k_{\o+1}-k_\o) }.
\end{split}
\end{align}
The regularity of $qq$-character states that
\begin{align}
    \left \langle \left[ X^{-1} \right] \left[ \frac{\EX_\o(X)}{\sqrt{X}} - \frac{1}{X} \underset{X=0}{\text{Res}} \frac{\EX_\o(X)}{\sqrt{X}} \right] \right \rangle_{\BZ_N} \Psi = 0, \ \o=0,\dots,N-1.
\end{align}

When the bulk Chern-Simons level is $\sk<N$, we assign the fractional Chern-Simons level $\sk_\o = \theta_{\o<\sk}$ and consider the linear combination
\begin{align}
    \sum_{\o=0}^{N-1} \hat{C}_\o \left \langle \left[ X^{-1} \right] \left[ \frac{\EX_\o(X)}{\sqrt{X}} - \frac{1}{X} \underset{X=0}{\text{Res}} \frac{\EX_\o(X)}{\sqrt{X}} \right] \right \rangle_{\BZ_N} \Psi = 0
\end{align}
with coefficients $\hat{C}_\o$ chosen to cancel the unwanted $e^{c_\Box}$ terms. This would require
\begin{align}
\begin{split}
    & \hat{C}_{\overline{\o-1}} \left\langle e^{-\frac{a_{\o}-\ve_1}{2}} e^{\frac{\ve_1}{2}(k_{\o-1}-k_{\o})} \sum_{\Box\in \EK_\o} e^{c_\Box} \right \rangle _{\BZ_N} \Psi
    - \theta_{\o<\sk} \hat{C}_{\o} \left\langle \hat\kq_\o e^{\frac{a_\o}{2}} e^{\frac{\ve_1}{2}(k_{\o}-k_{\o-1})} q_1 \sum_{\Box\in \EK_\o} e^{c_\Box} \right \rangle_{\BZ_N} \Psi  \\
    & = \hat{C}_{\o} \left\langle e^{-\frac{a_{\o+1}-\ve_1}{2}} e^{\frac{\ve_1}{2}(k_{\o}-k_{\o+1})} \sum_{\Box\in \EK_\o} e^{c_\Box} \right \rangle _{\BZ_N} \Psi
    - \theta_{\overline{\o+1}<\sk} \hat{C}_{\overline{\o+1}}  \left\langle \hat\kq_{\o+1} e^{\frac{a_{\o+1}}{2}} e^{\frac{\ve_1}{2}(k_{\o+1}-k_\o)} q_1 \sum_{\Box\in \EK_\o} e^{c_\Box} \right \rangle _{\BZ_N} \Psi .
\end{split}
\end{align}
We can rewrite
$$
    \langle \ve_1(k_{\o-1}-k_{\o}) - a_\o \rangle_{\BZ_N} \Psi = \ve_1 \nabla^z_\o \Psi
$$
with $\nabla^z_\o = z_\o \p_{z_\o}$. The canceling condition can be rewritten as differential operators acting on expectation value of a common observable $\bK_\o = \sum_{\Box \in \EK_\o} e^{c_\Box}$
\begin{align}
\begin{split}
    & \left[ \hat{C}_{\overline{\o-1}} e^{\frac{\ve_1}{2}\nabla^z_{\o}+\frac{1}{2}} - \theta_{\o<\sk} \hat{C}_{\o} \hat\kq_\o e^{-\frac{\ve_1}{2}\nabla^z_\o+\ve_1} \right] \langle \bK_\o \rangle_{\BZ_N} \Psi \\
    & = \left[ \hat{C}_\o e^{\frac{\ve_1}{2}\nabla^z_{\o+1}+\frac{\ve_1}{2}} - \theta_{\overline{\o+1}<\sk} \hat{C}_{\overline{\o+1}} \hat\kq_{\o+1} e^{-\frac{\ve_1}{2}\nabla^z_{\o+1}+\ve_1 } \right] \langle \bK_\o \rangle_{\BZ_N} \Psi
\end{split}
\end{align}
We defined $\hat{D}_\o = \hat\kq_\o q_1^{ -\frac{1}{2} \nabla^z_\o - \frac{1}{2}\nabla^z_{\o+1}} = \hat{D}_{\o+N}$ in \eqref{def:hatD}. We find the solution when $\sk<N$ to be (up to an overscale of scalar)
\begin{subequations}
\begin{align}
    & \hat{C}_{{\o-1}} = \left[ 1 + \hat{D}_\o + \hat{D}_{\o+1} \hat{D}_{\o} + \cdots + \hat{D}_{\sk-1} \cdots \hat{D}_\o \right]  e^{-\frac{\ve_1}{2}\nabla^z_{\o}-\frac{\ve_1}{2}} , \ \o=1,\dots,\sk-1; \\
    & \hat{C}_{\o-1} = e^{-\frac{\ve_1}{2}\nabla^z_{\o}-\frac{\ve_1}{2}}, \ \o=\sk,\dots,N-1; \\
    & \hat{C}_{N-1} = \left[ 1 + \hat{D}_0 + \hat{D}_1 \hat{D}_0 + \cdots + \hat{D}_{\sk-1} \cdots \hat{D}_0 \right] e^{-\frac{\ve_1}{2}\nabla^z_{0}-\frac{\ve_1}{2}}. 
\end{align}
\end{subequations}
as an operator acting on $\Psi$
such that
\begin{align}
    \hat{C}_{\overline{\o-1}} e^{\frac{\ve_1}{2}\nabla^z_{\o}+\frac{\ve_1}{2}} - \theta_{\o<\sk} \hat{C}_\o  e^{\frac{\ve_1}{2}\nabla^z_{\o}+\frac{\ve_1}{2}} \hat{D}_\o = 1, \ \o=0,\dots,N-1.
\end{align}
The linear combination gives
\begin{align}
\begin{split}
    0 & = \sum_{\o=0}^{N-1} \hat{C}_\o \left \langle \left[ X^{-1} \right] \left[ \frac{\EX_\o(X)}{\sqrt{X}} - \frac{1}{X} \underset{X=0}{\text{Res}} \frac{\EX_\o(X)}{\sqrt{X}} \right] \right \rangle_{\BZ_N} \Psi  \\ 
    & = \sum_{\o=0}^{N-1} \hat{C}_{\o} \left[ e^{-\frac{\ve_1}{2}\nabla^z_{\o+1}-\frac{\ve_1}{2}} - e^{\frac{\ve_1}{2}\nabla^z_{\o+1}-\frac{\ve_1}{2}} e^{a_\o}
    + \theta_{\o<\sk} \hat\kq_\o e^{-\frac{\ve_1}{2}\nabla^z_\o} + (1-\theta_{\o<\sk}) \hat\kq_\o e^{-\frac{\ve_1}{2}\nabla^z_\o} \right] \Psi \\
    & \qquad + (q_2-1)(1-q_1^{-1}) \left \langle \sum_{\Box\in \EK_{N-1}} e^{c_\Box} \right\rangle_{\BZ_N} \Psi \\
    & = \sum_{\o=0}^{N-1}  \hat{C}_{\o} \left[ e^{-\frac{\ve_1}{2}\nabla^z_{\o+1}-\frac{\ve_1}{2}} + (1-\theta_{\o<\sk}) \hat\kq_\o e^{a_\o} e^{-\frac{\ve_1}{2}\nabla^z_\o} \right] \Psi - \langle q_1^{-1} \bS \rangle_{\BZ_N} \Psi \\
    & = q_1^{-1} \left[ \hat{\rm H}|_{N,\sk} - \langle \bS \rangle_{\BZ_N} \right] \Psi
\end{split}
\end{align}
The Hamiltonian $\hat{\rm H}|_{N,\sk}$ is
\begin{align}
\begin{split}
    \hat{\rm H}|_{N,\sk} = 
    & \sum_{\o=0}^{N-1} \hat{C}_\o q_1^{\frac{1}{2}\nabla^z_{\o+1}+\frac{1}{2}} \left(  q_1^{-\nabla^z_{\o+1}} + (1-\theta_{\o<\sk})  \hat{D}_\o \right), \quad \hat{D}_\o = \hat\kq_\o q_1^{-\frac{1}{2}\nabla^z_{\o}-\frac{1}{2} \nabla^z_{\o+1} },  \\
    = &  \sum_{\o=0}^{\sk-1} \left( 1 + \hat{D}_{\o} + \cdots + \hat{D}_{\sk-1}\cdots \hat{D}_{\o} \right) q_1^{-\nabla^z_{\o}} + \sum_{\o=\sk}^{N-2} q_1^{-\nabla^z_{\o}} + \hat{D}_{\o} \\
    & + q_1^{-\nabla^z_{N-1}} + \hat{D}_{N-1} + \hat{D}_0 \hat{D}_{N-1} + \cdots + \hat{D}_{\sk-1} \cdots \hat{D}_0 \hat{D}_{N-1}.
\end{split}
\end{align}

\paragraph{}
When the bulk Chern-Simons level $\sk=N$, the fractional Chern-Simons levels are $\sk_\o=1$ for all $\o=0,\dots,N-1$. We consider the linear combination
\begin{align}
    \sum_{\o=0}^{N-1} \hat{C}_\o \left \langle \left[ X^{-1} \right] \left[ \frac{\EX_\o(X)}{\sqrt{X}} - \frac{1}{X} \underset{X=0}{\text{Res}} \frac{\EX_\o(X)}{\sqrt{X}} \right] \right \rangle_{\BZ_N} \Psi = 0
\end{align}
with coefficients $\hat{C}_\o$ chosen to cancel the unwanted $e^{c_\Box}$ terms:
\begin{align}
\begin{split}
    & \hat{C}_{\overline{\o-1}} \left\langle e^{-\frac{a_{\o}-\ve_1}{2}} e^{\frac{\ve_1}{2}(k_{\o-1}-k_{\o})} \sum_{\Box\in \EK_\o} e^{c_\Box} \right \rangle - \hat{C}_{\o} \left\langle \hat\kq_\o e^{\frac{a_\o}{2}} e^{\frac{\ve_1}{2}(k_{\o}-k_{\o-1})} q_1 \sum_{\Box\in \EK_\o} e^{c_\Box} \right \rangle  \\
    & = \hat{C}_{\o} \left\langle e^{-\frac{a_{\o+1}-\ve_1}{2}} e^{\frac{\ve_1}{2}(k_{\o}-k_{\o+1})} \sum_{\Box\in \EK_\o} e^{c_\Box} \right \rangle 
    - \hat{C}_{\overline{\o+1}}  \left\langle \hat\kq_{\o+1} e^{\frac{a_{\o+1}}{2}} e^{\frac{\ve_1}{2}(k_{\o+1}-k_\o)} q_1 \sum_{\Box\in \EK_\o} e^{c_\Box} \right \rangle.
\end{split}
\end{align}
which can be organized into 
\begin{align}
\begin{split}
    & \left[ \hat{C}_{\overline{\o-1}} e^{\frac{\ve_1}{2}\nabla^z_\o+\frac{\ve_1}{2}} - \hat{C}_\o \hat\kq_\o e^{-\frac{\ve_1}{2}\nabla^z_\o} \right] \langle \bK_\o \rangle_{\BZ_N} \Psi \\
    & = \left[ \hat{C}_\o e^{\frac{\ve_1}{2}\nabla^z_{\o+1}+\frac{\ve_1}{2}} - \hat{C}_{\overline{\o+1}} \hat{\kq}_{\o+1} e^{-\frac{\ve_1}{2}\nabla^z_{\o+1}} \right] \langle \bK_\o \rangle_{\BZ_N} \Psi
\end{split}
\end{align}
Up to an over all scaling, the solution is chosen as
\begin{align}
    \hat{C}_\o = \left[ 1 + \hat{D}_{\o+1} + \hat{D}_{\o+2} \hat{D}_{\o+1} + \cdots + \hat{D}_{\o+N-1} \cdots \hat{D}_{\o+1} \right] q_1^{-\frac{1}{2}\nabla^z_{\o+1} - \frac{1}{2}}
\end{align}
with $\hat{D}_\o = \hat{D}_{\o+N}$ defined in \eqref{def:hatD}. In particular 
\begin{align}
    \hat{D}_{\o+N-1} \cdots \hat{D}_{\o} \Psi = \kq e^{\sum_{\o=0}^{N-1} -\ve_1 \nabla^z_\o } q_1^{\frac{1-N}{2}} \Psi = \kq e^{a}q_1^{\frac{1-N}{2}}
\end{align}
The linear combination gives
\begin{align}
\begin{split}
    0 & = \sum_{\o=0}^{N-1} \hat{C}_\o \left \langle \left[ X^{-1} \right] \left[ \frac{\EX_\o(X)}{\sqrt{X}} - \frac{1}{X} \underset{X=0}{\text{Res}} \frac{\EX_\o(X)}{\sqrt{X}} \right] \right \rangle_{\BZ_N} \Psi  \\ 
    & = \sum_{\o=0}^{N-1} \hat{C}_{\o} \left[ e^{-\frac{\ve_1}{2}\nabla^z_{\o+1}-\frac{\ve_1}{2}} - e^{\frac{\ve_1}{2}\nabla^z_{\o+1}-\frac{\ve_1}{2}} e^{a_\o}
    + \hat\kq_\o e^{-\frac{\ve_1}{2}\nabla^z_\o} \right] \Psi \\
    & \qquad + (1 - \hat{D}_{N-1}\dots\hat{D}_0) (q_2-1)(1-q_1^{-1}) \left \langle \sum_{\Box\in \EK_{N-1}} e^{c_\Box} \right\rangle_{\BZ_N} \Psi \\
    & = \sum_{\o=0}^{N-1}  \hat{C}_{\o} e^{-\frac{\ve_1}{2}\nabla^z_{\o+1}-\frac{\ve_1}{2}} \Psi - q_1^{-1} \left( 1-\kq e^{a}q_1^{\frac{1-N}{2}} \right) \langle \bS \rangle_{\BZ_N} \Psi \\
    & = q_1^{-1} \left[ \hat{\rm H}|_{N,\sk} - \left( 1-\kq e^{a}q_1^{\frac{1-N}{2}} \right) \langle \bS \rangle_{\BZ_N} \right] \Psi
\end{split}
\end{align}
The Hamiltonian take the form as in \eqref{eq:H-N-N}
\begin{align}
    \hat{\rm H} = \sum_{\o=0}^{N-1} \left[ 1 + \hat{D}_\o + \cdots + \hat{D}_{\o+N-2} \cdots \hat{D}_\o \right] e^{-\ve_1\nabla^z_\o}
\end{align}

\subsection{Detail calculation for section 5}\label{sec:comput-sec5}
The gauge origami for the derivation of fractional T-Q equation 
in section.~\NL{hyper link} reads
\begin{align}
\begin{split}
    \hat\bN_{12} & = \sum_{\o'=0}^{N-1} e^{{a}_{\o'}} \hat{q}_2^{\o'} \CalR_0 \otimes \fR_{\o'}, \ \hat\bK_{12} = \sum_{\o'=0}^{N-1} \bK_{12,\o} \hat{q}_2^{\o'} \CalR_0 \otimes \fR_{\o'} \\
    \hat{\bN}_{13} & = \sum_{\o'=0}^{N-1} e^{x'_{\o'}+\ve_1+\ve_3} \hat{q}_2^{\o'} \CalR_1 \otimes \fR_{\o'}, \\
    \hat{\bN}_{34} & = e^x \hat{q}_2^{\o} \CalR_0 \otimes \fR_{\o}, \ \hat{\bK}_{34} = \bK_{34,\o} \hat{q}_2^{\o} \CalR_0 \otimes \fR_{\o}.
\end{split}
\end{align}
Here we keep $N$ $x'_{\o'}$ arbitrary. There can 
The Gauge origami partition function is given by 
\begin{align}
\begin{split}
    \hat\CalZ_{\text{GO},\o} = \sum_{\hat{\boldsymbol{\lambda}},k_{34}=0,1} & \hat\kq_\o^{k_{34}} \det(\bK_{34,\o})^{\sk_\o} \prod_{\o'=0}^{N} \hat\kq_{\o'}^{k_{\o'}} \times \prod_{\o'=0}^{N-1} \det(\bK_{12,\o'})^{\sk_{\o'}} \\
    \times \BE &  \left[ -\frac{\hat{P}_3^*\hat{\bS}_{12}\hat{\bS}_{12}^*}{\hat{P}_{12}^*} - \frac{\hat{P}_{2}^*\hat{\bS}_{13}\hat{\bS}_{13}^*}{\hat{P}_{13}} - \frac{\hat{P}_{1}\hat{\bS}_{34}\hat{\bS}_{34}^*}{\hat{P}_{34}} -\hat{q}_{12} \bS_{34}^* \hat\bS_{12}^* \right. \\ 
    & \quad \left. + \hat{q}_{2} \hat{P}_4 \hat{\bS}_{13}\frac{\hat{\bS}_{12}^*}{\hat{P}_{1}^*} -\hat{q}_{13} \hat\bN_{34} \hat\bN_{13}^* - \hat{q}_{34} \hat\bN_{13} \hat\bN_{34}^* + \hat{P}_2 \hat{P_4} \hat\bN_{13} \hat{\bN}_{34}^* \right]^{\BZ_3 \times \BZ_N} \\  
\end{split}
\end{align}
The $qq$-character gives
\begin{align}
\begin{split}
    T_\o (x,\bx') {\EQ}(\bx') = & \ \sh(x-x'_\o) \EY_{\overline{\o+1}}(x+\ve_1+\ve_2 \d_{\o,N-1}) {\EQ}(\bx') \\
    & + \hat\kq_\o e^{\sk_\o x} \sh({x-x'_{\overline{\o+1}}+\ve_2\d_{\o,N-1} }) \frac{{\EQ}(\bx')}{\EY_\o(x)}
\end{split}
\end{align}
with
\begin{align}
    {\EQ}(\bx') = \prod_{\o=0}^{N-1} Q_\o(x_\o'), \quad \frac{Q_\o(x)}{Q_\o(x-\ve_1)} = \EY_\o(x). 
\end{align}
Unlike in the previous section, here we keep the assignment of fractional Chern-Simons levels $\sk_\o$ arbitrary with $\sum_{\o=0}^{N-1} \sk_\o = \sk$. 
We consider two special cases
\begin{subequations}
\begin{align}
    T_\o(x=x'_\o,\bx'){\EQ}(\bx') & = \hat\kq_\o e^{\sk_\o x_\o'} \sh({x'_\o-x'_{\overline{\o+1}} +\ve_2 \d_{\o,N-1} }) \frac{{\EQ}(\bx')}{\EY_{\o}(x'_\o)} \\
    & = \hat\kq_\o e^{\sk_\o x'_\o} \sh({x'_\o - x'_{\overline{\o+1}}+\ve_2 \d_{\o,N-1} }) {\EQ}(\bx' - \ve_1 e_{\o}) \nonumber\\
    T_\o(x=x'_{\overline{\o+1}}-\ve_2\d_{\o,N-1},\bx') {\EQ}(\bx')
    & = \sh(-{x'_\o + x'_{\overline{\o+1}}-\ve_2\d_{\o,N-1} }) \EY_{\overline{\o+1}}(x'_{\overline{\o+1}}+\ve_1) {\EQ} (\bx') \\
    & = -\sh({x'_\o - x'_{\overline{\o+1}} + \ve_2\d_{\o,N-1} }) {\EQ} (\bx'+\ve_1 e_{\overline{\o+1}}) \nonumber
\end{align}
\end{subequations}
and define
\begin{align}
    \langle \ET_\o(\bx') {\EQ}(\bx') \rangle_{\BZ_N} = \left \langle \frac{T_\o(x=x'_\o,\bx'){\EQ}(\bx') - T_\o(x=x'_{\overline{\o+1}}-\ve_2\d_{\o,N-1},\bx'){\EQ}(\bx')}{\sh({x'_\o-x'_{\overline{\o+1}} +\ve_2 \d_{\o,N-1} })} \right \rangle_{\BZ_N}
\end{align}
which satisfies the fractional T-Q equation
\begin{align}\label{eq:frac-TQ}
    \langle \ET_\o(\bx'){\EQ}(\bx') \rangle_{\BZ_N} = \langle {\EQ}(\bx'+\ve_1e_{\overline{\o+1}}) \rangle_{\BZ_N} + \hat\kq_\o e^{\sk_\o x'_\o} \langle {\EQ}(\bx'-\ve_1 e_\o) \rangle_{\BZ_N}
\end{align}
To calculate $\ET_\o(x)$, we write down the explicit form of $T_\o(x,\rx')$ in $\EY_\o(x)$:
\begin{align}
\begin{split}
    T_\o(x,\bx')
    & = \left( \frac{X}{e^{\frac{x'_\o}{2} }} - {e^{\frac{x_{\o}'}{2} }} \right) \left[  e^{\frac{-a_{\o+1}+\ve_1}{2}} \left( 1 - \frac{e^{a_{\o+1}-\ve_1}}{X} \right) \prod_{\Box \in \EK_{\o+1}} e^{-\frac{\ve_1}{2}} \frac{1-\frac{e^{c_\Box}}{X}}{ 1-\frac{e^{c_\Box-\ve_1}}{X} } \prod_{\Box \in \EK_{\o}} e^{\frac{\ve_1}{2}} \frac{1- \frac{e^{c_\Box-\ve_1}}{X} q_2^{\delta_{N-1,\o}} } {1-\frac{e^{c_\Box}}{X}q_2^{\delta_{N-1,\o}} } \right] \\
    & \quad - \left( \frac{X}{e^{ \frac{x'_{\overline{\o+1}}-\ve_2\d_{\o,N-1} }{2} }} - {e^{\frac{x'_{\overline{\o+1}}-\ve_2\d_{\o,N-1} }{2}}} \right) \left[  \frac{\hat\kq_\o X^{\sk_\o} e^{\frac{-a_\o}{2}} }{1 - \frac{X}{e^{a_\o}}} \prod_{\Box\in \EK_{\o}} e^{\frac{\ve_1}{2}} \frac{{\frac{e^{c_\Box}}{X}} - 1}{ \frac{e^{c_\Box+\ve_1} }{X} - 1 } \prod_{\Box \in \EK_{\o-1}} e^{-\frac{\ve_1}{2}} \frac{\frac{e^{c_\Box+\ve_1}}{X}q_2^{\delta_{0,\o}}-1}{\frac{e^{c_\Box}}{X}q_2^{\delta_{0,\o}}-1} \right]
\end{split}
\end{align}
where $X=e^x$. Since $\langle T_\o(x,\bx') \rangle$ is analytic in $x$, it is analytic in $X$ except for potential poles coming from $X=0$ (corresponds to $x \to -\infty$). Once we removed the poles at $X=0$, 
\begin{align}
\begin{split}
    & T_\o(X,\bx') - \frac{1}{X} \underset{X=0}{\text{Res}}T_\o(X,\bx') \\
    & = \left( \frac{X}{e^{\frac{x'_\o}{2} }} - {e^{\frac{x_{\o}'}{2} }} \right) \left[  e^{\frac{-a_{\o+1}+\ve_1}{2}} \left( 1 - \frac{e^{a_{\o+1}-\ve_1}}{X} \right) \prod_{\Box \in \EK_{\o+1}} e^{-\frac{\ve_1}{2}} \frac{1-\frac{e^{c_\Box}}{X}}{ 1-\frac{e^{c_\Box-\ve_1}}{X} } \prod_{\Box \in \EK_{\o}} e^{\frac{\ve_1}{2}} \frac{1- \frac{e^{c_\Box-\ve_1}}{X} q_2^{\delta_{N-1,\o}} } {1-\frac{e^{c_\Box}}{X}q_2^{\delta_{N-1,\o}} } \right] \\
    & \quad - \left( \frac{X}{e^{ \frac{x'_{\overline{\o+1}}-\ve_2\d_{\o,N-1} }{2} }} - {e^{\frac{x'_{\overline{\o+1}}-\ve_2\d_{\o,N-1} }{2}}} \right) \left[  \frac{\hat\kq_\o X^{\sk_\o} e^{\frac{-a_\o}{2}} }{1 - \frac{X}{e^{a_\o}}} \prod_{\Box\in \EK_{\o}} e^{\frac{\ve_1}{2}} \frac{{\frac{e^{c_\Box}}{X}} - 1}{ \frac{e^{c_\Box+\ve_1} }{X} - 1 } \prod_{\Box \in \EK_{\o-1}} e^{-\frac{\ve_1}{2}} \frac{\frac{e^{c_\Box+\ve_1}}{X}q_2^{\delta_{0,\o}}-1}{\frac{e^{c_\Box}}{X}q_2^{\delta_{0,\o}}-1} \right] \\
    & \quad - \frac{1}{X} e^{\frac{x'_\o}{2}+\frac{a_{\o+1}-\ve_1}{2} - \frac{\ve_1}{2}(k_\o - k_{\o+1}) }
\end{split}
\end{align}
is a polynomial in $X$. For the case $\sk_\o=1$: 
\begin{align}
\begin{split}
    & T_\o(X,\bx') - \frac{1}{X} \underset{X=0}{\text{Res}}T_\o(X,\bx') \\
    = & \left( \frac{X}{e^{\frac{x'_\o}{2} }} - {e^{\frac{x_{\o}'}{2} }} \right) e^{\frac{-a_{\o+1}+\ve_1}{2} + \frac{\ve_1}{2}(k_\o-k_{\o+1}) } + \left( \frac{X}{e^{ \frac{x'_{\overline{\o+1}}-\ve_2\d_{\o,N-1} }{2} }} - {e^{\frac{x'_{\overline{\o+1}}-\ve_2\d_{\o,N-1} }{2}}} \right) \hat\kq_\o e^{\frac{a_\o}{2} - \frac{\ve_1}{2}(k_{\o-1}-k_\o) } \\
    & - e^{\frac{-x_\o'-a_{\o+1}+\ve_1}{2} + \frac{\ve_2}{2} (k_\o-k_{\o+1}) } \left[ e^{a_{\o+1}-\ve_1} + \sum_{\Box \in \EK_{\o+1}} (1-q_1^{-1}) e^{c_\Box} + \sum_{\Box \in \EK_\o} (q_1^{-1}-1) e^{c_\Box} q_2^{\d_{N-1,\o}} \right] \\
    & - \hat\kq_\o e^{-\frac{x'_{\overline{\o+1}}-\ve_2\d_{\o,N-1} }{2}}  e^{\frac{a_\o}{2}-\frac{\ve_1}{2}(k_{\o-1}-k_\o)} \left[ e^{a_\o} + \sum_{\Box\in \EK_{\o}} e^{c_\Box} (1-q_1) + \sum_{\Box \in \EK_{\o-1}} (q_1-1) e^{c_\Box} q_2^{\d_{\o,0}} \right]
\end{split}
\end{align}
This gives
\begin{align}\label{eq:ET-k_o=1}
    \ET_\o(\bx')|_{\sk_\o=1} =  e^{\frac{x'_{\overline{\o+1}}}{2}} e^{\frac{\ve_1}{2}\nabla^z_{\o+1} +\frac{\ve_1}{2} } - e^{- \frac{x'_{\overline{\o+1}}}{2}} e^{-\frac{\ve_1}{2}\nabla^z_{\o+1} - \frac{\ve_1}{2} }   +  \hat\kq_\o e^{\frac{x'_{\o}}{2}} e^{-\frac{\ve_1}{2}\nabla^z_{\o} } 
\end{align}
as an operator acting on $\prod_{\o=0}^{N-1} z_\o^{-\frac{a_\o}{\ve_1}} \langle\EQ(\bx') \rangle_{\BZ_N} \hat\CalZ_{c,\sk_\o} $. 
For the case $\sk_\o=0$:
\begin{align}
\begin{split}
    & T_\o(X,\bx') - \frac{1}{X} \underset{X=0}{\text{Res}}T_\o(X,\bx') \\
    = & \left( \frac{X}{e^{\frac{x'_\o}{2} }} - {e^{\frac{x_{\o}'}{2} }} \right) e^{-\frac{a_{\o+1}+\ve_1}{2} + \frac{\ve_1}{2}(k_\o-k_{\o+1}) } + \hat\kq_\o e^{-\frac{x'_{\o+1}-\ve_2\d_{\o,N-1} }{2}} e^{\frac{a_\o}{2} - \frac{\ve_1}{2}(k_{\o-1}-k_\o)} \\
    & - e^{\frac{-x_\o'-a_{\o+1}+\ve_1}{2} + \frac{\ve_2}{2} (k_\o-k_{\o+1}) } \left[ e^{a_{\o+1}-\ve_1} + \sum_{\Box \in \EK_{\o+1}} (1-q_1^{-1}) e^{c_\Box} + \sum_{\Box \in \EK_\o} (q_1^{-1}-1) e^{c_\Box} q_2^{\d_{N-1,\o}} \right]
\end{split}
\end{align}
This gives
\begin{align}\label{eq:ET-k_o=0}
    \ET_\o(\bx')|_{\sk_\o=0} = e^{\frac{x'_{\overline{\o+1}}}{2}} e^{\frac{\ve_1}{2}\nabla^z_{\o+1} +\frac{\ve_1}{2} } - e^{- \frac{x'_{\overline{\o+1}}}{2}} e^{-\frac{\ve_1}{2}\nabla^z_{\o+1} - \frac{\ve_1}{2} } 
\end{align}
The final result for the $\ET_\o(\bx')$ function is 
\begin{align}
    \ET_\o(\bx') = \sh(x'_{\overline{\o+1}} + \ve_1 + \ve_1\nabla^z_{\o+1} ) + \sk_\o \hat\kq_\o e^{\frac{x'_{\o}}{2}} e^{-\frac{\ve_1}{2}\nabla^z_{\o} }
\end{align}

\newpage
\bibliographystyle{utphys}
\bibliography{R-Toda}

\providecommand{\href}[2]{#2}\begingroup\raggedright\begin{thebibliography}{10}

\bibitem{Nikita-Shatashvili}
N.~A. Nekrasov and S.~L. Shatashvili, \href{http://dx.doi.org/10.1142/9789814304634_0015}{``{Quantization of Integrable Systems and Four Dimensional Gauge Theories},''} in {\em {XVIth International Congress on Mathematical Physics}}, pp.~265--289.
\newblock 2009.
\newblock
\href{http://arxiv.org/abs/0908.4052}{{\ttfamily arXiv:0908.4052 [hep-th]}}.
\newblock

\bibitem{NRS2011}
N.~Nekrasov, A.~Rosly, and S.~Shatashvili, ``{Darboux coordinates, Yang-Yang functional, and gauge theory},'' \href{http://dx.doi.org/10.1016/j.nuclphysbps.2011.04.150}{{\em Nucl.Phys.Proc.Suppl.} {\bfseries 216} (2011) 69--93},
\href{http://arxiv.org/abs/1103.3919}{{\ttfamily arXiv:1103.3919 [hep-th]}}.

\bibitem{Nikita:I}
N.~Nekrasov, ``{BPS/CFT correspondence: non-perturbative Dyson--Schwinger equations and $qq$-characters},'' \href{http://dx.doi.org/10.1007/JHEP03(2016)181}{{\em JHEP} {\bfseries 1603} (2016) 181},
\href{http://arxiv.org/abs/1512.05388}{{\ttfamily arXiv:1512.05388 [hep-th]}}.

\bibitem{Nekrasov:2012xe}
N.~Nekrasov and V.~Pestun, ``{Seiberg-Witten geometry of four dimensional N=2 quiver gauge theories},'' \href{http://arxiv.org/abs/1211.2240}{{\ttfamily arXiv:1211.2240 [hep-th]}}.

\bibitem{Nekrasov:1996cz}
N.~Nekrasov, ``{Five dimensional gauge theories and relativistic integrable systems},'' \href{http://dx.doi.org/10.1016/S0550-3213(98)00436-2}{{\em Nucl. Phys. B} {\bfseries 531} (1998) 323--344}, \href{http://arxiv.org/abs/hep-th/9609219}{{\ttfamily arXiv:hep-th/9609219}}.

\bibitem{goncharov2011dimers}
A.~B. Goncharov and R.~Kenyon, ``Dimers and cluster integrable systems,'' 2011.

\bibitem{Eager:2011dp}
R.~Eager, S.~Franco, and K.~Schaeffer, ``{Dimer Models and Integrable Systems},'' \href{http://dx.doi.org/10.1007/JHEP06(2012)106}{{\em JHEP} {\bfseries 06} (2012) 106}, \href{http://arxiv.org/abs/1107.1244}{{\ttfamily arXiv:1107.1244 [hep-th]}}.

\bibitem{Benvenuti:2004wx}
S.~Benvenuti, A.~Hanany, and P.~Kazakopoulos, ``{The Toric phases of the Y**p,q quivers},'' \href{http://dx.doi.org/10.1088/1126-6708/2005/07/021}{{\em JHEP} {\bfseries 07} (2005) 021}, \href{http://arxiv.org/abs/hep-th/0412279}{{\ttfamily arXiv:hep-th/0412279}}.

\bibitem{Franco:2005rj}
S.~Franco, A.~Hanany, K.~D. Kennaway, D.~Vegh, and B.~Wecht, ``{Brane dimers and quiver gauge theories},'' \href{http://dx.doi.org/10.1088/1126-6708/2006/01/096}{{\em JHEP} {\bfseries 01} (2006) 096}, \href{http://arxiv.org/abs/hep-th/0504110}{{\ttfamily arXiv:hep-th/0504110}}.

\bibitem{Benvenuti:2004dy}
S.~Benvenuti, S.~Franco, A.~Hanany, D.~Martelli, and J.~Sparks, ``{An Infinite family of superconformal quiver gauge theories with Sasaki-Einstein duals},'' \href{http://dx.doi.org/10.1088/1126-6708/2005/06/064}{{\em JHEP} {\bfseries 06} (2005) 064}, \href{http://arxiv.org/abs/hep-th/0411264}{{\ttfamily arXiv:hep-th/0411264}}.

\bibitem{Chen:2012we}
H.-Y. Chen, T.~J. Hollowood, and P.~Zhao, ``{A 5d/3d duality from relativistic integrable system},'' \href{http://dx.doi.org/10.1007/JHEP07(2012)139}{{\em JHEP} {\bfseries 07} (2012) 139}, \href{http://arxiv.org/abs/1205.4230}{{\ttfamily arXiv:1205.4230 [hep-th]}}.

\bibitem{HYC:2011}
H.-Y. Chen, N.~Dorey, T.~J. Hollowood, and S.~Lee, ``{A New 2d/4d Duality via Integrability},'' \href{http://dx.doi.org/10.1007/JHEP09(2011)040}{{\em JHEP} {\bfseries 09} (2011) 040},
\href{http://arxiv.org/abs/1104.3021}{{\ttfamily arXiv:1104.3021 [hep-th]}}.

\bibitem{Dorey:2011pa}
N.~Dorey, S.~Lee, and T.~J. Hollowood, ``{Quantization of Integrable Systems and a 2d/4d Duality},'' \href{http://dx.doi.org/10.1007/JHEP10(2011)077}{{\em JHEP} {\bfseries 10} (2011) 077},
\href{http://arxiv.org/abs/1103.5726}{{\ttfamily arXiv:1103.5726 [hep-th]}}.

\bibitem{Kim:2016qqs}
H.-C. Kim, ``{Line defects and 5d instanton partition functions},'' \href{http://dx.doi.org/10.1007/JHEP03(2016)199}{{\em JHEP} {\bfseries 03} (2016) 199}, \href{http://arxiv.org/abs/1601.06841}{{\ttfamily arXiv:1601.06841 [hep-th]}}.

\bibitem{Tong:2014cha}
D.~Tong and K.~Wong, ``{Instantons, Wilson lines, and D-branes},'' \href{http://dx.doi.org/10.1103/PhysRevD.91.026007}{{\em Phys. Rev. D} {\bfseries 91} no.~2, (2015) 026007}, \href{http://arxiv.org/abs/1410.8523}{{\ttfamily arXiv:1410.8523 [hep-th]}}.

\bibitem{Nikita:II}
N.~Nekrasov, ``{BPS/CFT correspondence II: Instantons at crossroads, Moduli and Compactness Theorem},'' \href{http://dx.doi.org/10.4310/ATMP.2017.v21.n2.a4}{{\em Adv. Theor. Math. Phys.} {\bfseries 21} (2017) 503--583},
\href{http://arxiv.org/abs/1608.07272}{{\ttfamily arXiv:1608.07272 [hep-th]}}.

\bibitem{Nekrasov:2009ui}
N.~A. Nekrasov and S.~L. Shatashvili, ``{Quantum integrability and supersymmetric vacua},'' \href{http://dx.doi.org/10.1143/PTPS.177.105}{{\em Prog. Theor. Phys. Suppl.} {\bfseries 177} (2009) 105--119}, \href{http://arxiv.org/abs/0901.4748}{{\ttfamily arXiv:0901.4748 [hep-th]}}.

\bibitem{nekrasov20042d}
N.~Nekrasov, ``2d cft-type equations from 4d gauge theory,'' in {\em Lecture at the IAS conference “Langlands Program and Physics}.
\newblock 2004.

\bibitem{Gukov:2008sn}
S.~Gukov and E.~Witten, ``{Rigid Surface Operators},'' \href{http://dx.doi.org/10.4310/ATMP.2010.v14.n1.a3}{{\em Adv. Theor. Math. Phys.} {\bfseries 14} no.~1, (2010) 87--178}, \href{http://arxiv.org/abs/0804.1561}{{\ttfamily arXiv:0804.1561 [hep-th]}}.

\bibitem{Gukov:2006jk}
S.~Gukov and E.~Witten, ``{Gauge Theory, Ramification, And The Geometric Langlands Program},'' \href{http://arxiv.org/abs/hep-th/0612073}{{\ttfamily arXiv:hep-th/0612073}}.

\bibitem{Nikita:III}
N.~Nekrasov, ``{BPS/CFT Correspondence III: Gauge Origami partition function and $qq$-characters},'' \href{http://dx.doi.org/10.1007/s00220-017-3057-9}{{\em Commun. Math. Phys.} {\bfseries 358} (2017) 863--894},
\href{http://arxiv.org/abs/1701.00189}{{\ttfamily arXiv:1701.00189 [hep-th]}}.

\bibitem{Nikita:V}
N.~Nekrasov, ``{BPS/CFT correspondence V: BPZ and KZ equations from $qq$-characters},''
\href{http://arxiv.org/abs/1711.11582}{{\ttfamily arXiv:1711.11582 [hep-th]}}.

\bibitem{Jeong:2023qdr}
S.~Jeong, N.~Lee, and N.~Nekrasov, ``{Parallel surface defects, Hecke operators, and quantum Hitchin system},'' \href{http://arxiv.org/abs/2304.04656}{{\ttfamily arXiv:2304.04656 [hep-th]}}.

\bibitem{jeong2021intersecting}
S.~Jeong, N.~Lee, and N.~Nekrasov, ``{Intersecting defects in gauge theory, quantum spin chains, and Knizhnik-Zamolodchikov equations},'' \href{http://dx.doi.org/10.1007/JHEP10(2021)120}{{\em JHEP} {\bfseries 10} (2021) 120}, \href{http://arxiv.org/abs/2103.17186}{{\ttfamily arXiv:2103.17186 [hep-th]}}.

\bibitem{Chen:2019vvt}
H.-Y. Chen, T.~Kimura, and N.~Lee, ``{Quantum Elliptic Calogero-Moser Systems from Gauge Origami},'' \href{http://dx.doi.org/10.1007/JHEP02(2020)108}{{\em JHEP} {\bfseries 02} (2020) 108}, \href{http://arxiv.org/abs/1908.04928}{{\ttfamily arXiv:1908.04928 [hep-th]}}.

\bibitem{Chen:2020rxu}
H.-Y. Chen, T.~Kimura, and N.~Lee, ``{Quantum Integrable Systems from Supergroup Gauge Theories},'' \href{http://dx.doi.org/10.1007/JHEP09(2020)104}{{\em JHEP} {\bfseries 09} (2020) 104}, \href{http://arxiv.org/abs/2003.13514}{{\ttfamily arXiv:2003.13514 [hep-th]}}.

\bibitem{Lee:2020hfu}
N.~Lee and N.~Nekrasov, ``{Quantum spin systems and supersymmetric gauge theories. Part I},'' \href{http://dx.doi.org/10.1007/JHEP03(2021)093}{{\em JHEP} {\bfseries 03} (2021) 093}, \href{http://arxiv.org/abs/2009.11199}{{\ttfamily arXiv:2009.11199 [hep-th]}}.

\bibitem{Jeong:2017pai}
S.~Jeong, ``{Splitting of surface defect partition functions and integrable systems},'' \href{http://dx.doi.org/10.1016/j.nuclphysb.2018.12.007}{{\em Nucl. Phys.} {\bfseries B938} (2019) 775--806},
\href{http://arxiv.org/abs/1709.04926}{{\ttfamily arXiv:1709.04926 [hep-th]}}.

\bibitem{Jeong:2024hwf}
S.~Jeong, N.~Lee, and N.~Nekrasov, ``{di-Langlands correspondence and extended observables},'' \href{http://arxiv.org/abs/2402.13888}{{\ttfamily arXiv:2402.13888 [hep-th]}}.

\bibitem{Jeong:2024mxr}
S.~Jeong and N.~Lee, ``{Bispectral duality and separation of variables from surface defect transition},'' \href{http://arxiv.org/abs/2402.13889}{{\ttfamily arXiv:2402.13889 [hep-th]}}.

\bibitem{ishii2016dimer}
A.~Ishii and K.~Ueda, ``Dimer models and the special mckay correspondence,'' {\em Geometry \& Topology} {\bfseries 19} no.~6, (2016) 3405--3466.

\bibitem{Hanany:2005ve}
A.~Hanany and K.~D. Kennaway, ``{Dimer models and toric diagrams},'' \href{http://arxiv.org/abs/hep-th/0503149}{{\ttfamily arXiv:hep-th/0503149}}.

\bibitem{Eager:2010ji}
R.~Eager, ``{Brane Tilings and Non-Commutative Geometry},'' \href{http://dx.doi.org/10.1007/JHEP03(2011)026}{{\em JHEP} {\bfseries 03} (2011) 026}, \href{http://arxiv.org/abs/1003.2862}{{\ttfamily arXiv:1003.2862 [hep-th]}}.

\bibitem{kenyon2003introduction}
R.~Kenyon, ``An introduction to the dimer model,'' {\em arXiv preprint math/0310326} (2003) .

\bibitem{treumann2019kasteleyn}
D.~Treumann, H.~Williams, and E.~Zaslow, ``Kasteleyn operators from mirror symmetry,'' {\em Selecta Mathematica} {\bfseries 25} (2019) 1--36.

\bibitem{Huang:2020neq}
M.-x. Huang, Y.~Sugimoto, and X.~Wang, ``{Quantum periods and spectra in dimer models and Calabi-Yau geometries},'' \href{http://dx.doi.org/10.1007/JHEP09(2020)168}{{\em JHEP} {\bfseries 09} (2020) 168}, \href{http://arxiv.org/abs/2006.13482}{{\ttfamily arXiv:2006.13482 [hep-th]}}.

\bibitem{DW1}
R.~Donagi and E.~Witten, ``{Supersymmetric Yang--Mills theory and integrable systems},'' \href{http://dx.doi.org/10.1016/0550-3213(95)00609-5}{{\em Nucl. Phys.} {\bfseries B460} (1996) 299--334},
\href{http://arxiv.org/abs/hep-th/9510101}{{\ttfamily arXiv:hep-th/9510101 [hep-th]}}.

\bibitem{Gorsky:1994dj}
A.~Gorsky and N.~Nekrasov, ``{Elliptic Calogero-Moser system from two-dimensional current algebra},''
\href{http://arxiv.org/abs/hep-th/9401021}{{\ttfamily arXiv:hep-th/9401021 [hep-th]}}.

\bibitem{Nekrasov:2002qd}
N.~A. Nekrasov, ``{Seiberg--Witten prepotential from instanton counting},'' \href{http://dx.doi.org/10.4310/ATMP.2003.v7.n5.a4}{{\em Adv. Theor. Math. Phys.} {\bfseries 7} (2004) 831--864}, \href{http://arxiv.org/abs/hep-th/0206161}{{\ttfamily hep-th/0206161}}.

\bibitem{Pestun:2016zxk}
V.~Pestun {\em et~al.}, ``{Localization techniques in quantum field theories},'' \href{http://dx.doi.org/10.1088/1751-8121/aa63c1}{{\em J. Phys.} {\bfseries A50} no.~44, (2017) 440301},
\href{http://arxiv.org/abs/1608.02952}{{\ttfamily arXiv:1608.02952 [hep-th]}}.

\bibitem{nekrasov2006seiberg}
N.~A. Nekrasov and A.~Okounkov, ``Seiberg-witten theory and random partitions,'' in {\em The unity of mathematics}, pp.~525--596.
\newblock Springer, 2006.

\bibitem{Nekrasov:1998ss}
N.~Nekrasov and A.~S. Schwarz, ``{Instantons on noncommutative ${\BR}^{4}$ and $(2,0)$ superconformal six-dimensional theory},'' \href{http://dx.doi.org/10.1007/s002200050490}{{\em Commun. Math. Phys.} {\bfseries 198} (1998) 689--703}, \href{http://arxiv.org/abs/hep-th/9802068}{{\ttfamily arXiv:hep-th/9802068}}.

\bibitem{Haouzi:2020yxy}
N.~Haouzi and J.~Oh, ``{On the Quantization of Seiberg-Witten Geometry},'' \href{http://arxiv.org/abs/2004.00654}{{\ttfamily arXiv:2004.00654 [hep-th]}}.

\bibitem{NO1}
N.~A. Nekrasov and A.~Okounkov, \href{http://dx.doi.org/10.1007/0-8176-4467-9_15}{``{Seiberg--Witten Theory and Random Partitions},''} in {\em The Unity of Mathematics}, P.~Etingof, V.~Retakh, and I.~M. Singer, eds., vol.~244 of {\em Progress in Mathematics}, pp.~525--596.
\newblock Birkh\"auser Boston, 2006.
\newblock
\href{http://arxiv.org/abs/hep-th/0306238}{{\ttfamily arXiv:hep-th/0306238 [hep-th]}}.
\newblock

\bibitem{NaveenNikita}
N.~Nekrasov and N.~S. Prabhakar, ``{Spiked Instantons from Intersecting D-branes},'' \href{http://dx.doi.org/10.1016/j.nuclphysb.2016.11.014}{{\em Nucl. Phys.} {\bfseries B914} (2017) 257--300},
\href{http://arxiv.org/abs/1611.03478}{{\ttfamily arXiv:1611.03478 [hep-th]}}.

\bibitem{Hwang:2014uwa}
C.~Hwang, J.~Kim, S.~Kim, and J.~Park, ``{General instanton counting and 5d SCFT},'' \href{http://dx.doi.org/10.1007/JHEP07(2015)063}{{\em JHEP} {\bfseries 07} (2015) 063}, \href{http://arxiv.org/abs/1406.6793}{{\ttfamily arXiv:1406.6793 [hep-th]}}. [Addendum: JHEP 04, 094 (2016)].

\bibitem{Kim:2012qf}
H.-C. Kim, J.~Kim, and S.~Kim, ``{Instantons on the 5-sphere and M5-branes},'' \href{http://arxiv.org/abs/1211.0144}{{\ttfamily arXiv:1211.0144 [hep-th]}}.

\bibitem{Nekrasov:1995nq}
N.~Nekrasov, ``{Holomorphic bundles and many body systems},'' \href{http://dx.doi.org/10.1007/BF02099624}{{\em Commun. Math. Phys.} {\bfseries 180} (1996) 587--604},
\href{http://arxiv.org/abs/hep-th/9503157}{{\ttfamily arXiv:hep-th/9503157 [hep-th]}}.

\bibitem{Nikita-Pestun-Shatashvili}
N.~Nekrasov, V.~Pestun, and S.~Shatashvili, ``{Quantum geometry and quiver gauge theories},'' \href{http://dx.doi.org/10.1007/s00220-017-3071-y}{{\em Commun. Math. Phys.} {\bfseries 357} (2018) 519--567}, \href{http://arxiv.org/abs/1312.6689}{{\ttfamily arXiv:1312.6689 [hep-th]}}.

\bibitem{Nekrasov:2009uh}
N.~A. Nekrasov and S.~L. Shatashvili, ``{Supersymmetric vacua and Bethe ansatz},'' \href{http://dx.doi.org/10.1016/j.nuclphysbps.2009.07.047}{{\em Nucl. Phys. B Proc. Suppl.} {\bfseries 192-193} (2009) 91--112}, \href{http://arxiv.org/abs/0901.4744}{{\ttfamily arXiv:0901.4744 [hep-th]}}.

\bibitem{SNN}
S.~Jeong, N.~Lee, and N.~Nekrasov, ``{$\hbar$-Langlands correspondence, bispectral duality, and separation of variables from surface defects in $\EuScript{N}=2$ gauge theory},'' {\em To appear} .

\bibitem{Nikita:IV}
N.~Nekrasov, ``{BPS/CFT correspondence IV: sigma models and defects in gauge theory},'' \href{http://dx.doi.org/10.1007/s11005-018-1115-7}{{\em Lett. Math. Phys.} {\bfseries 109} (2019) 579--622},
\href{http://arxiv.org/abs/1711.11011}{{\ttfamily arXiv:1711.11011 [hep-th]}}.

\bibitem{Nakajima:2011yq}
H.~Nakajima, ``{Handsaw quiver varieties and finite W-algebras},'' {\em Moscow Math. J.} {\bfseries 12} (2012) 633,
\href{http://arxiv.org/abs/1107.5073}{{\ttfamily arXiv:1107.5073 [math.QA]}}.

\bibitem{Kanno:2011fw}
H.~Kanno and Y.~Tachikawa, ``{Instanton counting with a surface operator and the chain-saw quiver},'' \href{http://dx.doi.org/10.1007/JHEP06(2011)119}{{\em JHEP} {\bfseries 06} (2011) 119}, \href{http://arxiv.org/abs/1105.0357}{{\ttfamily arXiv:1105.0357 [hep-th]}}.

\bibitem{JB2019}
J.-E. Bourgine and S.~Jeong, ``{New quantum toroidal algebras from 5d $\EuScript{N}=1$ instantons on orbifolds},'' \href{http://dx.doi.org/10.1007/JHEP05(2020)127}{{\em JHEP} {\bfseries 05} (2020) 127},
\href{http://arxiv.org/abs/1906.01625}{{\ttfamily arXiv:1906.01625 [hep-th]}}.

\bibitem{Finkelberg:2010JEMS}
M.~Finkelberg and L.~Rybnikov, ``{Quantization of Drinfeld Zastava in type $A$},'' \href{http://dx.doi.org/10.4171/JEMS/432}{{\em J. Euro. Math. Soc.} {\bfseries 16} no.~2, (2014) 235--271}, \href{http://arxiv.org/abs/1009.0676}{{\ttfamily arXiv:1009.0676 [math.AG]}}.

\bibitem{Feigin:2011SM}
B.~Feigin, M.~Finkelberg, A.~Negut, and L.~Rybnikov, ``{Yangians and cohomology rings of Laumon spaces},'' \href{http://dx.doi.org/10.1007/s00029-011-0059-x}{{\em Selecta Math.} {\bfseries 17} (2011) 573--607}, \href{http://arxiv.org/abs/0812.4656}{{\ttfamily arXiv:0812.4656 [math.AG]}}.

\bibitem{Bruzzo:2010fk}
U.~Bruzzo, W.~y. Chuang, D.~E. Diaconescu, M.~Jardim, G.~Pan, and Y.~Zhang, ``{D-branes, surface operators, and ADHM quiver representations},'' \href{http://dx.doi.org/10.4310/ATMP.2011.v15.n3.a6}{{\em Adv. Theor. Math. Phys.} {\bfseries 15} no.~3, (2011) 849--911}, \href{http://arxiv.org/abs/1012.1826}{{\ttfamily arXiv:1012.1826 [hep-th]}}.

\bibitem{Jeong:2018qpc}
S.~Jeong and N.~Nekrasov, ``{Opers, surface defects, and Yang-Yang functional},'' \href{http://dx.doi.org/10.4310/ATMP.2020.v24.n7.a4}{{\em Adv. Theor. Math. Phys.} {\bfseries 24} no.~7, (2020) 1789--1916}, \href{http://arxiv.org/abs/1806.08270}{{\ttfamily arXiv:1806.08270 [hep-th]}}.

\bibitem{Kimura:2022zsx}
T.~Kimura and N.~Lee, ``{Defect in gauge theory and quantum Hall states},'' \href{http://dx.doi.org/10.1016/j.nuclphysb.2023.116218}{{\em Nucl. Phys. B} {\bfseries 991} (2023) 116218}, \href{http://arxiv.org/abs/2210.05949}{{\ttfamily arXiv:2210.05949 [hep-th]}}.

\bibitem{Kim:2023qwh}
H.-C. Kim, M.~Kim, S.-S. Kim, and G.~Zafrir, ``{Superconformal indices for non-Lagrangian theories in five dimensions},'' \href{http://arxiv.org/abs/2307.03231}{{\ttfamily arXiv:2307.03231 [hep-th]}}.

\bibitem{bogoyavlensky1976perturbations}
O.~I. Bogoyavlensky, ``On perturbations of the periodic toda lattice,'' {\em Communications in Mathematical Physics} {\bfseries 51} (1976) 201--209.

\bibitem{Martinec:1995by}
E.~J. Martinec and N.~P. Warner, ``{Integrable systems and supersymmetric gauge theory},'' \href{http://dx.doi.org/10.1016/0550-3213(95)00588-9}{{\em Nucl. Phys.} {\bfseries B459} (1996) 97--112},
\href{http://arxiv.org/abs/hep-th/9509161}{{\ttfamily arXiv:hep-th/9509161 [hep-th]}}.

\bibitem{Gorsky:1999gx}
A.~Gorsky and A.~Mironov, ``{Solutions to the reflection equation and integrable systems for N=2 SQCD with classical groups},'' \href{http://dx.doi.org/10.1016/S0550-3213(99)00134-0}{{\em Nucl. Phys. B} {\bfseries 550} (1999) 513--530}, \href{http://arxiv.org/abs/hep-th/9902030}{{\ttfamily arXiv:hep-th/9902030}}.

\bibitem{Rui-dong1}
S.~Nawata and R.-D. Zhu, ``{Instanton counting and O-vertex},'' \href{http://dx.doi.org/10.1007/JHEP09(2021)190}{{\em JHEP} {\bfseries 09} (2021) 190}, \href{http://arxiv.org/abs/2107.03656}{{\ttfamily arXiv:2107.03656 [hep-th]}}.

\bibitem{Nawata:2023wnk}
S.~Nawata, K.~Zhang, and R.-D. Zhu, ``{ABCD of qq-characters},'' \href{http://arxiv.org/abs/2302.00525}{{\ttfamily arXiv:2302.00525 [hep-th]}}.

\bibitem{Aharony:1997bh}
O.~Aharony, A.~Hanany, and B.~Kol, ``{Webs of (p,q) five-branes, five-dimensional field theories and grid diagrams},'' \href{http://dx.doi.org/10.1088/1126-6708/1998/01/002}{{\em JHEP} {\bfseries 01} (1998) 002}, \href{http://arxiv.org/abs/hep-th/9710116}{{\ttfamily arXiv:hep-th/9710116}}.

\bibitem{Gomis:2006sb}
J.~Gomis and F.~Passerini, ``{Holographic Wilson Loops},'' \href{http://dx.doi.org/10.1088/1126-6708/2006/08/074}{{\em JHEP} {\bfseries 08} (2006) 074}, \href{http://arxiv.org/abs/hep-th/0604007}{{\ttfamily arXiv:hep-th/0604007}}.

\end{thebibliography}\endgroup

\end{document}